\documentclass[a4paper, amsfonts, amssymb, amsmath, reprint, showkeys, nofootinbib, twoside, superscriptaddress, twocolumn]{revtex4-1}
\usepackage{float}
\usepackage{xr}
\usepackage[english]{babel}
\usepackage[utf8]{inputenc}
\usepackage[colorinlistoftodos, color=green!40, prependcaption]{todonotes}
\usepackage{placeins}
\usepackage{mhchem} 
\usepackage{xr}

\usepackage[pdftex, pdftitle={Article}, pdfauthor={Author}]{hyperref} % For hyperlinks in the PDF
\usepackage{siunitx}
\usepackage{xr-hyper}
\usepackage{upgreek}
\usepackage{hyperref}
\usepackage{cleveref}
\begin{document}

%\externaldocument{SI_text.tex}

\title{\textcolor{black}{Quantum Spin Hall Effect in Magnetic Graphene}}

\author{Talieh S. Ghiasi }
    \email[Correspondence email address:]{t.s.ghiasi@tudelft.nl}% Your name
    \affiliation{Kavli Institute of Nanoscience,
Delft University of Technology, Lorentzweg 1, 2628 CJ Delft, The
Netherlands}
\affiliation{Department of Physics, Harvard University, Cambridge, Massachusetts 02138, USA}
\author{Davit Petrosyan}
    \affiliation{Kavli Institute of Nanoscience,
Delft University of Technology, Lorentzweg 1, 2628 CJ Delft, The
Netherlands}
\author{Josep Ingla-Ayn\'{e}s}
    \affiliation{Kavli Institute of Nanoscience,
Delft University of Technology, Lorentzweg 1, 2628 CJ Delft, The
Netherlands}
\author{Tristan Bras}
    \affiliation{Kavli Institute of Nanoscience,
Delft University of Technology, Lorentzweg 1, 2628 CJ Delft, The
Netherlands}

\author{Kenji Watanabe}\affiliation{Research Center for Electronic and Optical Materials, National Institute for Materials Science, 1-1 Namiki, Tsukuba 305-0044, Japan}
\author{Takashi Taniguchi}\affiliation{Research Center for Materials Nanoarchitectonics, National Institute for Materials Science,  1-1 Namiki, Tsukuba 305-0044, Japan}
\author{Samuel~Ma\~{n}as-Valero}\affiliation{Kavli Institute of Nanoscience, Delft University of Technology, Lorentzweg 1, 2628 CJ Delft, The Netherlands}\affiliation{Institute of Molecular Science, University of Valencia, Catedrático José Beltrán 2, Paterna 46980, Spain}
\author{Eugenio~Coronado}\affiliation{Institute of Molecular Science, University of Valencia, Catedrático José Beltrán 2, Paterna 46980, Spain}

\author{Klaus Zollner} \affiliation{Institute for Theoretical Physics, University of Regensburg, 93040 Regensburg, Germany}

\author{Jaroslav Fabian} \affiliation{Institute for Theoretical Physics, University of Regensburg, 93040 Regensburg, Germany}

\author{Philip Kim} \affiliation{Department of Physics, Harvard University, Cambridge, Massachusetts 02138, USA}

\author{Herre~S.~J. van der Zant} \affiliation{Kavli Institute of Nanoscience,
Delft University of Technology, Lorentzweg 1, 2628 CJ Delft, The
Netherlands}

\date{\today} % Leave empty to omit a date

\begin{abstract}

%A promising approach to attain long-distance coherent spin propagation is accessing quantum Hall (QH) topological spin-polarized edge states in graphene.~Achieving this without external magnetic fields necessitates engineering graphene band structure, obtainable through proximity to magnetic materials. In particular, due to proximity effects, a topological bulk gap is expected to form in graphene, along with gapless spin-polarized edge states that are robust against disorder. In this work, we detect spin-polarized helical edge transport in graphene at zero external magnetic field, allowed by the proximity of CrPS$_4$. The emergence of the quantum anomalous spin Hall (ASH) state is assisted by the induced spin-orbit and exchange couplings in the graphene that also give rise to a large anomalous Hall (AH) effect signal. The observation of spin-polarized helical edge states at zero external magnetic field, together with the robustness of the AH effect up to room temperature, opens the route for practical applications of magnetic graphene in quantum spintronic circuitries. 

\textcolor{black}{A promising approach to attain long-distance coherent spin propagation is accessing topological spin-polarized edge states in graphene.~Achieving this without external magnetic fields necessitates engineering graphene band structure, obtainable through proximity effects in van der Waals heterostructures. In particular, proximity-induced staggered potentials and spin-orbit coupling are expected to form a topological bulk gap in graphene with gapless helical edge states that are robust against disorder. In this work, we detect the spin-polarized helical edge transport in graphene at zero external magnetic field, allowed by the proximity of an interlayer antiferromagnet, CrPS$_4$. We show the coexistence of the quantum spin Hall (QSH) states and magnetism in graphene, where the induced spin-orbit and exchange couplings also give rise to a large anomalous Hall (AH) effect. The detection of the QSH states at zero external magnetic field, together with the AH signal that persists up to room temperature, opens the route for practical applications of magnetic graphene in quantum spintronic circuitries.} %\textcolor{black}{in the absence of an applied magnetic field}.

\end{abstract}

\maketitle

Decades of research in graphene have unveiled its remarkable charge and spin-related properties, such as distinctive quantum Hall (QH) charge transport and long-distance spin propagation~\cite{neto2009electronic, han2014graphene}. Yet, harnessing \textcolor{black}{topologically protected spin transport }for the development of practical quantum spintronic devices necessitates targeted modifications to the graphene band structure. This band structure engineering has been recently achieved non-invasively by bringing graphene in the proximity of other two-dimensional (2D) materials~\cite{sierra2021van}, leading to spin Hall~\cite{wei2016strong, safeer2019room}, Rashba-Edelstein~\cite{mendes2015spin, ghiasi2019charge, benitez2020tunable}, AH~\cite{wang2015proximity}, and spin-dependent Seebeck~\cite{ghiasi2021electrical} effects in graphene. The realization of these phenomena in proximitized graphene is due to induced spin-orbit and exchange interactions that \textcolor{black}{result in} spin-splitting of the graphene band structure~\cite{geim2013van, yang2013proximity, dyrdal2017anomalous}. These spin-related effects, however, have been experimentally addressed in the proximitized graphene majorly by diffusive charge/spin transport, where the spin relaxation length is limited by spin and momentum scattering mechanisms. On the contrary, if the transport occurs through topological spin-polarized edge states, their topological protection allows for long-distance quantum coherent spin propagation.

\begin{figure*} 
    \centering
    \includegraphics[width=0.92\textwidth]{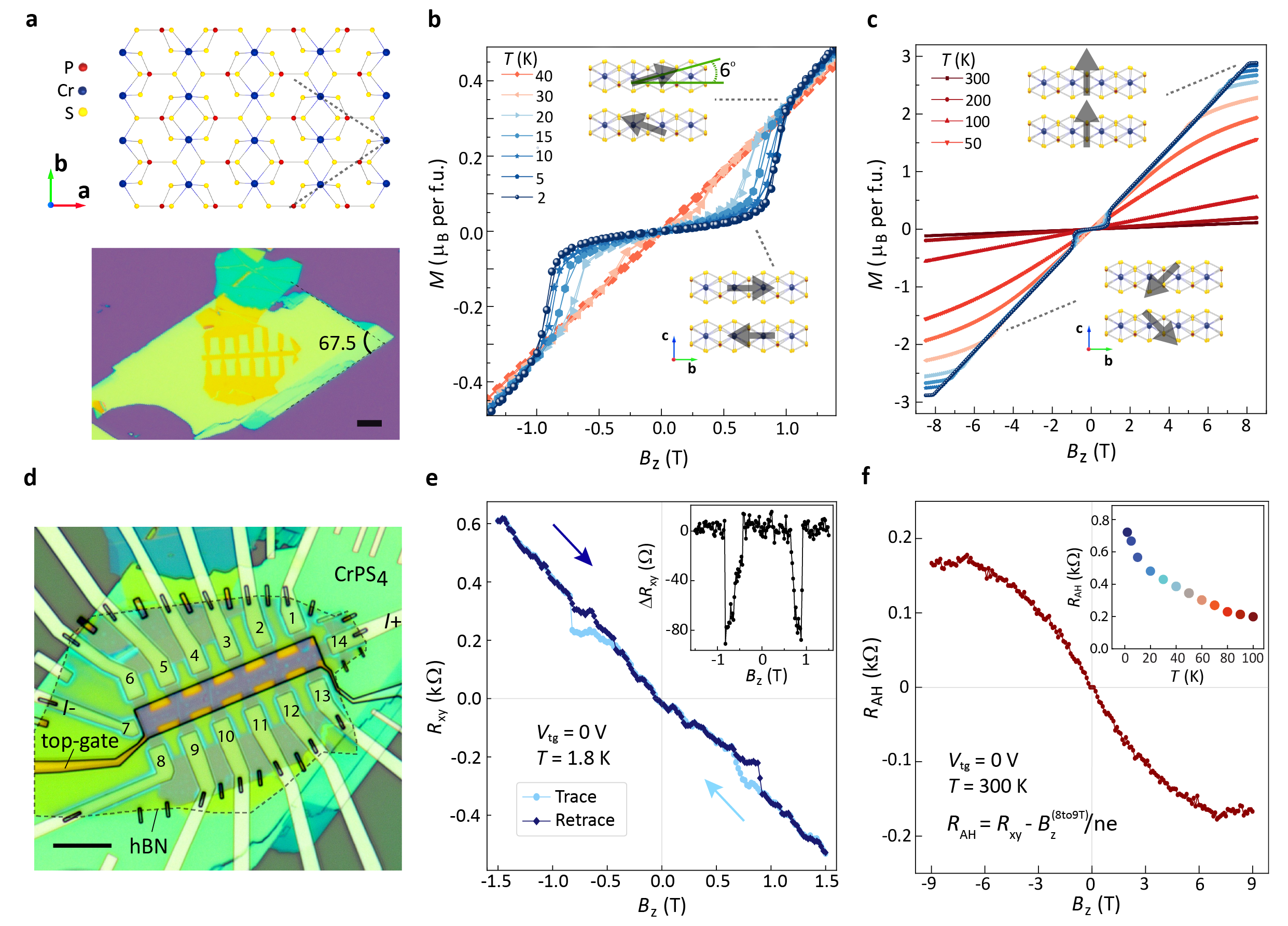}
    \caption{\textbf{CrPS$_4$ (CPS) magnetic ordering and induced magnetism in graphene}. (a) Top-view ($a-b$ plane) CPS crystal structure schematics showing monoclinic symmetry with space group C2 and lattice constants $a$ = \SI{10.871}{\r{A}}, $b$ = \SI{7.254}{\r{A}}, and $c$ = \SI{6.140}{\r{A}}~\cite{diehl1977crystal}. The \textcolor{black}{optical micrograph (with the scale bar of $\SI{10}{\micro m}$) shows a \SI{38}{nm} thick }CPS \textcolor{black}{flake }covered with an hBN/graphene Hall-bar, as described in the supplementary information (SI), \Cref{Fig:fab}. (b,c) SQUID magnetometry of a CPS bulk crystal at various temperatures, respectively shown in panels b and c for the small and large range of magnetic field $B_\mathrm{z}$ applied perpendicular to the ($a-b$) plane. \textcolor{black}{The $B_\mathrm{z}$ cycling in the SQUID measurements shows a minimal hysteresis of the bulk CPS magnetization.} The insets are side-view schematics of CPS layers representing the layers' magnetic ordering in bulk CPS for each indicated range of $B_\mathrm{z}$. (d) Optical micrograph of device A fabricated with an hBN/graphene/CPS heterostructure on a SiO$_2$/Si substrate with Ti-Pd electrodes. The scale bar is $\SI{10}{\micro m}$. The CPS crystal used for the device fabrication is from the same source that is explored by the SQUID magnetometry (of panels b and c). The thickness of the hBN and SiO$_2$, used as top-gate and back-gate dielectrics, is 73 and \SI{285}{nm}, respectively. (e) Transverse resistance $R_\mathrm{xy} = V_\mathrm{xy}/I$ measured vs.~$B_\mathrm{z}$ in the graphene Hall-bar at $T = \SI{1.8}{K}$ with $I = \SI{0.5}{\micro A}$. The light blue curve is measured from 1.5 to \SI{-1.5}{T} (trace) and the dark blue curve is measured from $-1.5$ to \SI{1.5}{T} (retrace). This measurement is performed at back-gate and top-gate voltages of $V_\mathrm{bg} = \SI{50}{V}$ and $V_\mathrm{tg} = \SI{0}{V}$. Note that the non-linearity of $R_\mathrm{xy}$ at $B_\mathrm{z}< \SI{1}{T}$ is due to the non-linear magnetization behaviour of the CPS in that range \textcolor{black}{(see \Cref{fig:Rxy_nonlinear})}. The inset is the subtraction of the trace and retrace measurements ($\Delta R_\mathrm{xy}$) showing the modulation of $R_\mathrm{xy}$ due to the spin-flop transition. (f) Non-linear component of $R_\mathrm{xy}$ associated with the AH effect ($R_\mathrm{AH}$), measured at $T = \SI{300}{K}$, $V_\mathrm{bg} = \SI{50}{V}$ and $V_\mathrm{tg} = \SI{0}{V}$. $R_\mathrm{AH}$ is acquired by subtraction of a linear background (related to the ordinary Hall effect) from $R_\mathrm{xy}$ anti-symmetrized vs.~$B_\mathrm{z}$ (also see SI \Cref{Fig:RTAHE}). For the linear fit, the $R_\mathrm{xy}$ data in the range of $\SI{8}{T} < B_\mathrm{z}\le\SI{9}{T}$ is considered, assuming saturation of the magnetization in that range. Inset: $R_\mathrm{AH}$ at various temperatures extracted for $V_\mathrm{tg}=\SI{0}{V}$.}
    \label{Fig:Fig.1}
\end{figure*}

Spin-polarized \textcolor{black}{chiral and helical }QH edge states have been detected in graphene through the Zeeman splitting of the bands by applying large magnetic fields~\cite{young2012spin, young2014tunable, veyrat2020helical}\textcolor{black}{, which is conceived to be practically challenging for applications. Thus, numerous theoretical and experimental reports so far focus on resolving spin-polarized edge states in graphene without the need for any external magnetic field, possible through the emergence of quantum spin Hall (QSH) or quantum anomalous Hall (QAH) states in a modified graphene band structure~\cite{haldane1988model, kane2005quantum, qiao2010quantum, yang2011time,  Qiao2014:PRL, Kaloni2014:APL, hatsuda2018evidence, offidani2018anomalous, hogl2020quantum, vila2021valley, bora2022magnetic, obata2024coexistence}. These spin-polarized gapless edge states can form within the bulk gap of graphene, attainable by inducing staggered potentials, spin-orbit coupling and/or magnetic exchange interactions~\cite{haldane1988model, kane2005quantum, qiao2010quantum}. \textcolor{black}{Depending on the respective magnitude of the spin-orbit vs. exchange interactions,} these edge states can be \textcolor{black}{chiral or helical~\cite{yang2011time}, allowing for} topologically protected spin transport that is expected to be robust against disorder~\cite{kane2005quantum}.  }

Despite several recent experimental efforts in addressing the topological transport in various graphene-based van der Waals heterostructure~\cite{song2018electrical,song2018asymmetric, wu2020large, wu2021magnetic, chau2022two, hu2023tunable}, direct experimental evidence for the detection of the QAH and QSH effects in proximitized graphene is still missing. One major obstacle \textcolor{black}{has been} the presence of dominant interfacial charge transfer~\cite{wang2022quantum, tseng2022gate} \textcolor{black}{that }may hinder the exploration of the proximity effects. 
\textcolor{black}{However, the suppression of the interfacial charge transfer in graphene-CrPS$_4$ heterostructure allows us to study the transport in the presence of magnetic exchange and spin-orbit interactions, evidenced by the detection of a large AH signal up to room temperature. Remarkably, we experimentally realize the presence of helical states at zero external magnetic field, indicating the emergence of the QSH effect despite the breaking of time-reversal symmetry by the induced magnetism~\cite{yang2011time, wang2017dirac}. The unprecedented zero-magnetic-field detection of the QSH state in this graphene-based magnetic heterostructure, coexisting with the AH effect, makes this system intriguing for the development of quantum spintronic circuitries.}%\mbox{~\cite{qiao2010quantum, zhang2015robust, hogl2020quantum}} 

We bring graphene in the van der Waals proximity of the interlayer antiferromagnet,~CrPS$_4$ (CPS), which is an air-stable semiconductor with a bandgap of $\sim$ 1.3~eV~\cite{lee2017structural} and a N\'eel temperature $T_\mathrm{N}\approx$ 38~K~\cite{peng2020magnetic}. Figure~\ref{Fig:Fig.1}a (top panel) illustrates a top view of the CPS crystal structure with the dashed lines indicating the crystallographic directions along which the crystals preferentially cleave, as shown in the optical micrograph of an exfoliated flake. As a result, the CPS flakes acquire a corner angle of \SI{67.5}{\degree}, the bisector of which is along the magnetic $a$-axis~\cite{lee2017structural} (see also supplementary information (SI), \Cref{Fig:CPSmag}). The presence of this characteristic angle assists with the identification of the CPS crystallographic orientation and guides the alignment of the graphene Hall bar with one of the CPS magnetic axes (see Figure 1a, lower panel). In accordance with previous reports~\cite{peng2020magnetic}, CPS has an anisotropic magnetic behavior which is evaluated using a superconducting quantum interference device (SQUID) at various temperatures, shown in Figure~\ref{Fig:Fig.1}b and c (see Methods for more details on CPS). %The magnetic ordering of a bulk CPS crystal at $B_\mathrm{z}=$ \SI{0}{T} is expected to be mainly out-of-plane \textcolor{black}{and canted} \SI{20}{\degree} towards the in-plane $a$-axis.
The SQUID measurements at the small-\textcolor{black}{range} $B_\mathrm{z}$ (panel b) for $T<\SI{40}{K}$, show the expected spin-flop transition when the magnetization of the layers rotates from canted out-of-plane towards the in-plane $b$-axis, while holding their respective antiparallel alignment (lower inset \textcolor{black}{in Figure~\ref{Fig:Fig.1}b}). Increasing $B_\mathrm{z}$ above the spin-flop transition field ($B_\mathrm{sf}\sim \SI{0.8}{T}$ at \SI{2}{K}) results in canting of the magnetic moments towards the $c$-axis, starting with $\sim6^{\circ}$ canting (upper inset in panel b) until full saturation ($90^{\circ}$) at $B_\mathrm{z} \approx \SI{8}{T}$, as shown in Figure~\ref{Fig:Fig.1}c.

\begin{figure*} 
    \centering
    \includegraphics[width=\textwidth]{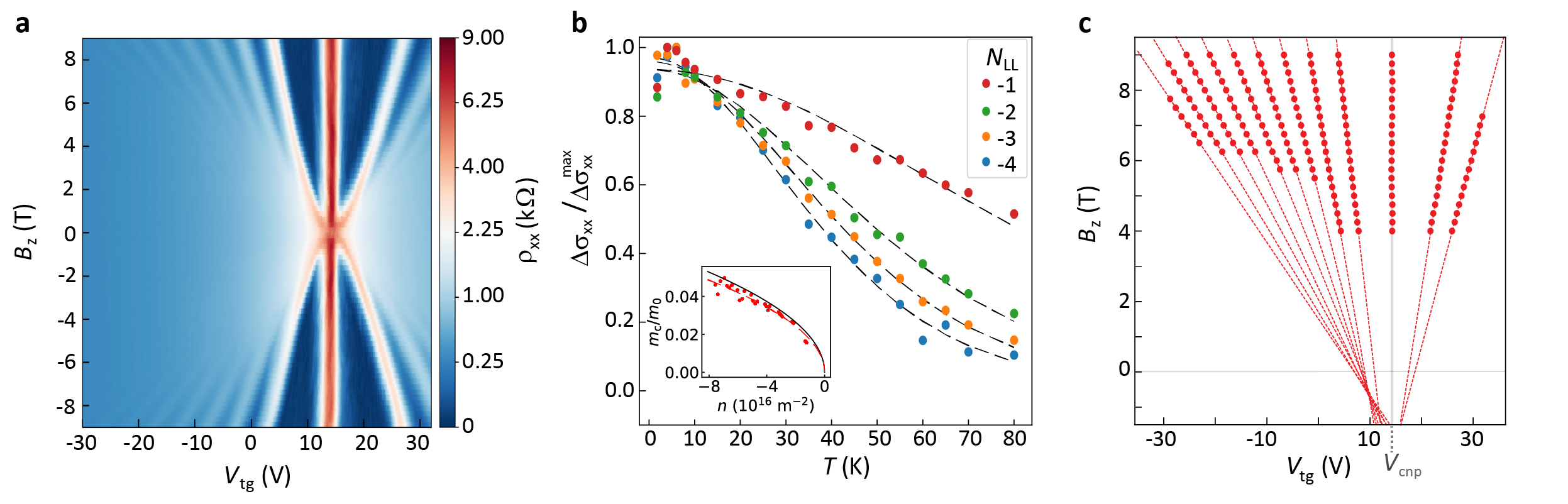}
    \caption{\textbf{Quantum Hall transport in magnetized graphene}. (a) Shubnikov–de Haas oscillations of the longitudinal resistivity $\rho_\mathrm{xx}$ in graphene, measured vs.~applied top-gate voltage and magnetic field \textcolor{black}{(using voltage probes V$_5$ and V$_6$, shown in \Cref{Fig:Fig.1}d)}. %Inset: Maxima of the $\rho_\mathrm{xx}$ vs. carrier density ($n$) with the red lines as the linear fits.
    (b) Amplitude of the SdHOs ($\Delta \sigma_\mathrm{xx}$) for LL with $N_\mathrm{LL} = -1$ to $-4$ as a function of $T$, normalized to their maxima. Dashed lines are the fits to the data using $\Delta\sigma_\mathrm{xx}=T/\sinh(2 \pi^2 k_\mathrm{B} T m_\mathrm{c} / \hbar e B_\mathrm{z})$. Inset: cyclotron mass, $m_\mathrm{c}$, extracted from all the \textcolor{black}{hole-like} LLs, plotted as the ratio with respect to the electron rest mass ($m_0$) vs.~carrier density ($n$). The red curve is a fit to the extracted $m_\mathrm{c}/m_\mathrm{0}$ vs.~$n$, considering a linear dispersion relation, resulting in \textcolor{black}{Fermi velocity,} $v_\mathrm{F} \approx 1.196 \times 10^6~\mathrm{m/s}$. The black curve is the theoretically expected behavior for pristine graphene with $v_\mathrm{F}= 1.1 \times 10^6~\mathrm{m/s}$. The effective $n$ in the graphene channel is extracted from Hall measurements (for $B_\mathrm{z}\geq\SI{8}{T}$), and separately from the dependence of $B_\mathrm{z}$ vs.~$V_\mathrm{tg}$ for each LL. These two approaches give rise to similar extracted values for $n$ (see SI, Section 8). (c) Landau fan diagram: the red dots are the maxima of the $\rho_\mathrm{xx}$ vs.~$V_\mathrm{tg}$ extracted from panel a for gate voltages slightly away from the charge neutrality point ($|V_\mathrm{tg}-V_\mathrm{cnp}|>\SI{5}{V}$), with the red dashed lines as linear fits. }
    \label{Fig:QHE}
\end{figure*}

\begin{figure*} 
    \centering
    \includegraphics[width=0.92\textwidth]{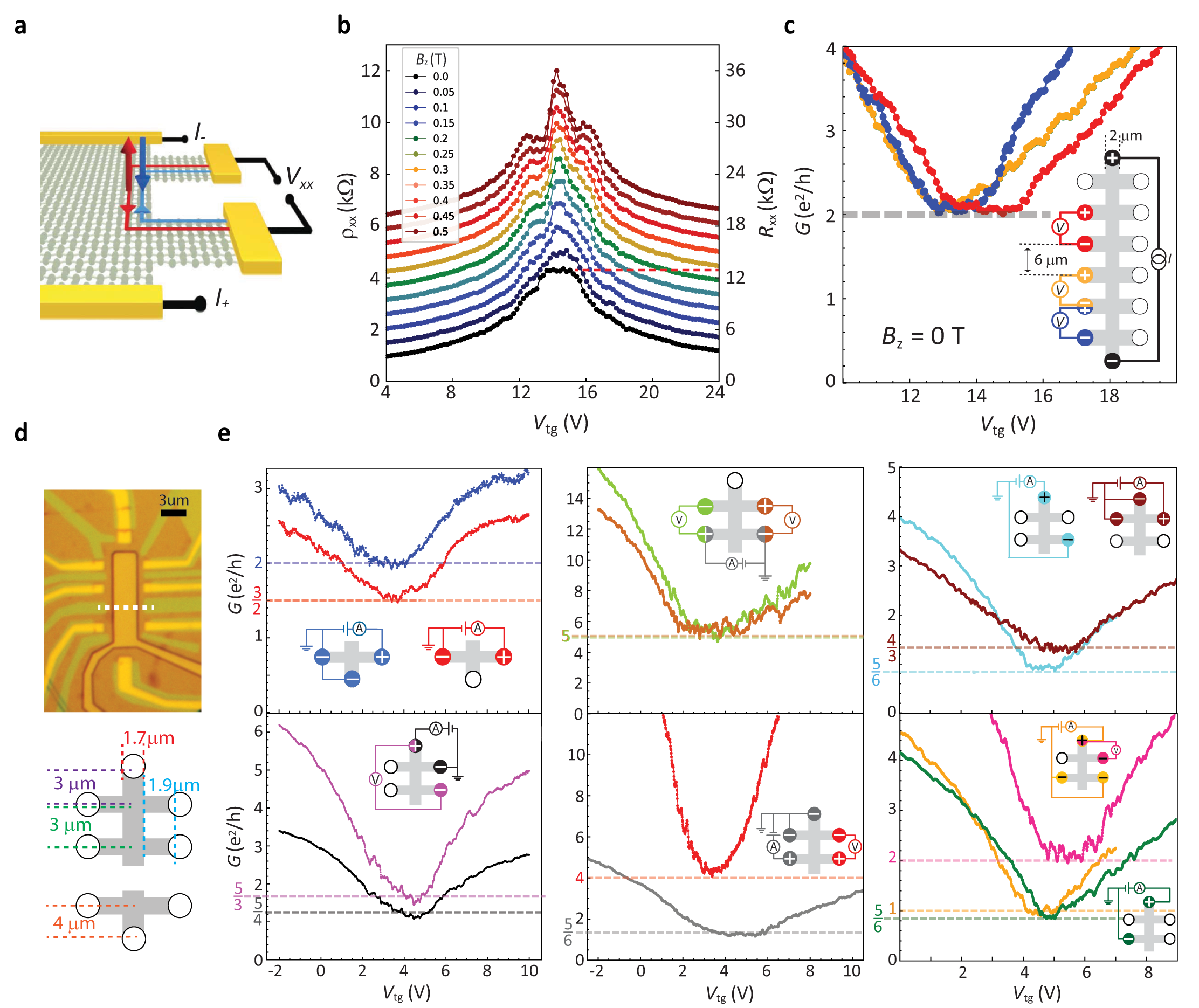}
    \caption{\textbf{Quantum spin Hall states in magnetized graphene near zero energy.} (a) A schematic of the device and a four-terminal measurement geometry are shown with the propagation direction of the helical states at the edge of the graphene channel. (b) $V_\mathrm{tg}$-dependence of the \textcolor{black}{graphene resistivity, $\rho_\mathrm{xx}$, (left axis), and resistance, $R_\mathrm{xx}$ (right axis),} at $B_\mathrm{z} = 0$ to \SI{0.5}{T}. The measurements are performed using the first voltage pair as in \Cref{Fig:QHE}a. The plots are offset (except for $B_\mathrm{z} = \SI{0}{T}$) for a clearer visualization of the SdHOs. \textcolor{black}{The red dashed line is to highlight the value of the $R_\mathrm{xx}\approx 13~$k$\Omega$ at the charge neutrality point ($V_\mathrm{cnp}$).} (c) Conductance ($G= I/V_\mathrm{xx}$) vs. $V_\mathrm{tg}$ close to the charge neutrality point at $B_\mathrm{z}= \SI{0}{T}$, measured in the four-terminal configurations shown in the schematics of the device in the inset \textcolor{black}{(with the channel length and width of 6 and 2 $\mu m$, respectively)}. (d) Optical micrograph and schematics of device C that has \textcolor{black}{two separate} graphene channels with a different aspect ratio compared with that of devices A and B \textcolor{black}{(the dimensions of the graphene channels are indicated in the schematics)}. The white dashed line in the optical image indicates the position where the graphene channel is cut to form two electrically isolated regions with 3 and 5 electrodes. (e) Two-, three- and four-terminal conductance in device C, measured at various configurations shown in the schematics of the device in the insets. All measurements are performed at $B_\mathrm{z}=\SI{0}{T}$ at $T = \SI{2.3}{K}$. The dashed lines indicate the theoretically expected values for conductance considering helical states, color-coded with respect to the measurement geometry indicated in the inset.}
    \label{fig:QSHS}
\end{figure*}

We evaluate the possibility of inducing the CPS magnetic properties in graphene by transport measurements. We report on \textcolor{black}{four} devices (\textcolor{black}{A-D}), each fabricated with a monolayer graphene Hall-bar on a multilayer CPS flake, fully covered with hBN that is used as a dielectric for top-gating (Figure~\ref{Fig:Fig.1}d). We apply current along the graphene channel (with $I+$ and $I-$ electrodes shown in panel d) while measuring the longitudinal and transverse voltages ($V_\mathrm{xx}$ and $V_\mathrm{xy}$). In Figure~\ref{Fig:Fig.1}e, we show the modulation of the transverse resistance $R_\mathrm{xy}\,=\,V_\mathrm{xy}/I$ as a function of $B_\mathrm{z}$, at $V_\mathrm{tg} = \SI{0}{V}$. \textcolor{black}{In addition to the slight non-linear behaviour at $|B_\mathrm{z}|< \SI{1}{T}$ (also see \Cref{fig:Rxy_nonlinear}), we observe switches in} $R_\mathrm{xy}$ at $\sim\pm\SI{0.8}{T}$, which is the magnetic field at which the spin-flop transition occurs in CPS ($B_\mathrm{sf}$). This is a direct indication that the magnetic behavior of the CPS influences the charge transport in graphene, \textcolor{black}{which is detected here by the AH effect}.
 
The trace and retrace measurements of $R_\mathrm{xy}$ show no hysteresis vs.~$B_\mathrm{z}$, except for $B_\mathrm{z}$ close to $B_\mathrm{sf}$. \textcolor{black}{This hysteresis is absent in the SQUID measurements of the bulk crystal, implying the sensitivity of the AH effect to the behaviour of the magnetization of the outer-most layer of the CPS ($\vec{M}_\mathrm{CPS}$). Under applied $B_\mathrm{z}$,} the behavior of $\vec{M}_\mathrm{CPS}$ is \textcolor{black}{expected to be} ruled by its antiferromagnetic coupling with respect to the other CPS layers, and yet, the lack of a neighboring CPS layer on top can cause the $\vec{M}_\mathrm{CPS}$ to lag behind the magnetization dynamics of the bulk of CPS, leading to the observed hysteresis. By decreasing $B_\mathrm{z}$, $\vec{M}_\mathrm{CPS}$ \textcolor{black}{undergoes a spin-flop transition at} $\sim\,$0.5 T, by which $R_\mathrm{xy}$ in the retrace retrieves its initial value as in the trace measurement. The difference of $R_\mathrm{xy}$ in the trace and retrace measurements is about 80$\,\Omega$ (shown in the inset of Figure~\ref{Fig:Fig.1}e), \textcolor{black}{which is the AH effect signal generated by} the spin-flop transition of the $\vec{M}_\mathrm{CPS}$.

%These measurements not only provide direct evidence for the induced magnetism and spin-orbit coupling in graphene, but also establish AH effect as a sensitive tool to probe the magnetization behaviour of the outermost layer of the layered magnetic materials.
 
The AH \textcolor{black}{resistance ($R_\mathrm{AH}$) is dependent on} the projection of the magnetization $\vec{M}_\mathrm{CPS}$ along the $c$-axis. From the comparison of the magnetization values measured by SQUID at $B_\mathrm{z}\sim$ 0.8 and 8 T (in \Cref{Fig:Fig.1}c), one can tell that through the spin-flop transition, the $\vec{M}_\mathrm{CPS}$ cants towards the $c$-axis only by 6 degrees, which has led to $R_\mathrm{AH} \approx 80\,\Omega$. Considering that and the linear correlation between the $R_\mathrm{AH}$ and $z$-component of $\vec{M}_\mathrm{CPS}$, $R_\mathrm{AH}$ should be about 700 $\Omega$ when $\vec{M}_\mathrm{CPS}$ is fully saturated out-of-plane, consistent with the value we measure at large $B_\mathrm{z}$ (see SI, \Cref{fig:AHE_Vg0V_vsT}). Further evaluation of the AH effect at various temperatures is shown in \Cref{Fig:Fig.1}f (inset), indicating a considerable decay vs. $T$. Yet, there is a finite value of $R_\mathrm{AH}$ of about $150\,\Omega$ that persists up to $T = \SI{300}{K}$. Note that at elevated temperatures CPS is a paramagnet, thus a finite $B$ is required to induce magnetism in graphene. The detected $R_\mathrm{AH}$ in this heterostructure is considerably larger than that achieved so far in proximitized graphene~\cite{wang2015proximity,tang2018approaching,ghiasi2021electrical}. The observation of such a sizable $R_\mathrm{AH}$ is a signature of the co-presence of strong induced spin-orbit and exchange interactions~\cite{nagaosa2010anomalous} in the proximitized graphene up to room temperature, \textcolor{black}{which is consistent with the recently reported enhanced Curie temperature at the graphene-CPS interface~\cite{zhu2023interface}}. \textcolor{black}{The detection of the unprecedentedly large AH effect in this system is also expected from the strong hybridization between the graphene and CPS bands, shown by our density functional theory (DFT) calculations} \textcolor{black}{(see SI, Section 15).}

\textcolor{black}{Having established the proximity-induced magnetism in the graphene channel, we explore the QH regime of transport in this system. }By cooling down the device, signatures of QH transport become evident at finite $B_\mathrm{z}$, as shown in \Cref{Fig:QHE}. Panel a shows the modulation of the longitudinal resistivity ($\rho_\mathrm{xx}$) vs.~$B_\mathrm{z}$, i.e., the Shubnikov de-Haas oscillations (SdHOs). The back-gate voltage ($V_\mathrm{bg}$) is set to $\SI{50}{V}$ to shift the Dirac point such that the electron-like Landau levels (LLs) are reachable by the top-gating. We evaluate the back-gate vs.~top-gate dependence of the LLs in \Cref{fig:double_gate}, indicating only a shift of the LLs by electrostatic doping without any significant change in the Landau fan diagram with $V_\mathrm{bg}$ (SI,~\Cref{fig:Back-gate-effect}). \textcolor{black}{Moreover, there is a negligible shift of the Dirac peak vs.~$B_\mathrm{z}$ (SI,~\Cref{fig:shift_of_CNP_vs_B}) and there is a minimal hysteresis for the $V_\mathrm{tg}$-dependence of $\rho_\mathrm{xx}$, which decreases with raising temperature~(SI,~\Cref{fig:hysteresis3}). These observations are} in contrast with recent reports on graphene in the proximity of CrX$_3$~\cite{tseng2022gate} or CrOCl~\cite{wang2022quantum}, where gate-dependent interfacial charge transfer influences the transport in graphene. 

\textcolor{black}{From the QH transport, we obtain the cyclotron mass $m_\mathrm{c}$ and the Fermi velocity $v_\mathrm{F}$ of the electrons in the magnetized graphene.}~Having the temperature-dependence of \textcolor{black}{the modulations of the longitudinal conductivity} $\Delta\sigma_\mathrm{xx}$ (see SI, \Cref{fig:Rxypair}b), $m_\mathrm{c}$ is extracted by fitting $\Delta\sigma_\mathrm{xx}$ to the standard expression $T/\sinh(2 \pi^2 k_\mathrm{B} T m_\mathrm{c} / \hbar e B_\mathrm{z})$, with $k_\mathrm{B}$ the Boltzmann constant~\cite{novoselov2005two, zhang2005experimental},  shown for the first few hole-like LLs in Figure~\ref{Fig:QHE}b. The ratio of the cyclotron mass vs.~the rest mass of the electrons ($m_\mathrm{c}$/$m_\mathrm{0}$) extracted from the SdHOs at $\SI{8}{T}<|B_\mathrm{z}|\le \SI{9}{T}$ for different carrier density ($n$) is shown in the inset. The extracted $m_\mathrm{c}$/$m_\mathrm{0}$ ratio \textcolor{black}{and the $v_\mathrm{F}$ determined from the fit vs.~$n$ show consistency with} that expected for the case of pristine graphene with $v_\mathrm{F}= 1.1 \times 10^6\,\mathrm{m/s}$ (indicated by the black solid line in the inset)~\cite{novoselov2005two, zhang2005experimental}. %\textcolor{black}{We note that under the dominance of the interfacial charge transfer in such heterostructures, the $v_\mathrm{F}$ is expected to increase with respect to that in pristine graphene\cite{wang2022quantum}, which is not the case in graphene-CPS.}  

However, the Landau fan diagram in the \textcolor{black}{graphene-CPS} (Figure~\ref{Fig:QHE}a) appears to be distinctive from that of pristine graphene, showing \textcolor{black}{shifts} of the LLs in $V_\mathrm{tg}$ and minimal broadening of the zeroth LL (zLL) vs.~$B_\mathrm{z}$. A slight non-linearity of the first Landau fans is present close to zero energy which could be related to a small non-linear dependence of $n$ vs.~$V_\mathrm{tg}$ close to the charge neutrality point ($V_\mathrm{cnp}$). This nonlinearity is likely related to a small contribution of CPS localized midgap states in the total capacitance at low density of states in graphene, particularly at large positive $V_\mathrm{tg}$ when the Fermi energy $E_\mathrm{F}$ is expected to get closer to the conduction band edge in the CPS. In \Cref{Fig:QHE}c, we show the maxima of $\rho_\mathrm{xx}$ vs.~$V_\mathrm{tg}$ for $|V_\mathrm{tg}-V_\mathrm{cnp}|>\SI{5}{V}$, %and also vs.~$n$ (in the inset) 
together with the linear fits, clearly showing the linear behavior of the Landau fans. That also rules out the dominant contribution of interfacial charge transfer that is expected to cause nonlinearities in Landau fans due to its dependence on magnetic ordering~\cite{wang2022quantum,tseng2022gate}. \Cref{Fig:QHE}c further indicates that the Landau fans do not converge at one point at $V_\mathrm{cnp}$, implying finite shifts of the LLs in $V_\mathrm{tg}$. We can correlate the shift of the LLs to the opening of the bulk gap at the charge neutrality point in the graphene channel due to the induced staggered potential. The shift is pronounced in the $V_\mathrm{tg}$-dependence since the presence of the CPS midgap states within the graphene bulk gap and LL gaps slows down the tuning of the Fermi energy close to the $V_\mathrm{cnp}$ (further details in SI, Section 9). It is also worth noting that the zLL shows minimal broadening and a negligible increase in resistance vs.~$B_\mathrm{z}$, in contrast to that reported in pristine graphene~\cite{young2012spin}. This observation is suggestive of the presence of edge states at the zLL gap that shunt the transport at $V_\mathrm{cnp}$. The slow modulation of the $E_\mathrm{F}$ by $V_\mathrm{tg}$ within the zLL energy gap in this system can assist with resolving the edge states within the gap.  

On another note, we highlight that the spin-splitting of the LLs with $N_\mathrm{LL}\neq0$ is not resolved within the broadening of the SdHOs in the Landau fan diagram of \Cref{Fig:QHE}a. We attribute this to the presence of disorders in this system (with low charge carrier mobility) that hinders the resolution of the spin-split bands at higher energies. In fact, the absence of the spin-split Landau fan diagram indicates that the induced magnetism is not large in this heterostructure, and thus the observation of the strong AH effect should be mainly due to a sizable SOC induced in the graphene channel. The larger SOC, as compared with the exchange interaction, is expected to result in the emergence of helical (and not chiral) states in this heterostructure~\cite{yang2011time}.

\textcolor{black}{For addressing the topological edge states in the graphene bulk gap, we focus }on the transport close to zero energy. \textcolor{black}{Helical states are spin-polarized electron- and hole-like bands within the bulk gap of the proximitized graphene~\mbox{\cite{kane2005quantum}} that counter-propagate at the edges of the graphene channel, as shown in the schematics of \Cref{fig:QSHS}a. \Cref{fig:QSHS}b shows the gate-dependence of the graphene resistivity ($\rho_\mathrm{xx}$, left axis) and resistance ($R_\mathrm{xx}$, right axis) close to the charge neutrality point,} for small applied magnetic fields ($B_\mathrm{z} \le \SI{0.5}{T}$). \textcolor{black}{The $\rho_\mathrm{xx}$ shows oscillations that develop into the SdHOs in the Landau fan diagram of \Cref{Fig:QHE}a at higher $B_\mathrm{z}$.} \textcolor{black}{The $V_\mathrm{tg}$ dependence of $\rho_\mathrm{xx}$} at $B_\mathrm{z} = \SI{0}{T}$ (in \Cref{fig:QSHS}b) has a wide peak \textcolor{black}{at the charge neutrality point}. When such four-terminal longitudinal measurements are evaluated in terms of conductance ($G_\mathrm{xx}=1/R_\mathrm{xx}= I/V_\mathrm{xx}$), as shown in \Cref{fig:QSHS}c for different voltage probe pairs, it becomes evident that the broadened resistance peak at $V_\mathrm{cnp}$ at \SI{0}{T} is a plateau of conductance at $2\,e^2/h$. This quantized conductance at zero energy is evidence for the presence of counter-propagating spin-polarized helical states, as depicted in \Cref{fig:QSHS}a. %These states result in the QSH effect, originating from the SOC and the staggered potential that could open a gap in the bulk of graphene leaving the topologically protected helical states as the dominant contributors to the transport~\cite{haldane1988model, kane2005quantum}.   

\begin{figure}
    \centering
    \includegraphics[width=\columnwidth]{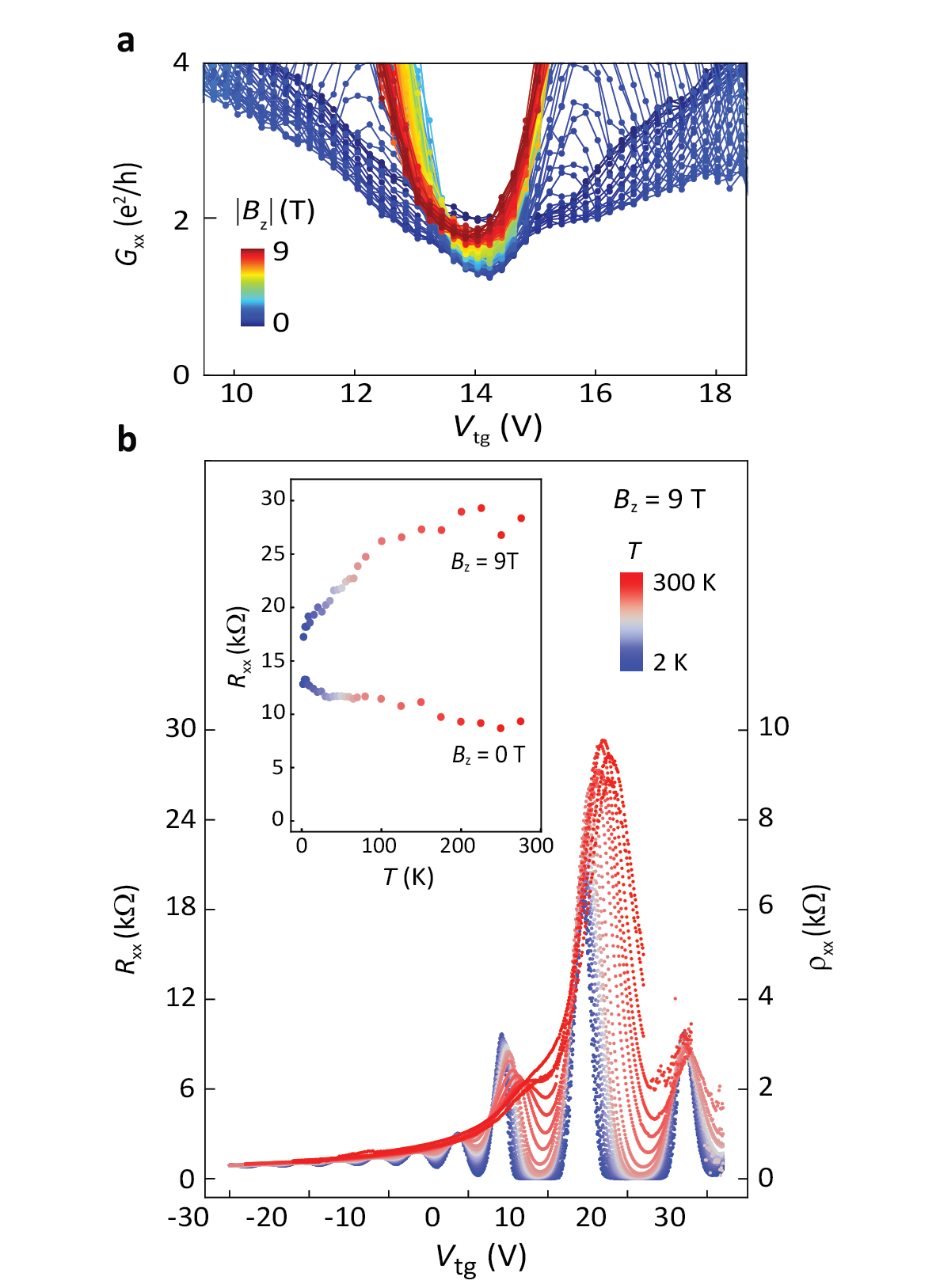}
    \caption{\textcolor{black}{\textbf{Magnetic field and temperature dependence of the QSH and QH transport.}} \textcolor{black}{(a) $V_\mathrm{tg}$-dependence of the conductance close to the $V_\mathrm{cnp}$, measured at  $0<|B_\mathrm{z}|<\SI{9}{T}$ in a four-terminal geometry with $V_5$ and $V_6$ voltage probes in device A. (b) Gate-dependence of the graphene resistance (left axis) and resistivity (right axis) at $B_\mathrm{z}=\SI{9}{T}$, measured at various $T$. Inset: Temperature-dependence of the maxima of the four-terminal resistance at $V_\mathrm{tg}=V_\mathrm{cnp}$, shown for $B_\mathrm{z}= 0, \SI{9}{T}$.}} %(d) Gate-dependence of the resistivity measured at $B_\mathrm{z}=\pm\SI{9}{T}$ at $T=\SI{275}{K}$. (e) $T-$dependence of $\sigma_\mathrm{xy}$ vs.~$V_\mathrm{tg}$, measured at $B_\mathrm{z}=\SI{9}{T}$ in device B.} The thickness of top-gate hBN in device A and B is 73 and \SI{18}{nm}, respectively.} %(e) Temperature-dependence of the transverse conductivity vs.~$B_\mathrm{z}$, measured at $V_\mathrm{tg}=\SI{18}{V}$.}
    \label{fig:Rxy}
\end{figure}

To further explore the helical nature of the edge transport at $B_\mathrm{z}\,=\,0\,$T, we investigate the effect of floating voltage probes on the conductance values at the $V_\mathrm{cnp}$ measured in various geometries. Helical charge transport occurs on both edges of the graphene channel with each edge hosting one spin-polarized state~\cite{buttiker2009edge}. In this case, the voltage probes present along the edges alter the conductance between the source and the drain, since the charges do not maintain their spin when entering the probes. %As a result, half of the incoming current is directed at each adjacent voltage probe. 
Thus, according to the Landauer-Buttiker formalism~\cite{buttiker1986} (SI, Section~14), the two-terminal conductance $G_\mathrm{2T}$ in the presence of two counter-propagating (helical) edge states is given by the relation: 
\begin{equation}
    G_\mathrm{2T} = \frac{e^2}{h}\left(\frac{1}{N_\mathrm{L}+1} + \frac{1}{N_\mathrm{R}+1}\right),
    \label{eq:G_2terminal}
\end{equation}
\noindent where $N_\mathrm{L}$ and $N_\mathrm{R}$ are the number of floating probes between the source and drain, along the left and right edges of the conductor. Similarly, the four-terminal conductance $G_\mathrm{4T}$ is given by \begin{equation}
    G_\mathrm{4T} = G_\mathrm{2T}\frac{N_\mathrm{I}+1}{N_\mathrm{V}+1},
    \label{eq:G_4terminal}
\end{equation}

\noindent where $N_\mathrm{I}$ is the number of floating electrodes in between the source and drain along the edge which hosts the voltage probes, and $N_\mathrm{V}$ is the (minimal) number of floating electrodes in between the voltage probes~\cite{roth2009nonlocal}. For the measurement of \Cref{fig:QSHS}c, $N_\mathrm{L} = N_\mathrm{R} = N_\mathrm{I} = 6$, and $N_\mathrm{V} = 0$, thus, $G_\mathrm{4T} = G_\mathrm{xx} = 2\,e^2/h$, which is \textcolor{black}{the value} measured at $B_\mathrm{z} = \SI{0}{T}$ at the zLL conductance plateau. Considering the equations above, we evaluate the transport in various two-\textcolor{black}{, three-} and four-terminal measurement geometries, shown \textcolor{black}{for device A in the SI, \Cref{fig:helicalstates_in_device A} and for device C in \Cref{fig:QSHS}e}. \textcolor{black}{The graphene channel in device C has been cut by an atomic force microscope tip, shaped into the Hall bar geometry with an aspect ratio different from that of device A and B, to rule out any coincidental geometrical factors in the measured conductance. Device C consists of two electrically disconnected regions with three and five electrodes connected to the channel (\Cref{fig:QSHS}d).} In panel e, the expected \textcolor{black}{$G$ values} considering the helical edge states following equations 1 and 2, are shown by the dashed lines that are color-coded with respect to each measurement geometry. We observe that, despite the distinct $G$ measured at $V_\mathrm{tg}\neq V_\mathrm{cnp}$, the conductance values at $V_\mathrm{cnp}$ in each geometry closely match the theoretical values expected in the presence of QSH effect, confirming that the helical states dominate the charge transport at $V_\mathrm{cnp}$ in the magnetized graphene in the absence of an external magnetic field.

\textcolor{black}{Under finite $B_\mathrm{z}$, the conductance at the charge neutrality point just slightly deviates from the quantized value of 2$e^2/h$, as shown in \Cref{fig:Rxy}a. Persistence of helical states at finite $B_\mathrm{z}$ depends on the evolution of the zLLs with applied $B_\mathrm{z}$ and can be disturbed by bands crossing at certain $B_\mathrm{z}$ ~\cite{bottcher2019survival}. In this case, the temperature dependence of the QH transport can provide further insight into the nature of the zLL gap~\cite{abanin2007dissipative}. In \Cref{fig:Rxy}b, we show the gate-dependence of the graphene resistance (and resistivity) at $B_\mathrm{z}=\SI{9}{T}$ up to room temperature. }We observe that by increasing $T$, the resistance at the zLL \textcolor{black}{increases}. This metallic $T$-dependence \textcolor{black}{is consistent with the presence of the gapless helical states at the zLL up to large $B_\mathrm{z}$.} \textcolor{black}{In the inset of panel b, the $T-$dependence of $R_\mathrm{xx}$ at zero energy is compared for $B_\mathrm{z} = 0$ and $\SI{9}{T}$. The slight decay of $R_\mathrm{xx}$ vs.~$T$ at $B_\mathrm{z} = 0$ can be related to the scattering of the spin-polarized edge states at higher $T$ due to random fluctuations of the magnetic moments \textcolor{black}{(which is eliminated at large $B_\mathrm{z}$)}. \textcolor{black}{Thus, at $B_\mathrm{z}=0$} the contribution of the bulk transport becomes more dominant at higher $T$ and the resistance drops as a result of the thermal broadening of the Dirac peak.}

\textcolor{black}{These observations introduce graphene-CPS heterostructure as a promising platform to exploit unconventional quantum phenomena for practical applications, as it hosts quantum coherent spin-polarized edge states without the need for any external magnetic field. The emergence of the QSH edge states is due to a strong hybridization at the graphene-CPS interface that induces spin-orbit and exchange interactions in graphene, further confirmed by the observation of a large AH effect. The coexistence of the AH and QSH effects in this system, experimentally confirms the possibility of the formation of QSH states despite the breaking of time-reversal symmetry. The observation of helical states in graphene-based van der Waals heterostructures without an external magnetic field, in addition to the persistence of AH signal up to room temperature, opens the route for the direct applications of magnetic graphene in 2D topological spintronic devices.}

\begin{flushleft}
{\large{\textbf{Methods}}}
\end{flushleft}

\textit{CrPS$_4$ Synthesis.} CrPS$_4$ crystals are grown following a solid-state reaction. Stoichiometric amounts of Cr (99.99 $\%$, Alfa-Aesar), P ($> 99.99\%$, Sigma-Aldrich) and S ($99.99 \%$, Sigma-Aldrich) are sealed inside an evacuated quartz tube (pressure $5\times10^{-5}$ mbar, length: \SI{50}{cm}, internal diameter: \SI{14}{mm}) and placed in a three-zone furnace. A temperature gradient of $750/650/700  ^{\circ}$C is kept for 21 days followed by a quench into water. The obtained crystals are analyzed by powder X-ray diffraction and energy-dispersive X-ray spectroscopy. The refinement of the X-ray pattern (ICSD 25059) reveals a monoclinic C face center crystal system with C121 space group and a unit cell determined by $\alpha$ = $\gamma$ = 90$^{\circ}$ and $\beta$ = 91.99(1)$^{\circ}$ and a = 10.841(9)$\AA$, b = 7.247(6)$\AA$ and c = 6.100(5)$\AA$. The amount of elements is Cr:$23.3\pm0.5\%$, P:$14.8\pm0.4\%$ and S:$61.8\pm1.5\%$, in good agreement with the expected ones (Cr:24.6$\%$, P: $14.7\%$ and S: $60.7\%$). The obtained results are in accordance with the ones previously reported in the literature~\cite{peng2020magnetic}. Note that the CPS flakes used in heterostructures with graphene in this work are from the same batch of crystal which is characterized by the SQUID magnetometry in \Cref{Fig:Fig.1}b and c.

\textit{Electrical characterization.} The \textcolor{black}{charge transport measurements are} performed using a DC current source, and voltages are measured using Keithleys. DC voltage source is used for the gates. The measurements are performed using a variable temperature insert at He atmosphere (attoDry2100 system).

\begin{flushleft}
{\textbf{\large{Data availability}}}
\end{flushleft}

All data contained in the figures will be available at zenodo.org upon publication. Additional data related to this paper may be requested from the corresponding author.\\

%\begin{flushleft}
%{\textbf{\large{Code availability}}}
%\end{flushleft}

%The codes used for data analysis are available upon request.

\begin{flushleft}
{\large {\textbf{Acknowledgements}}}
\end{flushleft}

We would like to acknowledge Yaroslav Blanter, Anton Akhmerov and Antonio Manesco for insightful discussions. This project has received funding from the European Union Horizon 2020 research and innovation program under grant agreement, No. 863098 (SPRING). T.S.G. acknowledges support from the Dutch Research Council (NWO) for a Rubicon grant (Project No.~019.222EN.013).~J.I.A. acknowledges support from the European Union’s Horizon 2020 research and innovation programme for a Marie Sklodowska–Curie individual fellowship, No. 101027187-PCSV. E.C and S.M.-V. acknowledge support from the European Union (ERC AdG Mol-2D 788222, FET OPEN SINFONIA 964396), and the Spanish MCIN (2D-HETEROS PID2020-117152RB-100 and Excellence Unit “María de Maeztu” CEX2019-000919-M). K.W. and T.T. acknowledge support from the JSPS KAKENHI (Grant Numbers 21H05233 and 23H02052) and World Premier International Research Center Initiative (WPI), MEXT, Japan. KZ and JF acknowledge support from DFG SFB1277 (Project No.~314695032) and SPP2244 (Project No.~443416183).\\

\begin{flushleft}
{\textbf{\large{Author Contributions}}}
\end{flushleft}

T.S.G. and H.S.J.v.d.Z. conceived and designed the experiments. T.S.G. performed device fabrication and measurements with the help of D.P. T.S.G. and D.P. analyzed the data with the help of J.I-A.. T.B. and J.I-A. assisted with measurements. S.M-V. and E.C. synthesized the CPS crystals and performed the CPS magnetometry measurements. T.S.G. and P.K. further investigated the observations through fabrication and measurements in device C. K.W. and T.T. synthesized the hBN crystals. K.Z. and J.F. performed the DFT calculations. T.S.G. wrote the manuscript with contributions from all co-authors.\\
\\
\begin{flushleft}
{\textbf{\large{Competing interests}}}
\end{flushleft}
The authors declare no competing financial or non-financial interests.

\bibliographystyle{naturemag}
\bibliography{sample}

\begin{thebibliography}{10}
\expandafter\ifx\csname url\endcsname\relax
  \def\url#1{\texttt{#1}}\fi
\expandafter\ifx\csname urlprefix\endcsname\relax\def\urlprefix{URL }\fi
\providecommand{\bibinfo}[2]{#2}
\providecommand{\eprint}[2][]{\url{#2}}

\bibitem{neto2009electronic}
\bibinfo{author}{Neto, A.~C.}, \bibinfo{author}{Guinea, F.}, \bibinfo{author}{Peres, N.~M.}, \bibinfo{author}{Novoselov, K.~S.} \& \bibinfo{author}{Geim, A.~K.}
\newblock \bibinfo{title}{The electronic properties of graphene}.
\newblock \emph{\bibinfo{journal}{Reviews of Modern Physics}} \textbf{\bibinfo{volume}{81}}, \bibinfo{pages}{109} (\bibinfo{year}{2009}).

\bibitem{han2014graphene}
\bibinfo{author}{Han, W.}, \bibinfo{author}{Kawakami, R.~K.}, \bibinfo{author}{Gmitra, M.} \& \bibinfo{author}{Fabian, J.}
\newblock \bibinfo{title}{Graphene spintronics}.
\newblock \emph{\bibinfo{journal}{Nature Nanotechnology}} \textbf{\bibinfo{volume}{9}}, \bibinfo{pages}{794--807} (\bibinfo{year}{2014}).

\bibitem{sierra2021van}
\bibinfo{author}{Sierra, J.~F.}, \bibinfo{author}{Fabian, J.}, \bibinfo{author}{Kawakami, R.~K.}, \bibinfo{author}{Roche, S.} \& \bibinfo{author}{Valenzuela, S.~O.}
\newblock \bibinfo{title}{Van der {W}aals heterostructures for spintronics and opto-spintronics}.
\newblock \emph{\bibinfo{journal}{Nature Nanotechnology}} \textbf{\bibinfo{volume}{16}}, \bibinfo{pages}{856--868} (\bibinfo{year}{2021}).

\bibitem{wei2016strong}
\bibinfo{author}{Wei, P.} \emph{et~al.}
\newblock \bibinfo{title}{Strong interfacial exchange field in the graphene/{E}u{S} heterostructure}.
\newblock \emph{\bibinfo{journal}{Nature Materials}} \textbf{\bibinfo{volume}{15}}, \bibinfo{pages}{711--716} (\bibinfo{year}{2016}).

\bibitem{safeer2019room}
\bibinfo{author}{Safeer, C.} \emph{et~al.}
\newblock \bibinfo{title}{Room-temperature spin {H}all effect in graphene/{M}o{S}$_2$ van der {W}aals heterostructures}.
\newblock \emph{\bibinfo{journal}{Nano Letters}} \textbf{\bibinfo{volume}{19}}, \bibinfo{pages}{1074--1082} (\bibinfo{year}{2019}).

\bibitem{mendes2015spin}
\bibinfo{author}{Mendes, J.} \emph{et~al.}
\newblock \bibinfo{title}{Spin-current to charge-current conversion and magnetoresistance in a hybrid structure of graphene and yttrium iron garnet}.
\newblock \emph{\bibinfo{journal}{Physical Review Letters}} \textbf{\bibinfo{volume}{115}}, \bibinfo{pages}{226601} (\bibinfo{year}{2015}).

\bibitem{ghiasi2019charge}
\bibinfo{author}{Ghiasi, T.~S.}, \bibinfo{author}{Kaverzin, A.~A.}, \bibinfo{author}{Blah, P.~J.} \& \bibinfo{author}{Van~Wees, B.~J.}
\newblock \bibinfo{title}{Charge-to-spin conversion by the rashba-edelstein effect in two-dimensional van der {W}aals heterostructures up to room temperature}.
\newblock \emph{\bibinfo{journal}{Nano Letters}} \textbf{\bibinfo{volume}{19}}, \bibinfo{pages}{5959--5966} (\bibinfo{year}{2019}).

\bibitem{benitez2020tunable}
\bibinfo{author}{Ben{\'\i}tez, L.~A.} \emph{et~al.}
\newblock \bibinfo{title}{Tunable room-temperature spin galvanic and spin {H}all effects in van der {W}aals heterostructures}.
\newblock \emph{\bibinfo{journal}{Nature Materials}} \textbf{\bibinfo{volume}{19}}, \bibinfo{pages}{170--175} (\bibinfo{year}{2020}).

\bibitem{wang2015proximity}
\bibinfo{author}{Wang, Z.}, \bibinfo{author}{Tang, C.}, \bibinfo{author}{Sachs, R.}, \bibinfo{author}{Barlas, Y.} \& \bibinfo{author}{Shi, J.}
\newblock \bibinfo{title}{Proximity-induced ferromagnetism in graphene revealed by the anomalous {H}all effect}.
\newblock \emph{\bibinfo{journal}{Physical Review Letters}} \textbf{\bibinfo{volume}{114}}, \bibinfo{pages}{016603} (\bibinfo{year}{2015}).

\bibitem{ghiasi2021electrical}
\bibinfo{author}{Ghiasi, T.~S.} \emph{et~al.}
\newblock \bibinfo{title}{Electrical and thermal generation of spin currents by magnetic bilayer graphene}.
\newblock \emph{\bibinfo{journal}{Nature Nanotechnology}} \textbf{\bibinfo{volume}{16}}, \bibinfo{pages}{788--794} (\bibinfo{year}{2021}).

\bibitem{geim2013van}
\bibinfo{author}{Geim, A.~K.} \& \bibinfo{author}{Grigorieva, I.~V.}
\newblock \bibinfo{title}{Van der {W}aals heterostructures}.
\newblock \emph{\bibinfo{journal}{Nature}} \textbf{\bibinfo{volume}{499}}, \bibinfo{pages}{419--425} (\bibinfo{year}{2013}).

\bibitem{yang2013proximity}
\bibinfo{author}{Yang, H.-X.} \emph{et~al.}
\newblock \bibinfo{title}{Proximity effects induced in graphene by magnetic insulators: first-principles calculations on spin filtering and exchange-splitting gaps}.
\newblock \emph{\bibinfo{journal}{Physical Review Letters}} \textbf{\bibinfo{volume}{110}}, \bibinfo{pages}{046603} (\bibinfo{year}{2013}).

\bibitem{dyrdal2017anomalous}
\bibinfo{author}{Dyrda{\l}, A.} \& \bibinfo{author}{Barna{\'s}, J.}
\newblock \bibinfo{title}{Anomalous, spin, and valley {H}all effects in graphene deposited on ferromagnetic substrates}.
\newblock \emph{\bibinfo{journal}{2D Materials}} \textbf{\bibinfo{volume}{4}}, \bibinfo{pages}{034003} (\bibinfo{year}{2017}).

\bibitem{diehl1977crystal}
\bibinfo{author}{Diehl, R.} \& \bibinfo{author}{Carpentier, C.-D.}
\newblock \bibinfo{title}{The crystal structure of chromium thiophosphate, {C}r{P}{S}$_4$}.
\newblock \emph{\bibinfo{journal}{Acta Crystallographica Section B: Structural Crystallography and Crystal Chemistry}} \textbf{\bibinfo{volume}{33}}, \bibinfo{pages}{1399--1404} (\bibinfo{year}{1977}).

\bibitem{young2012spin}
\bibinfo{author}{Young, A.~F.} \emph{et~al.}
\newblock \bibinfo{title}{Spin and valley quantum hall ferromagnetism in graphene}.
\newblock \emph{\bibinfo{journal}{Nature Physics}} \textbf{\bibinfo{volume}{8}}, \bibinfo{pages}{550--556} (\bibinfo{year}{2012}).

\bibitem{young2014tunable}
\bibinfo{author}{Young, A.} \emph{et~al.}
\newblock \bibinfo{title}{Tunable symmetry breaking and helical edge transport in a graphene quantum spin {H}all state}.
\newblock \emph{\bibinfo{journal}{Nature}} \textbf{\bibinfo{volume}{505}}, \bibinfo{pages}{528--532} (\bibinfo{year}{2014}).

\bibitem{veyrat2020helical}
\bibinfo{author}{Veyrat, L.} \emph{et~al.}
\newblock \bibinfo{title}{Helical quantum {H}all phase in graphene on {S}r{T}i{O}$_3$}.
\newblock \emph{\bibinfo{journal}{Science}} \textbf{\bibinfo{volume}{367}}, \bibinfo{pages}{781--786} (\bibinfo{year}{2020}).

\bibitem{haldane1988model}
\bibinfo{author}{Haldane, F. D.~M.}
\newblock \bibinfo{title}{Model for a quantum hall effect without landau levels: Condensed-matter realization of the" parity anomaly"}.
\newblock \emph{\bibinfo{journal}{Physical Review Letters}} \textbf{\bibinfo{volume}{61}}, \bibinfo{pages}{2015} (\bibinfo{year}{1988}).

\bibitem{kane2005quantum}
\bibinfo{author}{Kane, C.~L.} \& \bibinfo{author}{Mele, E.~J.}
\newblock \bibinfo{title}{Quantum spin {H}all effect in graphene}.
\newblock \emph{\bibinfo{journal}{Physical Review Letters}} \textbf{\bibinfo{volume}{95}}, \bibinfo{pages}{226801} (\bibinfo{year}{2005}).

\bibitem{qiao2010quantum}
\bibinfo{author}{Qiao, Z.} \emph{et~al.}
\newblock \bibinfo{title}{Quantum anomalous {H}all effect in graphene from rashba and exchange effects}.
\newblock \emph{\bibinfo{journal}{Physical Review B}} \textbf{\bibinfo{volume}{82}}, \bibinfo{pages}{161414} (\bibinfo{year}{2010}).

\bibitem{yang2011time}
\bibinfo{author}{Yang, Y.} \emph{et~al.}
\newblock \bibinfo{title}{Time-reversal-symmetry-broken quantum spin hall effect}.
\newblock \emph{\bibinfo{journal}{Physical Review Letters}} \textbf{\bibinfo{volume}{107}}, \bibinfo{pages}{066602} (\bibinfo{year}{2011}).

\bibitem{Qiao2014:PRL}
\bibinfo{author}{Qiao, Z.} \emph{et~al.}
\newblock \bibinfo{title}{{Quantum Anomalous Hall Effect in Graphene Proximity Coupled to an Antiferromagnetic Insulator}}.
\newblock \emph{\bibinfo{journal}{Physical Review Letters}} \textbf{\bibinfo{volume}{112}}, \bibinfo{pages}{116404} (\bibinfo{year}{2014}).

\bibitem{Kaloni2014:APL}
\bibinfo{author}{Kaloni, T.~P.}, \bibinfo{author}{Kou, L.}, \bibinfo{author}{Frauenheim, T.} \& \bibinfo{author}{Schwingenschl{\"{o}}gl, U.}
\newblock \bibinfo{title}{{Quantum spin Hall states in graphene interacting with WS$_2$ or WSe$_2$}}.
\newblock \emph{\bibinfo{journal}{Appl. Phys. Lett.}} \textbf{\bibinfo{volume}{105}}, \bibinfo{pages}{233112} (\bibinfo{year}{2014}).

\bibitem{hatsuda2018evidence}
\bibinfo{author}{Hatsuda, K.} \emph{et~al.}
\newblock \bibinfo{title}{Evidence for a quantum spin hall phase in graphene decorated with {B}i$_2${T}e$_3$ nanoparticles}.
\newblock \emph{\bibinfo{journal}{Science Advances}} \textbf{\bibinfo{volume}{4}}, \bibinfo{pages}{eaau6915} (\bibinfo{year}{2018}).

\bibitem{offidani2018anomalous}
\bibinfo{author}{Offidani, M.} \& \bibinfo{author}{Ferreira, A.}
\newblock \bibinfo{title}{Anomalous hall effect in 2d dirac materials}.
\newblock \emph{\bibinfo{journal}{Physical Review Letters}} \textbf{\bibinfo{volume}{121}}, \bibinfo{pages}{126802} (\bibinfo{year}{2018}).

\bibitem{hogl2020quantum}
\bibinfo{author}{H{\"o}gl, P.} \emph{et~al.}
\newblock \bibinfo{title}{Quantum anomalous {H}all effects in graphene from proximity-induced uniform and staggered spin-orbit and exchange coupling}.
\newblock \emph{\bibinfo{journal}{Physical Review Letters}} \textbf{\bibinfo{volume}{124}}, \bibinfo{pages}{136403} (\bibinfo{year}{2020}).

\bibitem{vila2021valley}
\bibinfo{author}{Vila, M.}, \bibinfo{author}{Garcia, J.~H.} \& \bibinfo{author}{Roche, S.}
\newblock \bibinfo{title}{Valley-polarized quantum anomalous hall phase in bilayer graphene with layer-dependent proximity effects}.
\newblock \emph{\bibinfo{journal}{Physical Review B}} \textbf{\bibinfo{volume}{104}}, \bibinfo{pages}{L161113} (\bibinfo{year}{2021}).

\bibitem{bora2022magnetic}
\bibinfo{author}{Bora, M.}, \bibinfo{author}{Behera, S.~K.}, \bibinfo{author}{Samal, P.} \& \bibinfo{author}{Deb, P.}
\newblock \bibinfo{title}{Magnetic proximity induced valley-contrasting quantum anomalous {H}all effect in a graphene-{C}r{B}r$_3$ van der {W}aals heterostructure}.
\newblock \emph{\bibinfo{journal}{Physical Review B}} \textbf{\bibinfo{volume}{105}}, \bibinfo{pages}{235422} (\bibinfo{year}{2022}).

\bibitem{obata2024coexistence}
\bibinfo{author}{Obata, R.} \emph{et~al.}
\newblock \bibinfo{title}{Coexistence of quantum-spin-hall and quantum-hall-topological-insulating states in graphene/hbn on {S}r{T}i{O}$_3$ substrate}.
\newblock \emph{\bibinfo{journal}{Advanced Materials}} \textbf{\bibinfo{volume}{36}}, \bibinfo{pages}{2311339} (\bibinfo{year}{2024}).

\bibitem{song2018electrical}
\bibinfo{author}{Song, H.-D.} \emph{et~al.}
\newblock \bibinfo{title}{Electrical control of magnetic proximity effect in a graphene/multiferroic heterostructure}.
\newblock \emph{\bibinfo{journal}{Applied Physics Letters}} \textbf{\bibinfo{volume}{113}} (\bibinfo{year}{2018}).

\bibitem{song2018asymmetric}
\bibinfo{author}{Song, H.-D.} \emph{et~al.}
\newblock \bibinfo{title}{Asymmetric modulation on exchange field in a graphene/{B}i{F}e{O}$_3$ heterostructure by external magnetic field}.
\newblock \emph{\bibinfo{journal}{Nano Letters}} \textbf{\bibinfo{volume}{18}}, \bibinfo{pages}{2435--2441} (\bibinfo{year}{2018}).

\bibitem{wu2020large}
\bibinfo{author}{Wu, Y.} \emph{et~al.}
\newblock \bibinfo{title}{Large exchange splitting in monolayer graphene magnetized by an antiferromagnet}.
\newblock \emph{\bibinfo{journal}{Nature Electronics}} \textbf{\bibinfo{volume}{3}}, \bibinfo{pages}{604--611} (\bibinfo{year}{2020}).

\bibitem{wu2021magnetic}
\bibinfo{author}{Wu, Y.} \emph{et~al.}
\newblock \bibinfo{title}{Magnetic exchange field modulation of quantum {H}all ferromagnetism in 2d van der {W}aals {C}r{C}l$_3$/graphene heterostructures}.
\newblock \emph{\bibinfo{journal}{ACS Applied Materials and Interfaces}} \textbf{\bibinfo{volume}{13}}, \bibinfo{pages}{10656--10663} (\bibinfo{year}{2021}).

\bibitem{chau2022two}
\bibinfo{author}{Chau, T.~K.}, \bibinfo{author}{Hong, S.~J.}, \bibinfo{author}{Kang, H.} \& \bibinfo{author}{Suh, D.}
\newblock \bibinfo{title}{Two-dimensional ferromagnetism detected by proximity-coupled quantum {H}all effect of graphene}.
\newblock \emph{\bibinfo{journal}{npj Quantum Materials}} \textbf{\bibinfo{volume}{7}}, \bibinfo{pages}{1--7} (\bibinfo{year}{2022}).

\bibitem{hu2023tunable}
\bibinfo{author}{Hu, J.} \emph{et~al.}
\newblock \bibinfo{title}{Tunable spin-polarized states in graphene on a ferrimagnetic oxide insulator}.
\newblock \emph{\bibinfo{journal}{Advanced Materials}} \bibinfo{pages}{2305763} (\bibinfo{year}{2023}).

\bibitem{wang2022quantum}
\bibinfo{author}{Wang, Y.} \emph{et~al.}
\newblock \bibinfo{title}{Quantum {H}all phase in graphene engineered by interfacial charge coupling}.
\newblock \emph{\bibinfo{journal}{Nature Nanotechnology}} \bibinfo{pages}{1--8} (\bibinfo{year}{2022}).

\bibitem{tseng2022gate}
\bibinfo{author}{Tseng, C.-C.} \emph{et~al.}
\newblock \bibinfo{title}{Gate-tunable proximity effects in graphene on layered magnetic insulators}.
\newblock \emph{\bibinfo{journal}{Nano Letters}}  (\bibinfo{year}{2022}).

\bibitem{wang2017dirac}
\bibinfo{author}{Wang, X.-L.}
\newblock \bibinfo{title}{{D}irac spin-gapless semiconductors: promising platforms for massless and dissipationless spintronics and new (quantum) anomalous spin {H}all effects}.
\newblock \emph{\bibinfo{journal}{National Science Review}} \textbf{\bibinfo{volume}{4}}, \bibinfo{pages}{252--257} (\bibinfo{year}{2017}).

\bibitem{lee2017structural}
\bibinfo{author}{Lee, J.} \emph{et~al.}
\newblock \bibinfo{title}{Structural and optical properties of single-and few-layer magnetic semiconductor {C}r{PS}$_4$}.
\newblock \emph{\bibinfo{journal}{ACS Nano}} \textbf{\bibinfo{volume}{11}}, \bibinfo{pages}{10935--10944} (\bibinfo{year}{2017}).

\bibitem{peng2020magnetic}
\bibinfo{author}{Peng, Y.} \emph{et~al.}
\newblock \bibinfo{title}{Magnetic structure and metamagnetic transitions in the van der {W}aals antiferromagnet {C}r{P}{S}$_4$}.
\newblock \emph{\bibinfo{journal}{Advanced Materials}} \textbf{\bibinfo{volume}{32}}, \bibinfo{pages}{2001200} (\bibinfo{year}{2020}).

\bibitem{tang2018approaching}
\bibinfo{author}{Tang, C.} \emph{et~al.}
\newblock \bibinfo{title}{Approaching quantum anomalous hall effect in proximity-coupled {YIG}/graphene/h-{BN} sandwich structure}.
\newblock \emph{\bibinfo{journal}{APL Materials}} \textbf{\bibinfo{volume}{6}} (\bibinfo{year}{2018}).

\bibitem{nagaosa2010anomalous}
\bibinfo{author}{Nagaosa, N.}, \bibinfo{author}{Sinova, J.}, \bibinfo{author}{Onoda, S.}, \bibinfo{author}{MacDonald, A.~H.} \& \bibinfo{author}{Ong, N.~P.}
\newblock \bibinfo{title}{Anomalous {H}all effect}.
\newblock \emph{\bibinfo{journal}{Reviews of Modern Physics}} \textbf{\bibinfo{volume}{82}}, \bibinfo{pages}{1539} (\bibinfo{year}{2010}).

\bibitem{zhu2023interface}
\bibinfo{author}{Zhu, W.} \emph{et~al.}
\newblock \bibinfo{title}{Interface-enhanced room-temperature curie temperature in {C}r{P}{S}$_4$/graphene van der waals heterostructure}.
\newblock \emph{\bibinfo{journal}{Physical Review B}} \textbf{\bibinfo{volume}{108}}, \bibinfo{pages}{L100406} (\bibinfo{year}{2023}).

\bibitem{novoselov2005two}
\bibinfo{author}{Novoselov, K.~S.} \emph{et~al.}
\newblock \bibinfo{title}{Two-dimensional gas of massless {D}irac fermions in graphene}.
\newblock \emph{\bibinfo{journal}{Nature}} \textbf{\bibinfo{volume}{438}}, \bibinfo{pages}{197--200} (\bibinfo{year}{2005}).

\bibitem{zhang2005experimental}
\bibinfo{author}{Zhang, Y.}, \bibinfo{author}{Tan, Y.-W.}, \bibinfo{author}{Stormer, H.~L.} \& \bibinfo{author}{Kim, P.}
\newblock \bibinfo{title}{Experimental observation of the quantum {H}all effect and {B}erry's phase in graphene}.
\newblock \emph{\bibinfo{journal}{Nature}} \textbf{\bibinfo{volume}{438}}, \bibinfo{pages}{201--204} (\bibinfo{year}{2005}).

\bibitem{buttiker2009edge}
\bibinfo{author}{B{\"u}ttiker, M.}
\newblock \bibinfo{title}{Edge-state physics without magnetic fields}.
\newblock \emph{\bibinfo{journal}{Science}} \textbf{\bibinfo{volume}{325}}, \bibinfo{pages}{278--279} (\bibinfo{year}{2009}).

\bibitem{buttiker1986}
\bibinfo{author}{B{\"u}ttiker, M.}
\newblock \bibinfo{title}{Four-terminal phase-coherent conductance}.
\newblock \emph{\bibinfo{journal}{Physical Review Letters}} \textbf{\bibinfo{volume}{57}}, \bibinfo{pages}{1761} (\bibinfo{year}{1986}).

\bibitem{roth2009nonlocal}
\bibinfo{author}{Roth, A.} \emph{et~al.}
\newblock \bibinfo{title}{Nonlocal transport in the quantum spin {H}all state}.
\newblock \emph{\bibinfo{journal}{Science}} \textbf{\bibinfo{volume}{325}}, \bibinfo{pages}{294--297} (\bibinfo{year}{2009}).

\bibitem{bottcher2019survival}
\bibinfo{author}{B{\"o}ttcher, J.}, \bibinfo{author}{Tutschku, C.}, \bibinfo{author}{Molenkamp, L.~W.} \& \bibinfo{author}{Hankiewicz, E.}
\newblock \bibinfo{title}{Survival of the quantum anomalous hall effect in orbital magnetic fields as a consequence of the parity anomaly}.
\newblock \emph{\bibinfo{journal}{Physical Review Letters}} \textbf{\bibinfo{volume}{123}}, \bibinfo{pages}{226602} (\bibinfo{year}{2019}).

\bibitem{abanin2007dissipative}
\bibinfo{author}{Abanin, D.~A.} \emph{et~al.}
\newblock \bibinfo{title}{Dissipative quantum {H}all effect in graphene near the {D}irac point}.
\newblock \emph{\bibinfo{journal}{Physical Review Letters}} \textbf{\bibinfo{volume}{98}}, \bibinfo{pages}{196806} (\bibinfo{year}{2007}).

\bibitem{zomer2014fast}
\bibinfo{author}{Zomer, P.}, \bibinfo{author}{Guimar{\~a}es, M.}, \bibinfo{author}{Brant, J.}, \bibinfo{author}{Tombros, N.} \& \bibinfo{author}{Van~Wees, B.}
\newblock \bibinfo{title}{Fast pick up technique for high quality heterostructures of bilayer graphene and hexagonal boron nitride}.
\newblock \emph{\bibinfo{journal}{Applied Physics Letters}} \textbf{\bibinfo{volume}{105}}, \bibinfo{pages}{013101} (\bibinfo{year}{2014}).

\bibitem{wu2023gate}
\bibinfo{author}{Wu, F.} \emph{et~al.}
\newblock \bibinfo{title}{Gate-controlled magnetotransport and electrostatic modulation of magnetism in 2{D} magnetic semiconductor {C}r{P}{S}$_4$}.
\newblock \emph{\bibinfo{journal}{Advanced Materials}} \bibinfo{pages}{2211653} (\bibinfo{year}{2023}).

\bibitem{qhe_klitzing}
\bibinfo{author}{von Klitzing, K.}
\newblock \bibinfo{title}{The quantized {H}all effect}.
\newblock \emph{\bibinfo{journal}{Reviews of Modern Physics}} \textbf{\bibinfo{volume}{58}}, \bibinfo{pages}{519--531} (\bibinfo{year}{1986}).

\bibitem{mobility_Vs_T1}
\bibinfo{author}{Gannett, W.} \emph{et~al.}
\newblock \bibinfo{title}{{Boron nitride substrates for high mobility chemical vapor deposited graphene}}.
\newblock \emph{\bibinfo{journal}{Applied Physics Letters}} \textbf{\bibinfo{volume}{98}} (\bibinfo{year}{2011}).
\newblock \bibinfo{note}{242105}.

\bibitem{mobility_vs_T2}
\bibinfo{author}{Zhu, W.}, \bibinfo{author}{Perebeinos, V.}, \bibinfo{author}{Freitag, M.} \& \bibinfo{author}{Avouris, P.}
\newblock \bibinfo{title}{Carrier scattering, mobilities, and electrostatic potential in monolayer, bilayer, and trilayer graphene}.
\newblock \emph{\bibinfo{journal}{Physical Review B}} \textbf{\bibinfo{volume}{80}}, \bibinfo{pages}{235402} (\bibinfo{year}{2009}).

\bibitem{zhu2015programmable}
\bibinfo{author}{Zhu, S.}, \bibinfo{author}{Stroscio, J.~A.} \& \bibinfo{author}{Li, T.}
\newblock \bibinfo{title}{Programmable extreme pseudomagnetic fields in graphene by a uniaxial stretch}.
\newblock \emph{\bibinfo{journal}{Physical Review Letters}} \textbf{\bibinfo{volume}{115}}, \bibinfo{pages}{245501} (\bibinfo{year}{2015}).

\bibitem{houmes2024highly}
\bibinfo{author}{Houmes, M.~J.} \emph{et~al.}
\newblock \bibinfo{title}{Highly anisotropic mechanical response of the van der waals magnet {C}r{PS}$_4$}.
\newblock \emph{\bibinfo{journal}{Advanced Functional Materials}} \textbf{\bibinfo{volume}{34}}, \bibinfo{pages}{2310206} (\bibinfo{year}{2024}).

\bibitem{ASE}
\bibinfo{author}{Bahn, S.~R.} \& \bibinfo{author}{Jacobsen, K.~W.}
\newblock \bibinfo{title}{An object-oriented scripting interface to a legacy electronic structure code}.
\newblock \emph{\bibinfo{journal}{Comput. Sci. Eng.}} \textbf{\bibinfo{volume}{4}}, \bibinfo{pages}{56} (\bibinfo{year}{2002}).

\bibitem{Lazic2015:CPC}
\bibinfo{author}{Lazic, P.}
\newblock \bibinfo{title}{Cellmatch: Combining two unit cells into a common supercell with minimal strain}.
\newblock \emph{\bibinfo{journal}{Computer Physics Communications}} \textbf{\bibinfo{volume}{197}}, \bibinfo{pages}{324 -- 334} (\bibinfo{year}{2015}).

\bibitem{Koda2016:JPCC}
\bibinfo{author}{Koda, D.~S.}, \bibinfo{author}{Bechstedt, F.}, \bibinfo{author}{Marques, M.} \& \bibinfo{author}{Teles, L.~K.}
\newblock \bibinfo{title}{Coincidence lattices of 2d crystals: heterostructure predictions and applications}.
\newblock \emph{\bibinfo{journal}{The Journal of Physical Chemistry C}} \textbf{\bibinfo{volume}{120}}, \bibinfo{pages}{10895--10908} (\bibinfo{year}{2016}).

\bibitem{Carr2020:NRM}
\bibinfo{author}{Carr, S.}, \bibinfo{author}{Fang, S.} \& \bibinfo{author}{Kaxiras, E.}
\newblock \bibinfo{title}{Electronic-structure methods for twisted moir{\'e} layers}.
\newblock \emph{\bibinfo{journal}{Nature Reviews Materials}} \textbf{\bibinfo{volume}{5}}, \bibinfo{pages}{748--763} (\bibinfo{year}{2020}).

\bibitem{Lee2017:ACS}
\bibinfo{author}{Lee, J.} \emph{et~al.}
\newblock \bibinfo{title}{Structural and optical properties of single-and few-layer magnetic semiconductor {C}r{PS}$_4$}.
\newblock \emph{\bibinfo{journal}{ACS nano}} \textbf{\bibinfo{volume}{11}}, \bibinfo{pages}{10935--10944} (\bibinfo{year}{2017}).

\bibitem{Hohenberg1964:PRB}
\bibinfo{author}{Hohenberg, P.} \& \bibinfo{author}{Kohn, W.}
\newblock \bibinfo{title}{Inhomogeneous electron gas}.
\newblock \emph{\bibinfo{journal}{Phys. Rev.}} \textbf{\bibinfo{volume}{136}}, \bibinfo{pages}{B864} (\bibinfo{year}{1964}).

\bibitem{Giannozzi2009:JPCM}
\bibinfo{author}{Giannozzi, P.} \& \bibinfo{author}{et~al.}
\newblock \bibinfo{title}{Quantum espresso: a modular and open-source software project for quantum simulations of materials}.
\newblock \emph{\bibinfo{journal}{J. Phys.: Cond. Mat.}} \textbf{\bibinfo{volume}{21}}, \bibinfo{pages}{395502} (\bibinfo{year}{2009}).

\bibitem{Joe2017:JP}
\bibinfo{author}{Joe, M.} \emph{et~al.}
\newblock \bibinfo{title}{A comprehensive study of piezomagnetic response in {C}r{PS}$_4$ monolayer: mechanical, electronic properties and magnetic ordering under strains}.
\newblock \emph{\bibinfo{journal}{Journal of Physics: Condensed Matter}} \textbf{\bibinfo{volume}{29}}, \bibinfo{pages}{405801} (\bibinfo{year}{2017}).

\bibitem{Zhuang2016:PRB2}
\bibinfo{author}{Zhuang, H.~L.} \& \bibinfo{author}{Zhou, J.}
\newblock \bibinfo{title}{Density functional theory study of bulk and single-layer magnetic semiconductor ${\mathrm{crps}}_{4}$}.
\newblock \emph{\bibinfo{journal}{Physical Review B}} \textbf{\bibinfo{volume}{94}}, \bibinfo{pages}{195307} (\bibinfo{year}{2016}).

\bibitem{Kresse1999:PRB}
\bibinfo{author}{Kresse, G.} \& \bibinfo{author}{Joubert, D.}
\newblock \bibinfo{title}{From ultrasoft pseudopotentials to the projector augmented-wave method}.
\newblock \emph{\bibinfo{journal}{Physical Review B}} \textbf{\bibinfo{volume}{59}}, \bibinfo{pages}{1758} (\bibinfo{year}{1999}).

\bibitem{Perdew1996:PRL}
\bibinfo{author}{Perdew, J.~P.}, \bibinfo{author}{Burke, K.} \& \bibinfo{author}{Ernzerhof, M.}
\newblock \bibinfo{title}{Generalized gradient approximation made simple}.
\newblock \emph{\bibinfo{journal}{Physical Review Letters}} \textbf{\bibinfo{volume}{77}}, \bibinfo{pages}{3865} (\bibinfo{year}{1996}).

\bibitem{Grimme2006:JCC}
\bibinfo{author}{Grimme, S.}
\newblock \bibinfo{title}{Semiempirical gga-type density functional constructed with a long-range dispersion correction}.
\newblock \emph{\bibinfo{journal}{J. Comput. Chem.}} \textbf{\bibinfo{volume}{27}}, \bibinfo{pages}{1787} (\bibinfo{year}{2006}).

\bibitem{Grimme2010:JCP}
\bibinfo{author}{Grimme, S.}, \bibinfo{author}{Antony, J.}, \bibinfo{author}{Ehrlich, S.} \& \bibinfo{author}{Krieg, H.}
\newblock \bibinfo{title}{{A consistent and accurate ab initio parametrization of density functional dispersion correction (DFT-D) for the 94 elements H-Pu}}.
\newblock \emph{\bibinfo{journal}{J. Chem. Phys.}} \textbf{\bibinfo{volume}{132}}, \bibinfo{pages}{154104} (\bibinfo{year}{2010}).

\bibitem{Barone2009:JCC}
\bibinfo{author}{Barone, V.} \emph{et~al.}
\newblock \bibinfo{title}{Role and effective treatment of dispersive forces in materials: Polyethylene and graphite crystals as test cases}.
\newblock \emph{\bibinfo{journal}{J. Comput. Chem.}} \textbf{\bibinfo{volume}{30}}, \bibinfo{pages}{934} (\bibinfo{year}{2009}).

\bibitem{Bengtsson1999:PRB}
\bibinfo{author}{Bengtsson, L.}
\newblock \bibinfo{title}{Dipole correction for surface supercell calculations}.
\newblock \emph{\bibinfo{journal}{Physical Review B}} \textbf{\bibinfo{volume}{59}}, \bibinfo{pages}{12301} (\bibinfo{year}{1999}).

\bibitem{Zollner2022:PRB}
\bibinfo{author}{Zollner, K.} \& \bibinfo{author}{Fabian, J.}
\newblock \bibinfo{title}{Proximity effects in graphene on monolayers of transition-metal phosphorus trichalcogenides $m\mathrm{P}{X}_{3}$ $(m:\mathrm{Mn}, \mathrm{Fe}, \mathrm{Ni}, \mathrm{Co}, \mathrm{and} x: \mathrm{S}, \mathrm{Se})$}.
\newblock \emph{\bibinfo{journal}{Physical Review B}} \textbf{\bibinfo{volume}{106}}, \bibinfo{pages}{035137} (\bibinfo{year}{2022}).

\end{thebibliography}

\onecolumngrid

\newpage
\setcounter{figure}{0}
\numberwithin{figure}{section}
\renewcommand\thesection{\arabic{section}}

\renewcommand{\thefigure}{S\arabic{figure}}

\textbf{\LARGE{Supplementary Information}}\\

\FloatBarrier
\textbf{\large{1.~Device Fabrication}}\\

\begin{figure*}[h] 
    \centering
    \includegraphics[width=0.9\textwidth]{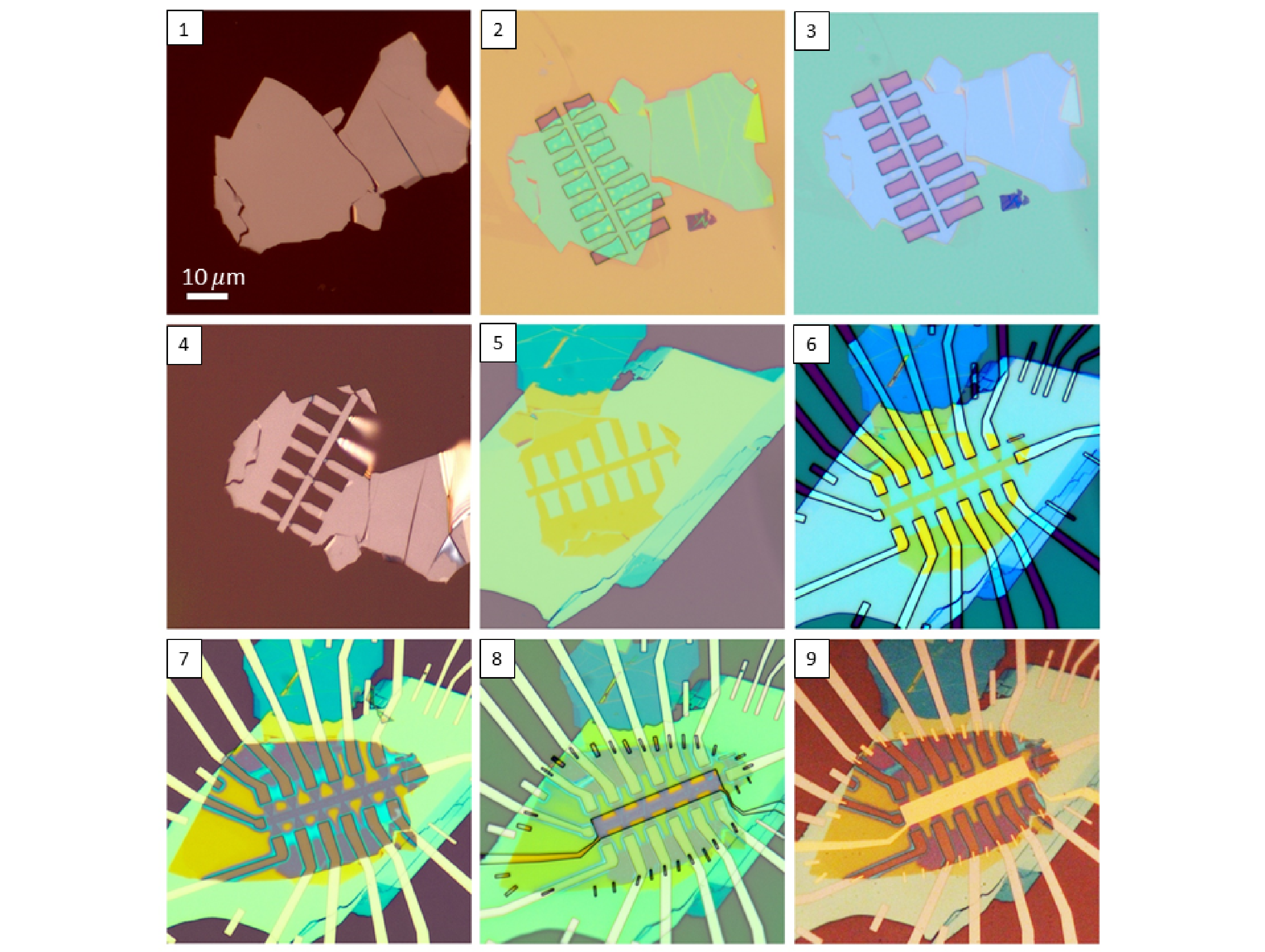}
    \caption{Device fabrication procedure. 1.~hBN flake exfoliated on PMMA (thickness of the hBN flake $\approx\,$\SI{32}{nm}, shown here for device A). The PMMA layer is initially spin-coated on a water-soluble layer to assist the suspension and transfer of the PMMA in the next step. 2.~hBN/PMMA transferred on graphene which is exfoliated in advance on a SiO$_2$/Si substrate. The rectangular-like openings in the PMMA are made by e-beam exposure, followed by cold developing in H$_2$O:IPA (1:3) solution. 3.~hBN/graphene, etched with O$_2$:CHF$_3$ plasma (power = \SI{40}{W}, bias= 100-150~V) for 80-\SI{110}{s}. PMMA is then dissolved in acetone. 4.~Pick up of the hBN/graphene Hall bar with a polycarbonate-PDMS stamp. 5.~Exfoliation of CrPS$_4$ (CPS) on a SiO$_2$/Si substrate and transfer of the hBN/graphene on the CPS flake, using conventional dry transfer technique~\cite{zomer2014fast}. 6.~PMMA-coated heterostructure, exposed by e-beam lithography of the electrodes, developed in MIBK:IPA (1:3) solution. After development, the contact regions are exposed to an O$_2$:CHF$_3$ plasma to enhance the possibility of making edge contact with graphene. 7.~Deposition of Ti(5~nm)-Pd(75~nm) electrodes, followed by lift-off in acetone. After annealing the device in Ar atmosphere for 1~hr at \SI{400}{\celsius}, another hBN flake is transferred on top for the top-gate dielectric (for device A the thickness of this hBN flake is about \SI{41}{nm}. Thus, the total thickness of hBN layers, including that of step 1, that would be counted for top-gating dielectric is about \SI{73}{nm}). 8.~The device is covered with PMMA and has the top-gate area exposed and developed. 9.~Deposition of Ti(5~nm)-Au(100~nm) for the top-gate electrode. The panels are all related to device A.}
    \label{Fig:fab}
\end{figure*}

\begin{figure*}[h] 
    \centering
    \includegraphics[width=\textwidth]{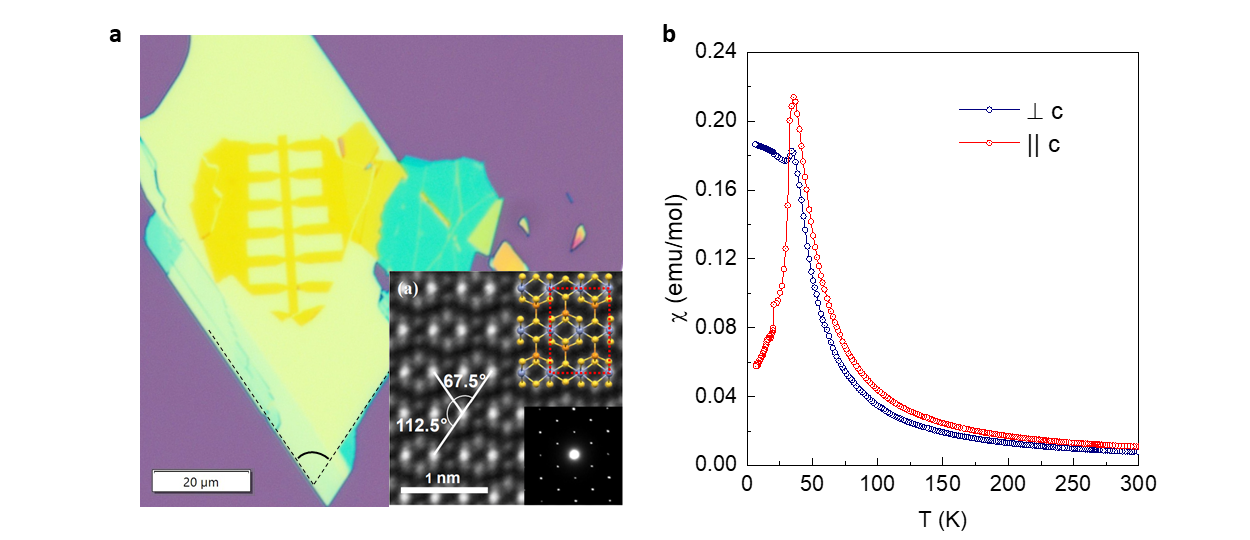}
    \caption{(a) Alignment of graphene channel with respect to the crystallographic orientation of the CPS. The magnetic $a$-axis is along the bisector of the 67.5$^{\circ}$ corner angle of the flake. Inset: STM image of CPS adapted from ref.~\cite{lee2017structural}. The white solid lines are along the Cr atoms. (b) Magnetic molar susceptibility ($\chi$) of CPS, measured parallel ($\|$) and perpendicular ($\perp$) to the $c-$axis (out-of-plane direction) of a bulk CPS crystal. }
    \label{Fig:CPSmag}
\end{figure*}

\Cref{Fig:fab} shows the steps for the fabrication of device A. Details of the fabrication steps are stated in the caption. Device B is also fabricated using the same procedure. Regarding the fabrication procedure of device C, we highlight that the graphene in this device was cut by the tip of an atomic force microscope. The cut graphene Hall bar is picked up by an hBN flake (with a polycarbonate mask) and is released on a CPS flake. This is important to highlight that we realize the presence of the helical states in all the devices, despite the different used techniques for shaping the graphene into the Hall-bar geometry, which is reactive ion etching for device A and B, and AFM cutting for device C. That indicates that the doping of the edges and shunting of the conductance cannot explain the detection of the quantized edge transport in these devices. 

\newpage
\FloatBarrier
\textbf{\large{2.~CPS magnetization anisotropy}}\\

\textbf{\large{3.~$B_\mathrm{z}$-dependence of the longitudinal and transverse resistivity at room temperature}}\\

\begin{figure*}[h] 
    \centering
    \includegraphics[width=\textwidth]{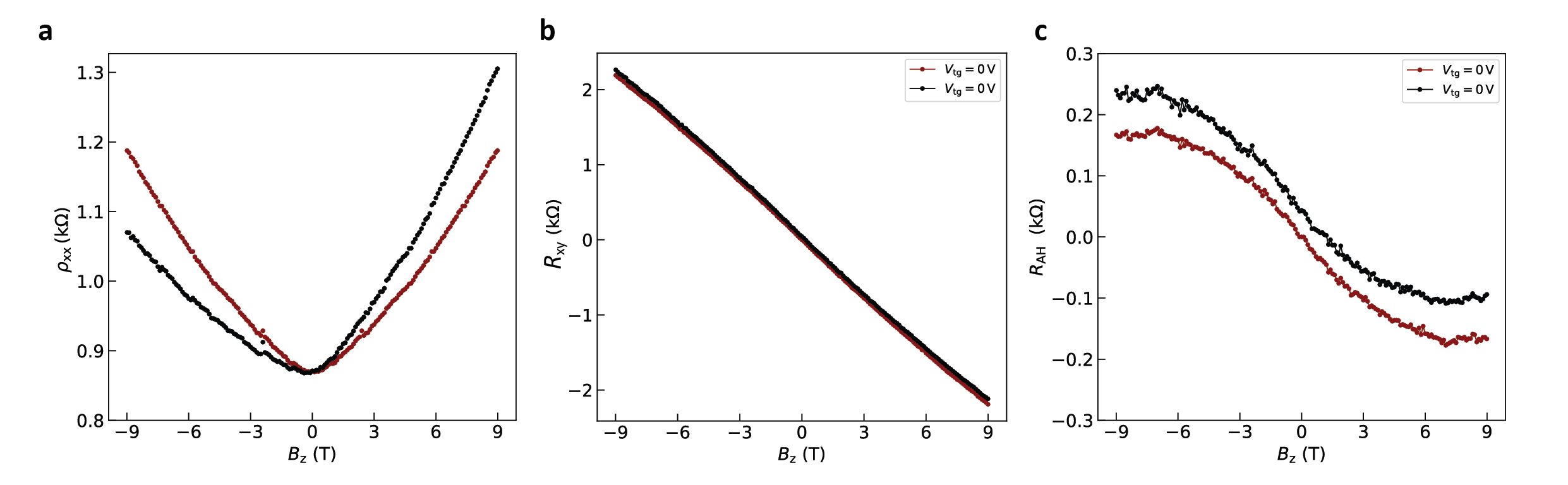}
    \caption{Longitudinal and transverse resistances measured in device A, at $T=\SI{300}{K}$, $V_\mathrm{bg} = \SI{50}{V}$ and $V_\mathrm{tg} = \SI{0}{V}$. (a) Data as measured for the longitudinal resistivity ($\rho_\mathrm{xx}$). (b) Transverse resistance ($R_\mathrm{xy}$), compared with its anti-symmetrized component versus $B_\mathrm{z}$. (c) Anomalous Hall effect component ($R_\mathrm{AH}$), acquired by subtraction of a linear fit from the raw $R_\mathrm{xy}$ data, and from the anti-symmetrized $R_\mathrm{xy}$ component (in panel b), done separately for comparison. The linear fit is the average of the linear fits to the $R_\mathrm{xy}$ data for $B_\mathrm{z} >\SI{8}{T}$ and $B_\mathrm{z} < \SI{-8}{T}$.}
    \label{Fig:RTAHE}
\end{figure*}

\newpage
\FloatBarrier
\textbf{\large{\textcolor{black}{4.~Anomalous Hall effect}}}\\

First, we investigate the $R_\mathrm{xy}$ measurements for the range of $|B_\mathrm{z}|< \SI{1}{T}$. The $R_\mathrm{xy}$ signal above 1 T is expected to linearly increase vs.~$B_\mathrm{z}$ due to the linear behavior of both ordinary Hall and anomalous Hall (AH) effects. Thus, we can linearly fit the $R_\mathrm{xy}$ data for the range of $|B_\mathrm{z}|> \SI{1}{T}$ as shown in \Cref{fig:Rxy_nonlinear}a. The subtraction of the linear fit from the measured $R_\mathrm{xy}$ highlights the non-linearity of the $R_\mathrm{xy}$ within the range of $|B_\mathrm{z}|< \SI{1}{T}$. This non-linearity can be related to the non-linear magnetization behavior of the outer-most CPS layer ($\vec{M}_\mathrm{CPS}$) vs.~$B_\mathrm{z}$. Despite the non-linearity, the magnitude of the $R_\mathrm{xy}$ at $B_\mathrm{z}\sim 0$ is much smaller than what one could expect from the $z-$component of the magnetization (which is expected to be mainly out-of-plane with 20 degrees of misalignment with respect to the $c-$axis~\cite{peng2020magnetic}). This discrepancy can be related to the fact that the outer-most CPS layer has only one neighboring CPS layer and thus the smaller exchange interactions can result in a different canting angle of the $\vec{M}_\mathrm{CPS}$ at $B_\mathrm{z}\sim \SI{0}{T}$, compared to the expected magnetic ordering in bulk CPS. Yet, the clear switches in the $R_\mathrm{xy}$, appearing precisely at the spin-flop transition field of $\SI{0.8}{T}$, reassure that the outermost CPS layer is following the expected bulk magnetization behavior at a finite $B_\mathrm{z}$.

\begin{figure*}[h]
    \centering
    \includegraphics[width=0.9\textwidth]{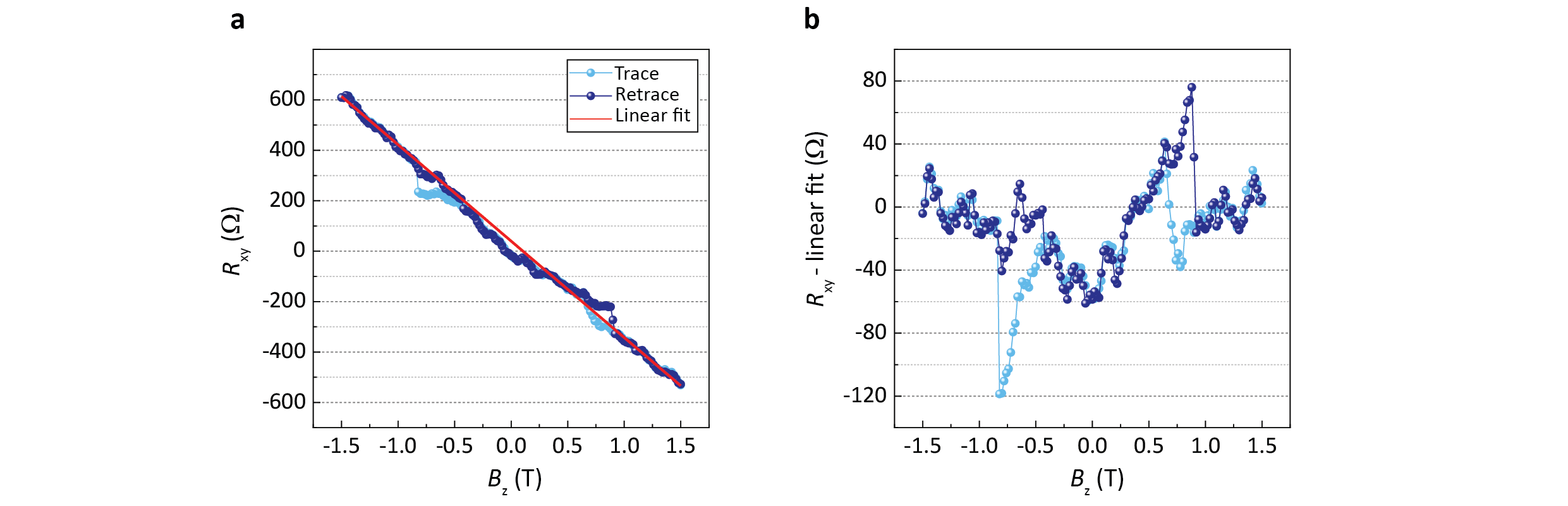}
    \caption{(a) $R_\mathrm{xy}$ vs. $B_\mathrm{z}$ at $V_\mathrm{tg} = 0$ and $T=\SI{1.8}{K}$, shown for trace (1.5 T to -1.5 T) and retrace measurements. The red line is the linear fit to the data for the range of $|B_\mathrm{z}|>\SI{1}{T}$. (b) $R_\mathrm{xy}$ with the linear fit in panel a subtracted from,  to highlight the non-linearity of the transverse resistance vs. $B_\mathrm{z}$. } 
    \label{fig:Rxy_nonlinear}
\end{figure*}

\textcolor{black}{We extract the AH effect component from the long-range $B_\mathrm{z}$-dependence of $R_\mathrm{xy}$ at various temperatures (shown in Figure 1f in the main manuscript). At low temperatures and large $B_\mathrm{z}$, $R_\mathrm{xy}$ is affected by the quantum Hall (QH) plateaus. However, it is yet possible to estimate the magnitude of the AH component, considering the extracted carrier density ($n$) as described in section 8. Having $n$, we estimate the Hall resistance vs $B_\mathrm{z}$ ($R_\mathrm{Hall}=B_\mathrm{z}/ne$) and we subtract that from the anti-symmetrized component of the measured $R_\mathrm{xy}$. In \Cref{fig:AHE_Vg0V_vsT}a, we show the comparison of the anti-symmetrized component of $R_\mathrm{xy}$, $R_\mathrm{Hall}$ and $R_\mathrm{AH}$ vs. $B_\mathrm{z}$. The $R_\mathrm{xy}$ vs $B_\mathrm{z}$ clearly changes slope at about $\SI{4.5}{T}$ due to approaching a QH plateau in the range of $\SI{6}{T}<B_\mathrm{z}<\SI{8.5}{T}$. Thus, the subtraction of the linear $R_\mathrm{Hall}$ from $R_\mathrm{xy}$ in the range of $B_\mathrm{z}>\SI{4.5}{T}$ results in the pronounced change in the extracted $R_\mathrm{AH}$. Yet, we can have an estimate for $R_\mathrm{AH}$ by taking an average of the signal at $B_\mathrm{z}>\SI{6}{T}$ which is about $\SI{700}{\Omega}$. In \Cref{fig:AHE_Vg0V_vsT}b we show the $R_\mathrm{AH}$ vs. $B_\mathrm{z}$ extracted by the explained procedure at various $T$ up to \SI{100}{K}. In panel c we estimate the maxima of $R_\mathrm{AH}$ as the average of the $R_\mathrm{AH}$ for the range of $B_\mathrm{z}>\SI{6}{T}$. Note, that this averaging results in slightly smaller values for the $R_\mathrm{AH}$ maxima, which can be considered as a lower bound for the estimation.        }

\begin{figure*}[h]
    \centering
    \includegraphics[width=0.9\textwidth]{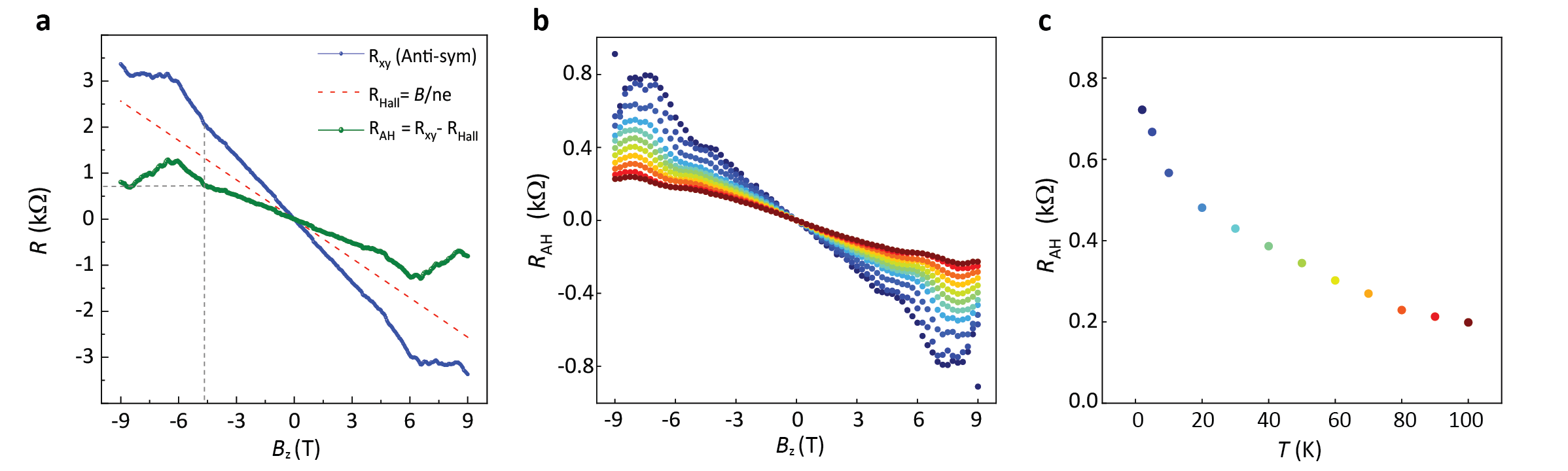}
    \caption{\textcolor{black}{(a) Anti-symmetrized component of $R_\mathrm{xy}$ vs. $B_\mathrm{z}$ (shown in blue). The dashed line shows the Hall resistance ($R_\mathrm{Hall} = B_\mathrm{z}/ne$) determined from the extracted carrier density (estimated in the procedure explained in section 8). The subtraction of $R_\mathrm{Hall}$ from the $R_\mathrm{xy}$ results in the green line, associated with $R_\mathrm{AH}$. (b) $R_\mathrm{AH}$ extracted at various temperatures (from 2 to \SI{100}{K}). (c) The maximum of $R_\mathrm{AH}$ is extracted for $B_\mathrm{z}>\SI{6}{T}$ at various temperatures. }} 
    \label{fig:AHE_Vg0V_vsT}
\end{figure*}

%\textcolor{black}{In \Cref{fig:AHE_Vg0V}, we show the symmetric and anti-symmetric components of $\rho_\mathrm{xx}$ and $R_\mathrm{xy}$ at room temperature, compared for $V_\mathrm{tg}= 0, \pm5$ V. The resistivity profiles clearly show that there is a finite magnetization of the CPS, persisting up to room temperature, which is also picked up by the AH effect, as extracted and shown in the main manuscript, Figure 1f.     }

%\begin{figure*}[h]
%    \centering
%    \includegraphics[width=1\textwidth]{Figures_supplementary/AHE_RoomT_Gatedep.png}
%    \caption{\textcolor{black}{(a) Symmetric component of the resistivity and (b) anti-symmetric component of the transverse resistance vs. $B_\mathrm{z}$, measured at $V_\mathrm{tg}= 0, \pm\SI{5}{V}$ and $T=\SI{300}{K}$.  }} 
%    \label{fig:AHE_Vg0V}
%\end{figure*}

\begin{figure*}[h]
    \centering
    \includegraphics[width=1\textwidth]{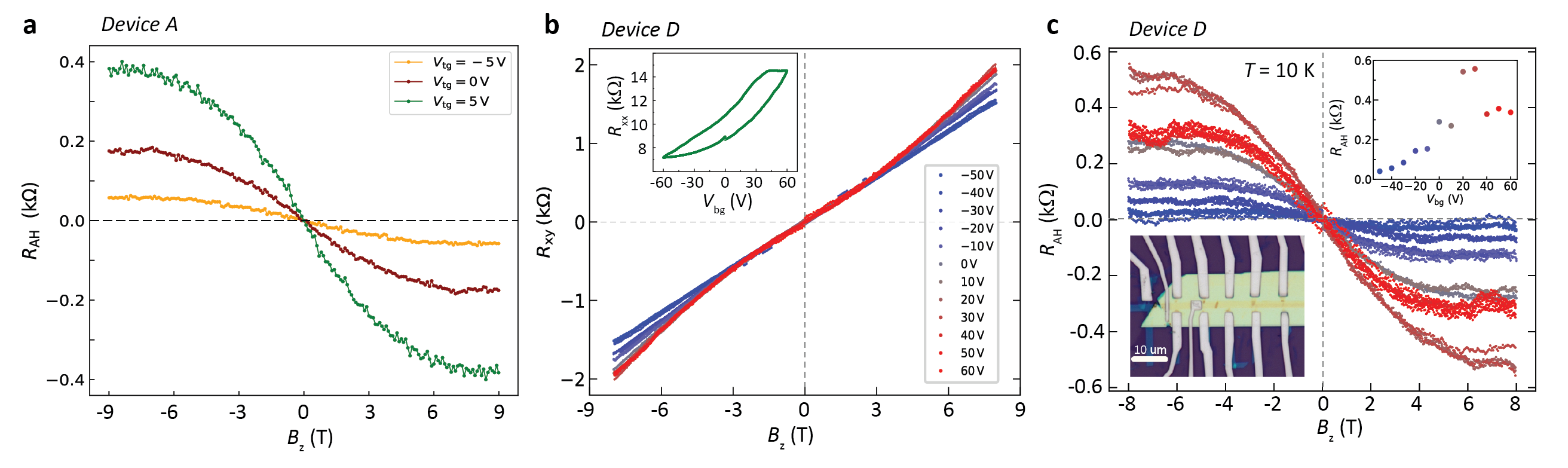}
    \caption{\textcolor{black}{(a) Anomalous Hall signal extracted at $V_\mathrm{tg}= -5 , 0, \SI{5}{V}$, at room temperature in device A. (b) Anti-symmetrized component of $R_\mathrm{xy}$ vs. $B_\mathrm{z}$ at various $V_\mathrm{bg}$ in device D, measured at $T= \SI{10}{K}$. Inset: $V_\mathrm{bg}$ dependence of $R_\mathrm{xx}$. (c) Non-linear component of $R_\mathrm{xy}$ extracted from panel b at various $V_\mathrm{bg}$, associated to the AH effect. Top-right inset: Back gate-dependence of the AH signal. Bottom left inset: Optical micrograph of device D.  }} 
   \label{fig:AHE_deviceA_deviceD}
\end{figure*}

\begin{figure*}[h]
    \centering
    \includegraphics[width=1\textwidth]{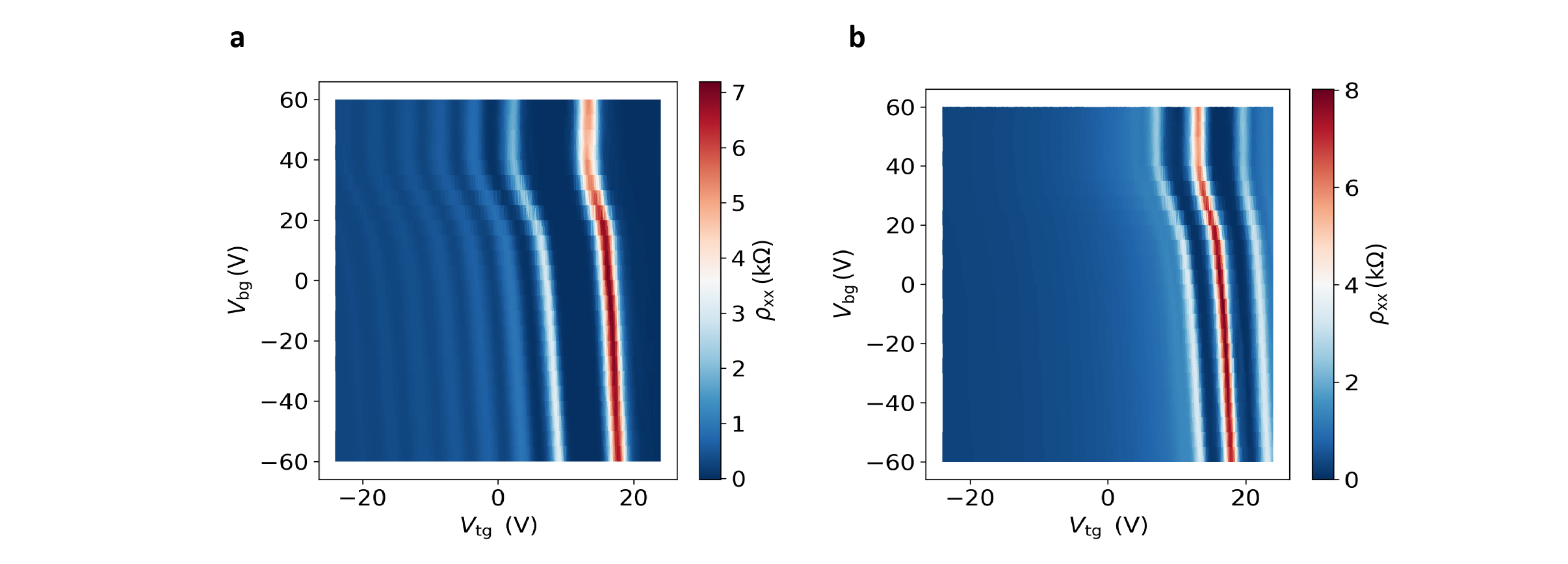}
    \caption{Top- and back-gate dependence of the longitudinal resistivity $\rho_\mathrm{xx}$, measured in device A at $B_\mathrm{z} =\SI{9}{T}$ (panel a), and at $B_\mathrm{z} =\SI{3}{T}$ (panel b). %(panel a,b, respectively) and in device B at $B_\mathrm{z} = \SI{3}{T}$ (panel c). 
    The sweeping of the back-gate voltage shifts the Dirac peak and so the Landau levels to lower top-gate voltages until the shift saturates at $V_\mathrm{bg}\geq\SI{40}{V}$.} %In device B, a second Dirac point emerges at $V_\mathrm{bg}\geq\SI{10}{V}$, the presence of which is associated to possible inhomogeneities in the graphene channel.} 
    \label{fig:double_gate}
\end{figure*}

\begin{figure*}[h]
    \centering
    \includegraphics[width=0.75\textwidth]{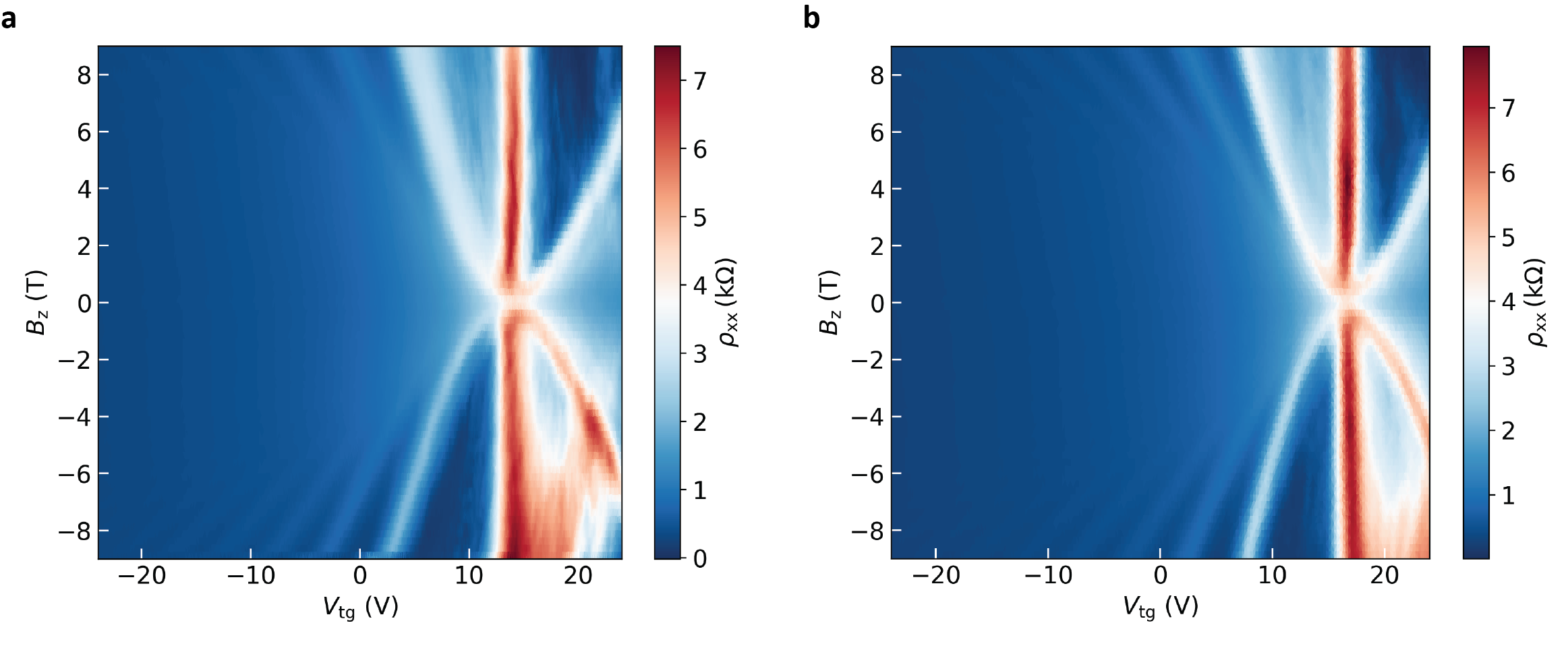}
    \caption{Landau fan diagrams measured in device A at $T= 1.8\,$K (a) at $V_\mathrm{bg}= 50\,$V (b), and at $V_\mathrm{bg}= 0\,$V. } 
    \label{fig:Back-gate-effect}
\end{figure*}

\begin{figure*}[h]
    \centering
    \includegraphics[width=1\textwidth]{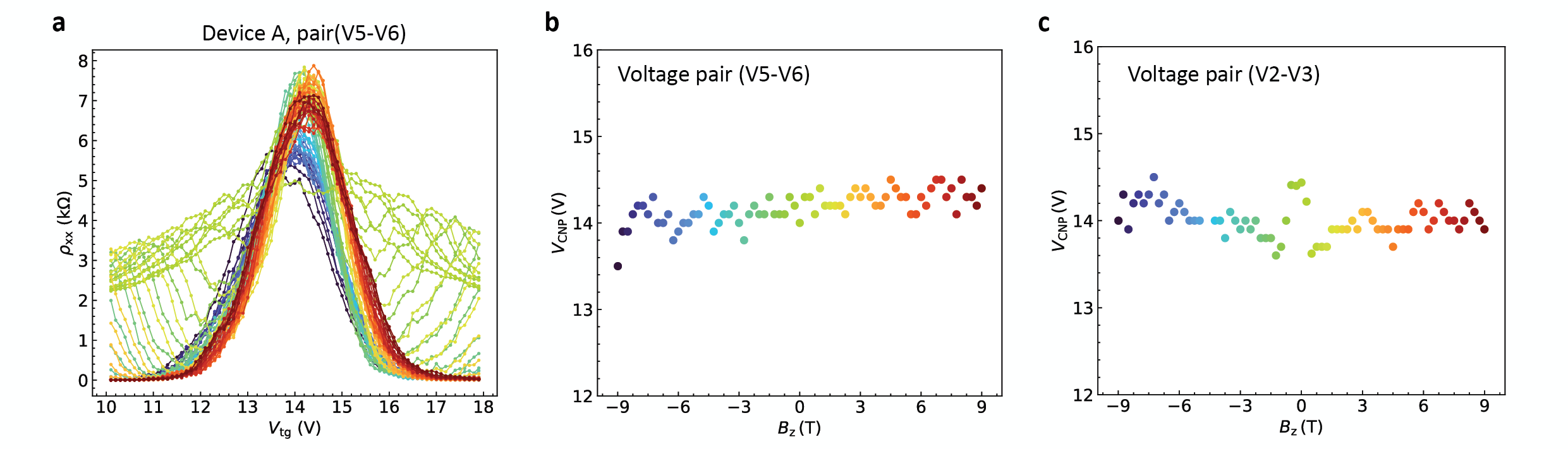}
    \caption{\textcolor{black}{(a) Modulation of the resistivity $\rho_\mathrm{xx}$ versus $V_\mathrm{tg}$ at various $B_\mathrm{z}$, measured in device A, with voltage probe pair V5-V6 (see \Cref{fig:differentpairs} for the device geometry).  (b, c) Modulation of the $V_\mathrm{cnp}$ (voltage of the charge neutrality point, associated with the maxima of $\rho_\mathrm{xx}$) is shown versus $B_\mathrm{z}$, measured with (b) V5-V6 pair, (c) and with V2-V3 pair.} }
    \label{fig:shift_of_CNP_vs_B}
\end{figure*}

\begin{figure*}[h]
    \centering
    \includegraphics[width=0.95\textwidth]{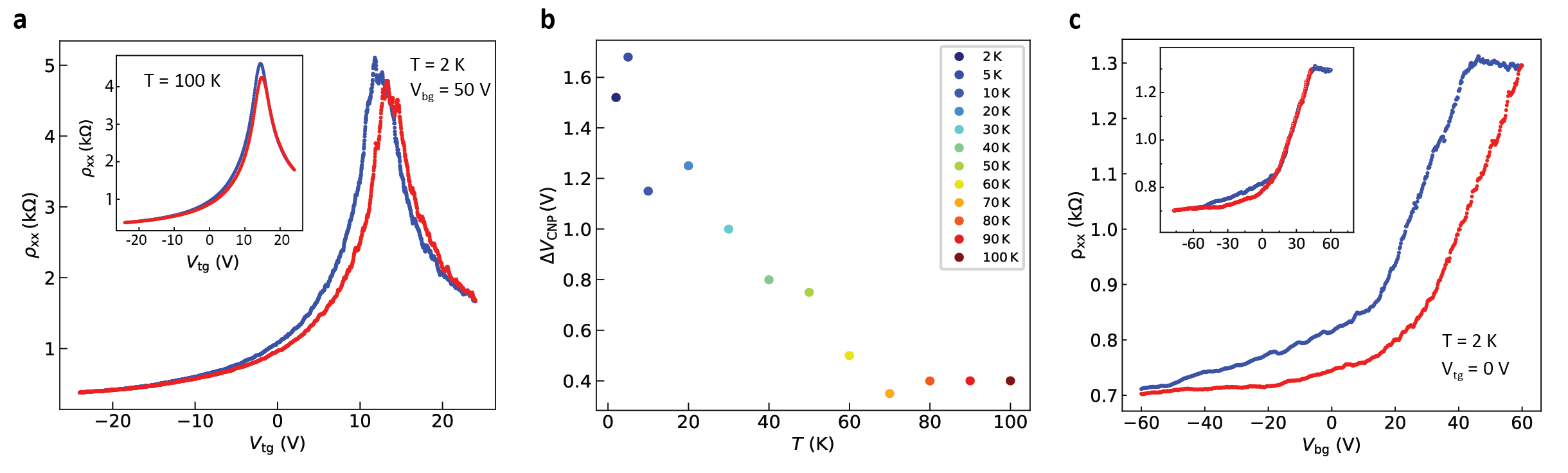}
    \caption{\textcolor{black}{(a) Top-gate dependence of graphene resistivity in device A, measured at $V_\mathrm{bg} = \SI{50}{V}$, $T \sim \SI{2}{K}$ and $B= 0$. Inset: The top-gate dependence measured at $T \sim \SI{100}{K}$. Blue and red curves represent the trace ($V_\mathrm{tg}; -24$ to $\SI{24}{V}$) and retrace measurements, respectively. (b) Temperature dependence of the magnitude of the hysteresis, determined at the charge neutrality point ($\Delta V_\mathrm{cnp}$). (c) Back-gate dependence of the graphene resistivity, measured at $V_\mathrm{tg} = \SI{0}{V}$, $T \sim \SI{2}{K}$ and $B= 0$. Inset: the trace and retrace curves, shown with an offset of the retrace (red) plot with about $V_\mathrm{bg}=\SI{-17}{V}$.} }
    \label{fig:hysteresis3}
\end{figure*}

\begin{figure*}[h]
    \centering
    \includegraphics[width=\textwidth]{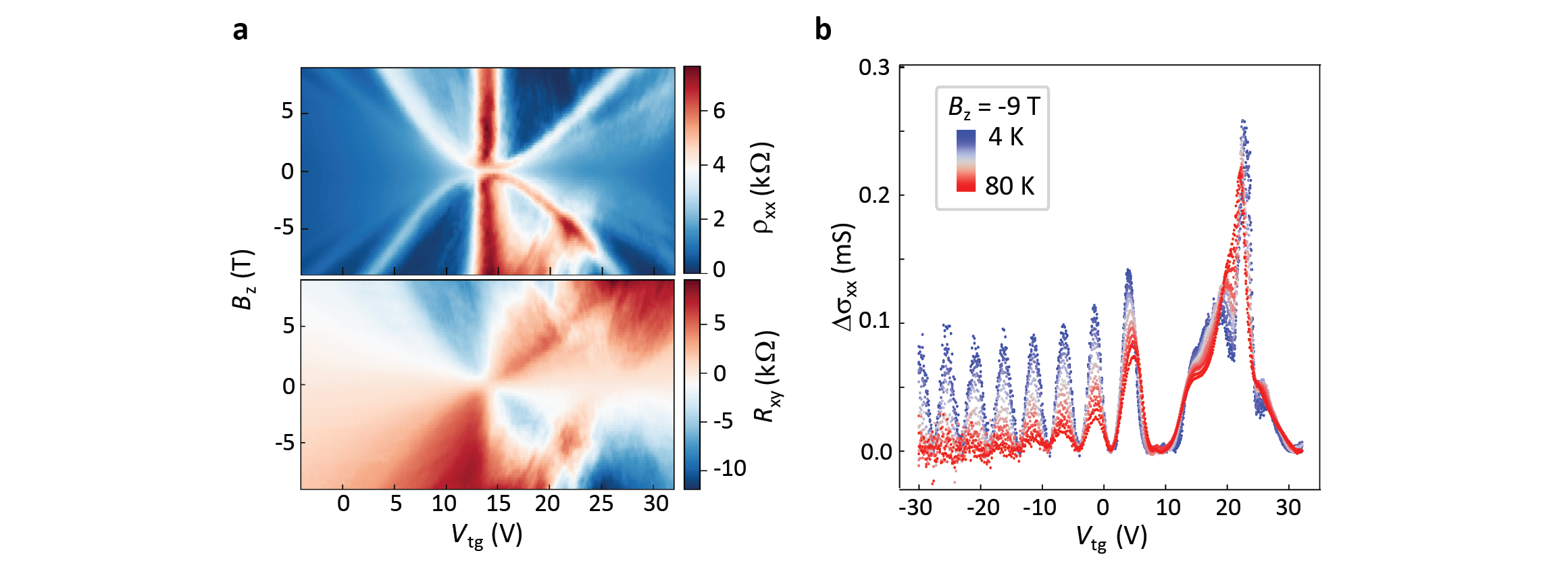}
    \caption{(a) Modulation of $\rho_\mathrm{xx}$ and transverse resistance ($R_\mathrm{xy}$) vs.~applied top-gate voltage and magnetic field, measured with another set of longitudinal voltage probe \textcolor{black}{(V$_2$,V$_3$ in \Cref{fig:differentpairs}) with} a transverse pair (\textcolor{black}{V$_3$,V$_{11}$}). (b) SdHO of the longitudinal conductivity $\sigma_\mathrm{xx} = \rho_\mathrm{xx}/(\rho_\mathrm{xx}^2+R_\mathrm{xy}^2)$ measured at $B_\mathrm{z}=\SI{-9}{T}$ for $T\,=\,4$ to \SI{80}{K}. The $\Delta\sigma_\mathrm{xx}$ plots are obtained by subtracting a (nearly) linear increase in $\sigma_\mathrm{xx}$ vs.~$V_\mathrm{tg}$. } 
    \label{fig:Rxypair}
\end{figure*}

We also perform the room-temperature $B_\mathrm{z}$-dependence of $R_\mathrm{xy}$ in device A with small steps in $B_\mathrm{z}$ at a few $V_\mathrm{tg}$, from which we extract the AH component, as shown in~\Cref{fig:AHE_deviceA_deviceD}a. The $R_\mathrm{AH}$ is extracted by subtracting a linear component related to the ordinary Hall effect from the measured $R_\mathrm{xy}$, considering the magnetization saturation field in CPS to be above \SI{8}{T}. \Cref{fig:AHE_deviceA_deviceD}a shows that the $R_\mathrm{AH}$ in device A clearly persists away from the charge neutrality point up to 300 K.

For low temperatures, as the $R_\mathrm{xy}$ in devices A, B, and C all have considerable contributions from QH plateaus, here we present the anomalous Hall signal at low temperature measured in device D which does not show dominant QH transport at low temperature. Device D only has a back-gate that can limitedly tune the Fermi energy which gets ineffective at $V_\mathrm{bg}>\SI{40}{V}$ due to the screening by the CPS midgap states (as also observed in the other devices, see \Cref{fig:hysteresis3}c). As \Cref{fig:AHE_deviceA_deviceD}b shows, there is a clear non-linearity in $R_\mathrm{xy}$ signal vs. $B_\mathrm{z}$, which is extracted in panel c for various $V_\mathrm{bg}$ in device D. The strong non-linearity of the Hall voltage at various gate voltages (far from the charge neutrality point) assures the presence of a strong anomalous Hall effect in the magnetized graphene. Note that the magnitude of the AH effect depends on the strength of the spin-orbit coupling and magnetism induced in graphene which is different in various samples, depending on the crystallographic orientations and relaxation of the graphene and CPS layers, which is not controlled in the sample fabrication process in this work.   

\FloatBarrier
\textbf{\large{5.~Evaluation of the interfacial charge transfer}}\\

In previous studies on graphene in the proximity of 2D magnetic materials, gate-dependent interfacial charge transfer from the 2D magnet to the graphene has been reported~\cite{tseng2022gate, wang2022quantum}. That has given rise to the observation of unconventional Landau fan diagrams that flatten off vs. gate due to the suppression of the gate action by the charge transfer. In this section, we provide several arguments against the dominant contribution of the interfacial charge transport in graphene-CrPS$_4$ which is important to ensure that the modulation of the LLs is due to the magnetic proximity effect, rather than charge-related effects.

To investigate the possible gate-dependent interfacial charge transfer in our graphene-CrPS$_4$ heterostructure, we conducted similar measurements where both top- and back-gates were swept. \Cref{fig:double_gate} shows the $\rho_\mathrm{xx}$ versus $V_\mathrm{tg}$ and $V_\mathrm{bg}$ measured at different magnetic fields in device A. We observe that increasing the $V_\mathrm{bg}$ shifts the position of the charge neutrality point, while the Landau gaps stay constant. %up to $V_\mathrm{bg} = \SI{40}{V}$, which could be the onset of Fermi level getting close to the conduction band of the CPS~\cite{wu2023gate}.  %In \Cref{fig:double_gate}a and b, Landau levels are present and show similar behavior. At around $V_\mathrm{bg} = \SI{20}{V}$, the 2D magnet seem to be, which results in a faster shift in the charge neutrality point,
In \Cref{fig:Back-gate-effect}, we further show the Landau fans measured at $V_\mathrm{bg}=\,0$ and \SI{50}{V}, indicating no significant change in the fan diagrams. As the $V_\mathrm{bg}$ is only shifting the position of the charge neutrality point, and the Landau gaps remain constant for the full $V_\mathrm{bg}$ and $V_\mathrm{tg}$ range, we can conclude that the charge transfer from the 2D magnet is minimal in our devices and does not affect the QH transport in the graphene channel.

\textcolor{black}{Another indication for the suppression of the gate-activated interfacial charge transfer in the graphene-CPS heterostructure is the absence of a shift of the charge neutrality point as a function of the external magnetic field, in contrast to the report on Gr-CrX$_3$ and Gr-CrOCl heterostructures~\cite{wang2022quantum, tseng2022gate}. In \Cref{fig:shift_of_CNP_vs_B} we show the behavior of the resistivity peak at the Dirac point at various $B_\mathrm{z}$. Panels b and c show the dependence of the voltage associated with the charge neutrality point ($V_\mathrm{cnp}$) on $B_\mathrm{z}$, measured with different voltage probe pairs, showing no correlation with the CPS magnetic ordering. This is an important observation that also rules out the contribution of interfacial charge transfer in determining the position of the Landau levels. Also note that the dependence of the interfacial charge transfer on magnetic ordering is expected to affect the position of all Landau levels (including the zeroth LL), as reported in Ref.~\cite{wang2022quantum, tseng2022gate}. However, Figure 2c in the main manuscript shows a fully linear behaviour of the Landau fans for the full range of $V_\mathrm{tg}$, slightly away from the charge neutrality point ($|V_\mathrm{tg} - V_\mathrm{cnp}|>\SI{5}{V}$), showing no correlation with the magnetization behaviour. We also refer to the Landau fan diagram of device C in particular, shown in \Cref{fig:QHtransportDeviceC}, indicating no non-linearity for the full range up to $B_\mathrm{z} = \SI{14}{T}$.}

\textcolor{black}{We further evaluate the hysteresis in the gate-dependence of the graphene resistivity, shown in \Cref{fig:hysteresis3}. In contrast to the large gate-dependent hysteresis observed in graphene-based heterostructures with interfacial charge transfer~\cite{wang2022quantum, tseng2022gate}, \Cref{fig:hysteresis3}a shows a small hysteresis present in the top-gate dependence which gets even smaller for the electron-doped range. This is while the electron-doped range of the gate-voltage is expected to bring the Fermi energy closer to the edge of the conduction band in the CPS, which should result in larger interfacial charge transfer and larger hysteresis. In \Cref{fig:hysteresis3}b, we evaluate the temperature-dependence of the hysteresis, determined at the Dirac point ($\Delta V_\mathrm{cnp}$), showing a decay of the hysteresis with temperature. This decay in hysteresis is also not expected if the transport is influenced by interfacial charge transfer. That is because the interfacial charge transfer with the semiconducting 2D magnet is a thermally activated process that is expected to enhance with increasing temperature. Thus, the minimal hysteresis and its decay with temperature are other proofs against the contribution of interfacial charge transfer in the transport measurements.}

\textcolor{black}{In \Cref{fig:hysteresis3}c, we also report the hysteresis in resistivity vs.~$V_\mathrm{bg}$. The trace measurement shows saturation of the resistivity for $V_\mathrm{bg}> \SI{40}{V}$, which indicates that the gate-induced charges populate the localized states in the CPS that accumulate at the CPS-SiO$_2$ interface and screen the back-gating action for the range of $V_\mathrm{bg}> \SI{40}{V}$. We see that the $V_\mathrm{bg}$-dependent trace and retrace profiles are quite similar (shown in the inset with an offset of the retrace measurements). That indicates an almost constant hysteresis for the full range of the applied $V_\mathrm{bg}$. The gate-independent hysteresis through the back-gating confirms that the hysteresis is not due to Fermi energy reaching the CPS conduction band (as that would change the size of the hysteresis for positive gate-voltages), but rather it is related to the screening
by the localized midgap states. }

\textcolor{black}{On a separate note on the interfacial charge transfer, we would like to highlight that, as explained in Ref.~\cite{wang2022quantum}, the interfacial charge transfer largely enhances the Fermi velocity to a few times greater than that of graphene. In contrast, the extracted Fermi velocity in magnetized graphene from our analysis, shown in Figure 2b (main manuscript), is very close to that of pristine graphene. }

%\textcolor{black}{Another indication that rules out the dominant contribution of the interfacial charge transfer is the linearity }

Moreover, a recent report on transport in CPS using graphite contacts~\cite{wu2023gate} shows the onset of current in CPS at gate electric field above \SI{0.25}{V/nm} at \SI{2}{K} with the resistivity of the bulk CPS to be at least 3-4 orders of magnitude larger than that of graphene. Additionally, in contrast to the large temperature-dependent charge transport reported in CPS~\cite{wu2023gate}, our QH measurements at various temperatures (shown in \Cref{fig:Rxypair}b) do not show any signature correlating with the CPS thermal activation. Thus, it confirms the suppression of the interfacial charge transfer between graphene and CPS in the heterostructure. 

Furthermore, we note that in our devices there have been several contacts that have been used to only address the transport in the CPS. Our 2T measurements in CPS up to RT do not show any current flow in the CPS above the noise level (pA). This confirms that there is no parallel transport in the CPS layer.\\

\textbf{\large{6.~SdHOs detected at different regions of device A}}\\

\begin{figure*}[h]
    \centering
    \includegraphics[width=\textwidth]{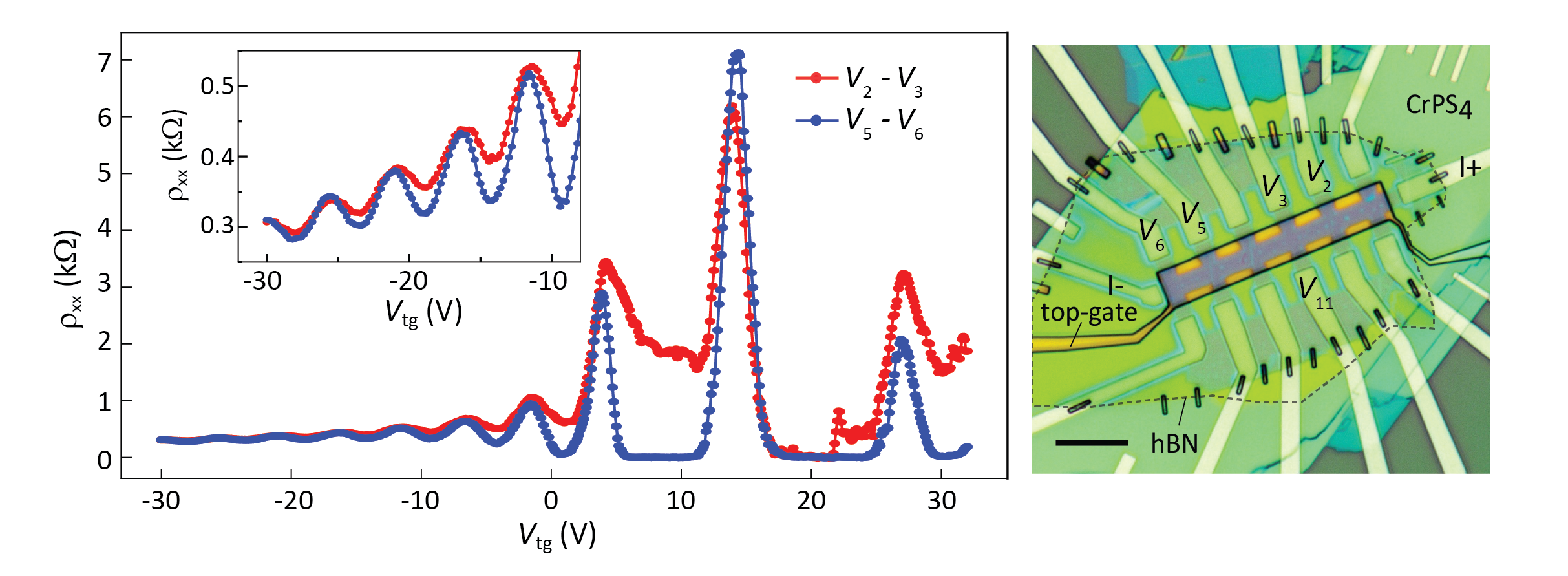}
    \caption{Shubnikov-de Haas oscillations (SdHOs) at $B_\mathrm{z}= \SI{9}{T}$ and $V_\mathrm{bg}= \SI{50}{V}$ detected by different voltage probes in device A. This comparison shows that the SdHOs are the same for different pairs, regardless of possible inhomogeneities in the sample. The inset shows the SdHOs more clearly for the range of $-30<V_\mathrm{tg}<-10\,$V. The used voltage and current probes are indicated in the optical micrograph of the sample (scale bar $= 10\,\mu m$).     } 
    \label{fig:differentpairs}
\end{figure*}

The behavior of the transverse voltage is addressed in a simultaneous measurement of $\rho_\mathrm{xx}$ and $R_\mathrm{xy}$, shown in Figure~\ref{fig:Rxypair}a for another pair of voltage probes than the one shown in Figure 2a of the main manuscript. Even though additional features appear in the Landau fan (absent in Figure 2a of the main manuscript), the position of the LLs vs.~$V_\mathrm{tg}$ is the same, as shown in \Cref{fig:differentpairs}. 

In \Cref{fig:Rxypair}b, we consider both longitudinal and transverse voltages in a conductivity tensor~$\sigma_\mathrm{xx} = \rho_\mathrm{xx}/(\rho_\mathrm{xx}^2+R_\mathrm{xy}^2)$\cite{qhe_klitzing} measured at various temperatures up to 80 K, obtained after standard background subtraction~\cite{novoselov2005two}.

\newpage
\FloatBarrier
\textbf{\large{7.~Landau fan diagram in device B and C }}\\

%\begin{figure*}[h]
%    \centering
%    \includegraphics[width=0.8\textwidth]{Figures_supplementary/GCPS17/Landaufan_DeviceAandB.png}
%    \caption{Landau fan diagram measured close to the charge neutrality point for (a) device A and (b) device B. } 
%    \label{fig:Lfd_DeviceAB}
%\end{figure*}
\textcolor{black}{The Landau fan diagram measured in device B shows the presence of a second Dirac peak that seems to be due to contributions from the inhomogeneous regions in the graphene channel with different doping. Note that the overlap of the two Landau fans is causing a slight non-linearity in the electron-like Landau fans and the zeroth LL which is absent on hole-like Landau fans.}

\begin{figure*}[h]
    \centering
    \includegraphics[width=0.75\textwidth]{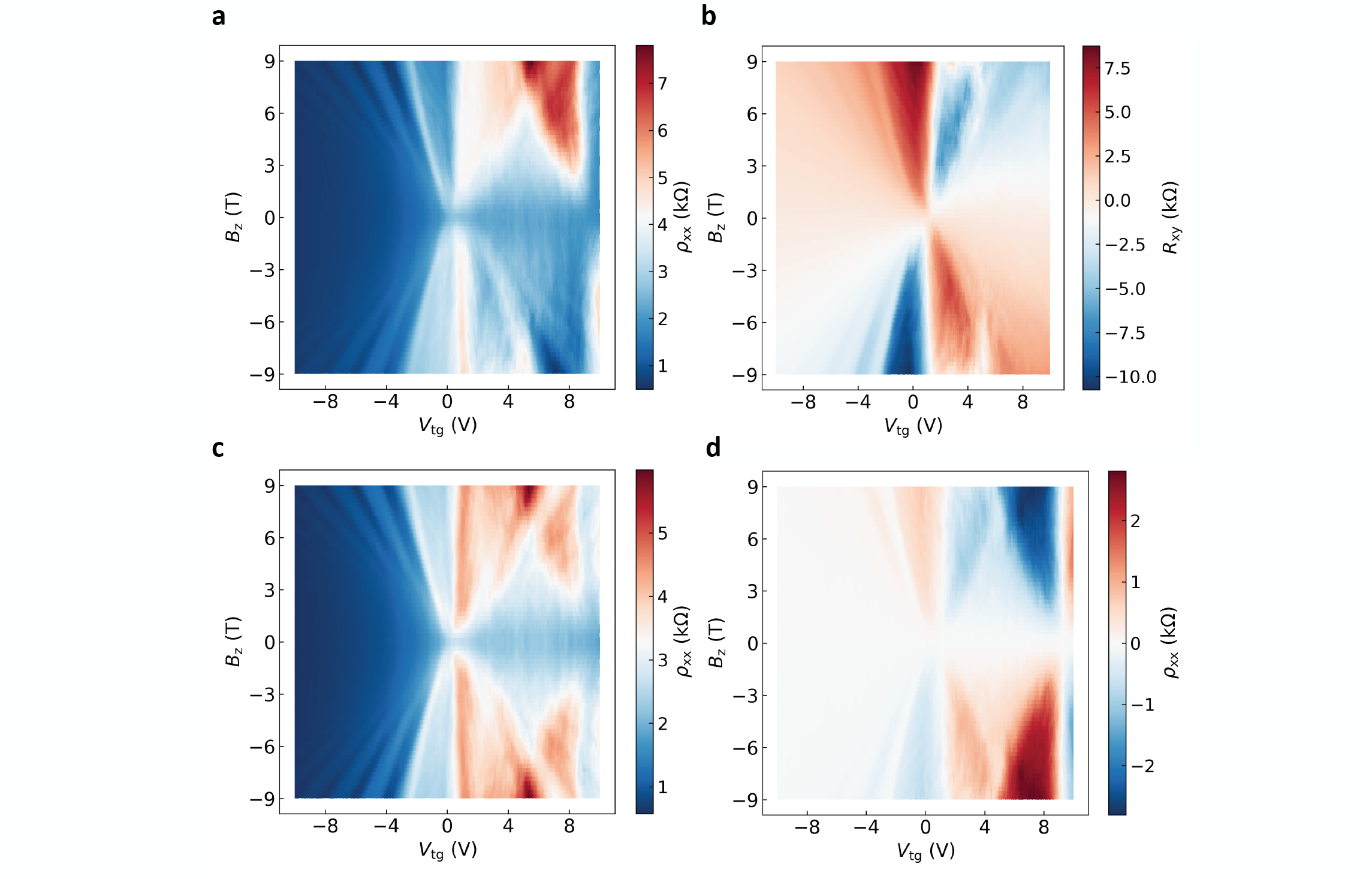}
    \caption{Landau fan diagram measured in device B. Modulation of longitudinal resistivity (a) and transverse resistance (b) vs. magnetic field and top-gate voltage, measured in 4-terminal geometry at $T=\SI{1.8}{K}$ and $V_\mathrm{bg}=\SI{50}{V}$. \textcolor{black}{(c) The Landau fan diagram shown in panel a, symmetrized vs. $B_\mathrm{z}$ and (d) anti-symmetrized vs. $B_\mathrm{z}$.}} 
    \label{fig:Lfd_DeviceAB}
\end{figure*}

The Landau fan diagram in device C is shown in \Cref{fig:QHtransportDeviceC}. This device is measured in a different setup with larger $B_\mathrm{z}$ up to $\SI{14}{T}$. The Landau fan diagram shows that for the full range of $B_\mathrm{z}$ up to \SI{14}{T} the Landau fans are linear, again reassuring that in the graphene-CrPS$_4$ heterostructure, the interfacial charge transfer is suppressed. The Landau fan in device C further shows a deviation of the resistance values at finite $B_\mathrm{z}$ from that expected for the helical states measured at $\SI{0}{T}$ at $V_\mathrm{cnp}$. As it is explained in the main text, depending on the evolution of the zLL gap with $B_\mathrm{z}$ and possible band crossing with respect to the helical states, the detection of the helical states can be disturbed at finite $B_\mathrm{z}$, thus, it is sample-dependent.

\begin{figure*}[h]
  \centering
  \includegraphics[width=0.9\textwidth]{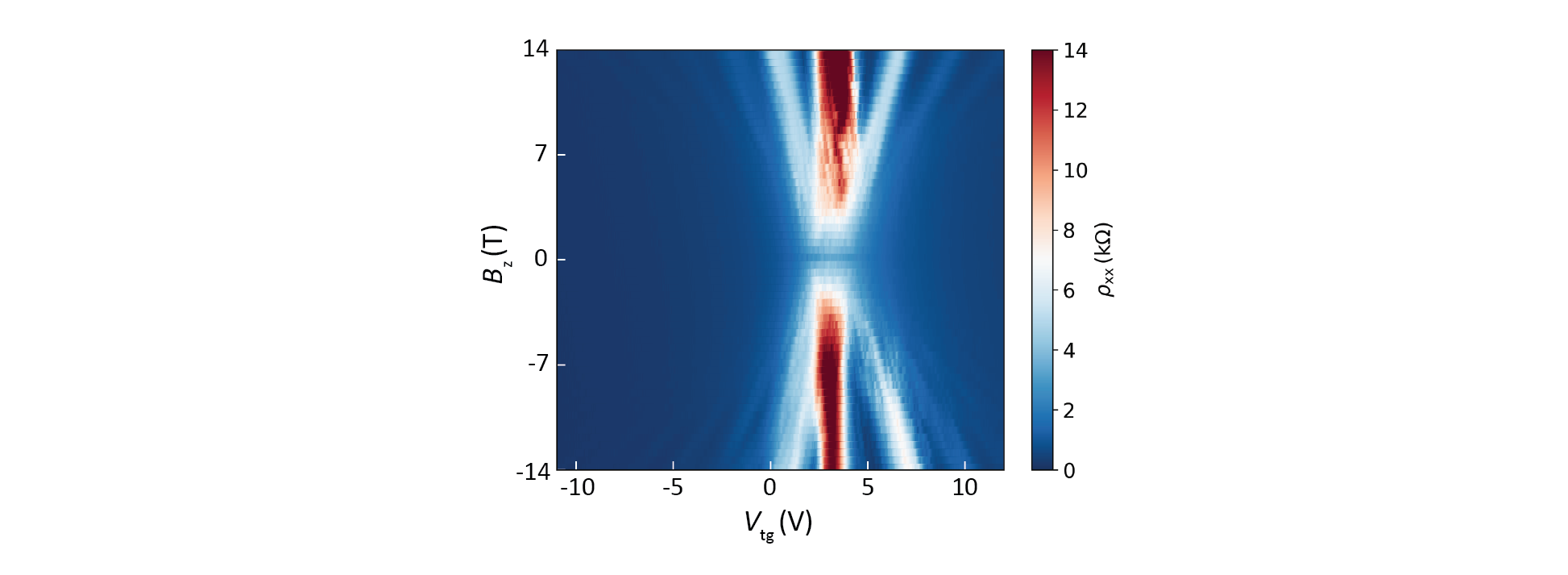}
  \caption{Landau fan diagram measured in device C. }
  \label{fig:QHtransportDeviceC}
\end{figure*}

\newpage
Note that the crystallographic orientation of the graphene and CPS is not controlled in any of the devices, thus the staggered potentials as well as the induced spin-orbit coupling and exchange interactions in each device can be different. Moreover, we also note that the heterostructure in device C is not annealed during the fabrication which may have an impact on the relaxation of the graphene and CPS crystals and affect the proximity effect. \\

\FloatBarrier
\textbf{\large{8.~Carrier density and mobility}}\\

\begin{figure*}[h]
    \centering
    \includegraphics[width=0.8\textwidth]{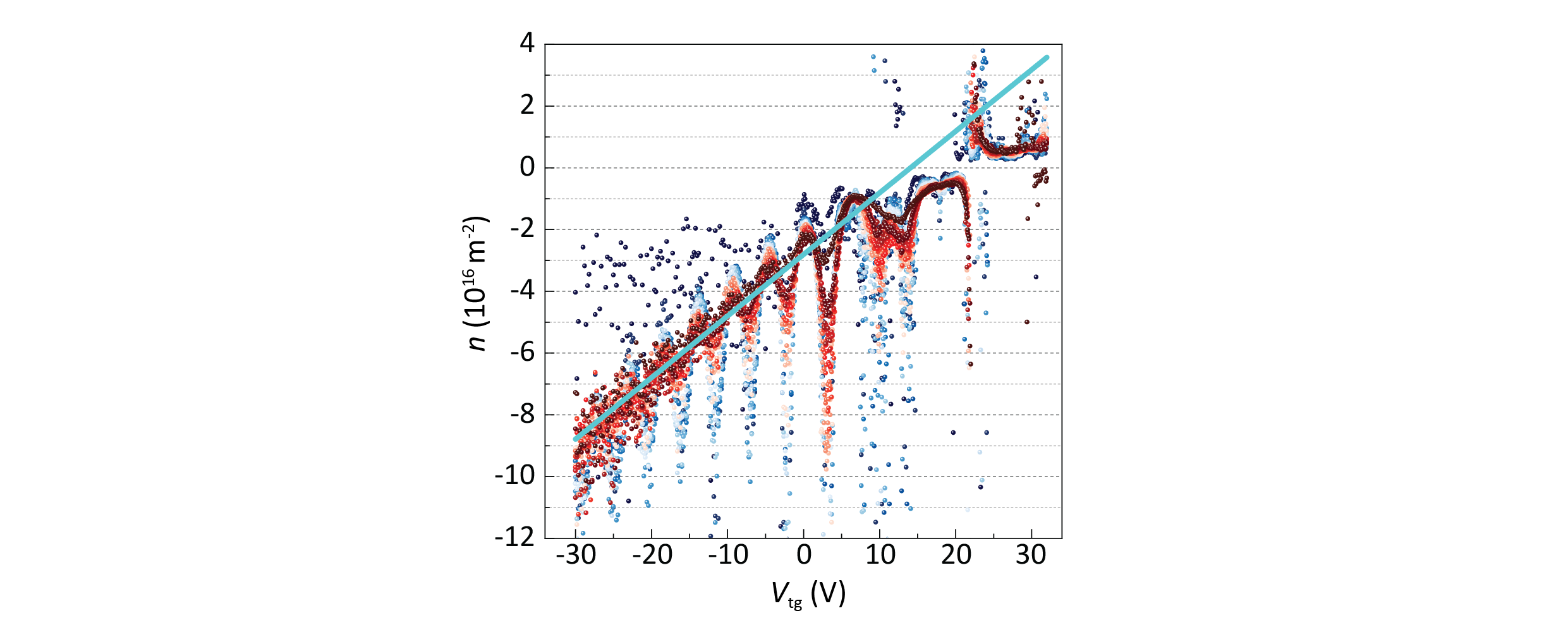}
    \caption{Carrier density ($n$) in device A, derived from the Hall measurements of $R_\mathrm{xy}$, anti-symmetrized vs. $B_\mathrm{z}$. The slopes to the anti-symmetrized $R_\mathrm{xy}$ are taken for $B_\mathrm{z}$ in the range of 8 to \SI{9}{T}. The relation between $n$ vs. $V_\mathrm{tg}$ is considered to be linear, obtained from the linear fit of $n$ vs. $V_\mathrm{tg}$ in the range of -30 to \SI{0}{V}, shown with the blue solid line. The derived relation for the gate-dependence of the carrier density for these measurements is $n [\mathrm{m}^{-2}]= 1.99\times10^{15}(V_\mathrm{tg}-14) \; $.} %We extract $n$ more accurately for higher temperatures, considering the slight shift of the Dirac point vs. $T$.  }
    \label{fig:n_vs_Vtg_device_A}
\end{figure*}

Encapsulation of the graphene between hBN and CPS can change the environmental dielectric constant which results in an effective capacitance that is different from the geometrical one. Thus, in a more reliable approach, we estimate the effective carrier density from the Hall voltage, measured vs.~$B_\mathrm{z}$ at various gate voltages and temperatures. At each top-gate voltage, the Hall component is separated from the AH component and is related to the carrier density by $R_{\mathrm{Hall}} = \frac{B_\mathrm{z}}{ne}$ where $R_\mathrm{Hall} = R_\mathrm{xy} - R_\mathrm{AH}$. We extract $n$ from the $B_\mathrm{z}$-dependence of $R_\mathrm{xy}$ at each $V_\mathrm{tg}$, by taking the slope of $R_\mathrm{xy}$ vs. $B_\mathrm{z}$ for $B_\mathrm{z}\,>\,$\SI{8}{T}. In \Cref{fig:n_vs_Vtg_device_A}%and \ref{fig:n_vs_Vtg_device_B} 
~we show the gate-dependence of the extracted $n$ from the Hall effect for device A. Using the linear fit, we obtain the gate-dependence of the effective $n$ in the graphene channel.  

\begin{figure*}[h]
    \centering
    \includegraphics[width=1.05\textwidth]{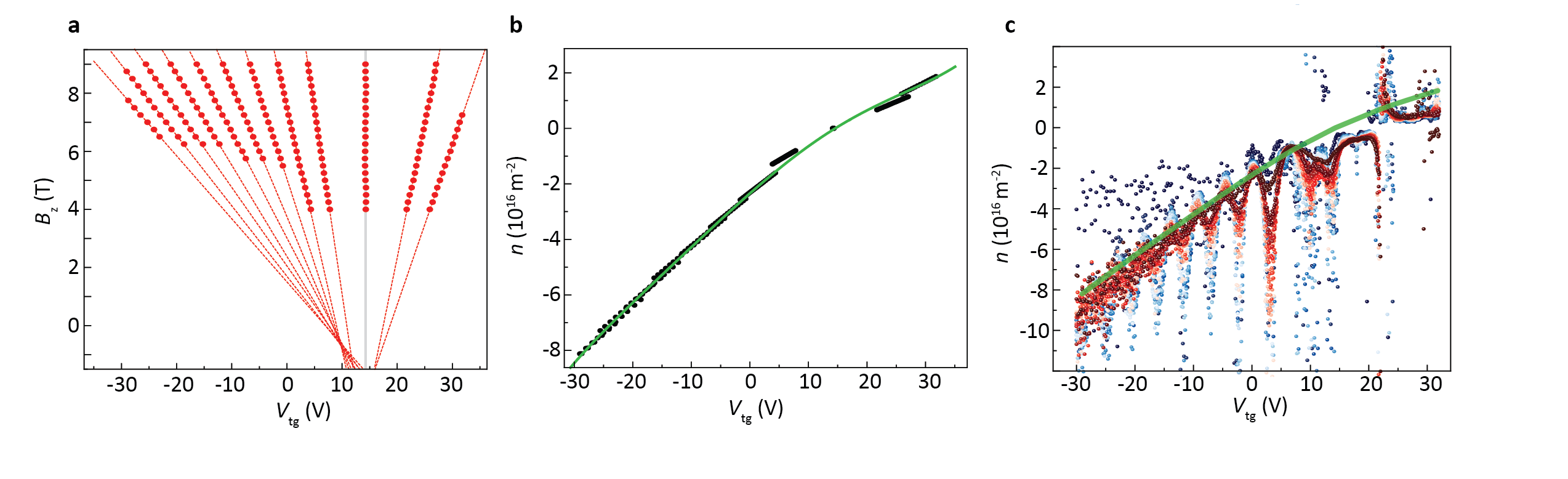}
    \caption{\textcolor{black}{(a) Position of the maxima of resistivity in the SdHOs (shown with red circles) for gate voltages slightly away from the charge neutrality point ($|V_\mathrm{tg}-V_\mathrm{cnp}|>\SI{5}{V}$), obtained from Figure 2a of the main manuscript for device A. The red dashed lines are the linear fits to the SdHO peaks in the magnetized graphene. (b) Extracted $n$ from the slopes of Landau fans in panel a, plotted vs $V_\mathrm{tg}$. The green line is a polynomial fit to the data. (c) Comparison of the polynomial fit (green line in panel b) with the $n$ vs. $V_\mathrm{tg}$ extracted from the Hall effect (as shown in \Cref{fig:n_vs_Vtg_device_A}).  }}
    \label{fig:GetBTforBlargerthan4T}
\end{figure*}

Another approach to extract $n$ is by considering the $B_\mathrm{z}$ vs.~$V_\mathrm{tg}$ dependence of the SdHO peaks for each LL, following the relation:
\begin{equation}
    \frac{c_\mathrm{ef}}{e} (V_\mathrm{tg}-V_\mathrm{cnp}) = \frac{2eN_\mathrm{LL}}{\hbar \pi B_\mathrm{z}}
    \label{eq:B_tot}
\end{equation}

\noindent with $c_\mathrm{ef}$ as the effective capacitance per unit area and $N_\mathrm{LL}$ as the Landau level number. From the slope of Landau fans ($B_\mathrm{z}$ vs $V_\mathrm{tg}$) for each LL, shown by the red line in \Cref{fig:GetBTforBlargerthan4T}a, one can extract the coefficients and determine $n$. The extracted $n$ per Landau level is plotted vs. $V_\mathrm{tg}$ in panel b. We observe that for $N_\mathrm{LL}\leq -2$, the $n$ vs. $V_\mathrm{tg}$ stays fully linear. However, for the $V_\mathrm{tg}>\SI{5}{V}$, the $n$ vs. $V_\mathrm{tg}$ dependence slightly deviates from the linear line. As mentioned in the main text, this slight non-linearity is likely to be related to a contribution of the CPS localized states in the effective capacitance when the Fermi energy gets closer to the CPS conduction band edge at large positive gate voltages. To account for the slight non-linearity we consider a polynomial fit to the $n$ vs. $V_\mathrm{tg}$ data, shown by the green line in panel b. In panel c we compare the $n$ vs. $V_\mathrm{tg}$ extracted from the slope of the Landau fans (in panel b) with that extracted from the ordinary Hall effect (in \Cref{fig:n_vs_Vtg_device_A}), showing good consistency and reliability of the extracted $n$ by the two described methods.

\begin{figure*}[h]
    \centering
    \includegraphics[width=0.9\textwidth]{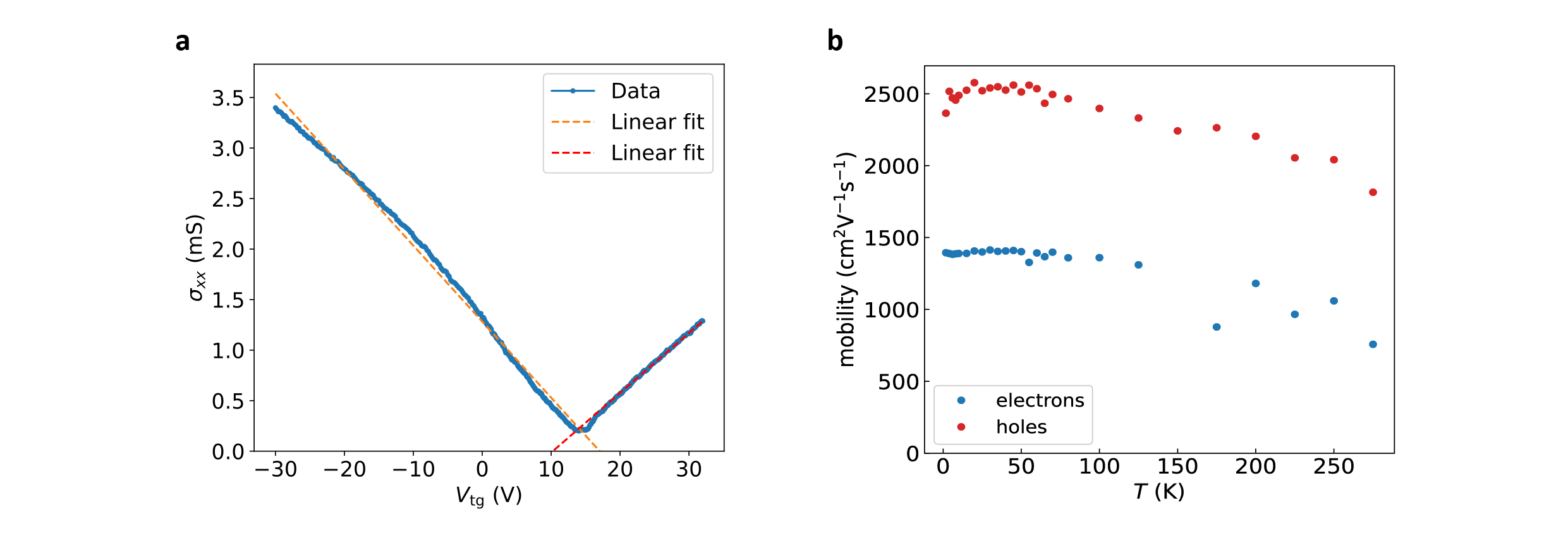}
    \caption{(a) Gate-dependence of the conductivity ($\sigma_{\mathrm{xx}}$) at $B_\mathrm{z} = \SI{0}{T}$, measured at $T=\,$\SI{1.8}{K} and $V_\mathrm{bg}=\SI{50}{V}$ in device A.% (a), and in device B (b). 
    ~We extract the mobility for the electrons and holes separately by the linear fits to the data above and below the charge neutrality point. (b) Temperature-dependence of the mobility extracted for each carrier type in device A.}
    \label{fig:extracting_mobility}
\end{figure*}

\newpage
%The mobility $\mu$ of the charges is an important quantity in transport measurements, as it qualitatively describes how disordered the sample is. 
We obtain the mobility $\mu$ of charge carriers in the graphene channel from $\sigma_{\mathrm{xx}} = ne\mu$. % and highlight that the mobility can be dependent on the charge carrier type i.e. $\sigma_{\mathrm{xx}} = ne(\mu_\mathrm{h} + \mu_\mathrm{e})$. To obtain the mobility in our device(s), 
The $\sigma_{\mathrm{xx}}$ is plotted versus $V_{\mathrm{tg}}$ in \Cref{fig:extracting_mobility}a, for device A.% and B respectively. 
~By considering the $n$ versus $V_\mathrm{tg}$ relation, described earlier, and by the linear fits to the $\sigma_{\mathrm{xx}} $ vs. $V_{\mathrm{tg}}$, we obtain $\mu_\mathrm{h} \sim \SI{2300}{cm^2/Vs}$ and  $\mu_\mathrm{e} \sim \SI{1450}{cm^2/Vs}$ for device A, and $\mu_\mathrm{h} \sim \SI{4100}{cm^2/Vs}$ and  $\mu_\mathrm{e} \sim \SI{920}{cm^2/Vs}$ for device B at $T \approx \SI{1.8}{K}$. We further extract the mobility at various temperatures, shown in \Cref{fig:extracting_mobility}b. In previous reports~\cite{mobility_Vs_T1, mobility_vs_T2}, pristine graphene encapsulated with hBN shows a slight decrease in mobility upon increasing the temperature, whereas, in graphene on SiO$_2$, the mobility decreases more rapidly due to higher electron-phonon interaction. In our samples, the mobility stays constant up to $T\approx \SI{90}{K}$ and decreases more rapidly at higher temperatures. Apart from the contribution from electron-phonon scattering, the temperature dependence in our devices could also be affected by the enhancement of the thermal fluctuations of the magnetic moments in CPS at temperatures above the Néel temperature. In device B, we extract the mobility for holes to be $\mu_\mathrm{h} \sim 4100$ and $\SI{835}{cm^2/Vs}$ at 1.8 K and RT, respectively.

\newpage
\FloatBarrier
\textbf{\large{9.~Energy shift of Landau levels}}\\

One can correlate the shift of the Landau levels to two possible effects: \\
1.~The finite staggered potential induced in graphene by the proximity of CPS that opens an orbital gap in the graphene band structure. That effectively shifts the Landau levels with respect to $E=0$. The shift of Landau levels in the measurement vs. $V_\mathrm{tg}$ is enhanced when there is a finite CPS midgap states at/close to the Dirac point. The reason is that the CPS midgap states can make the tuning of the Fermi energy at the Dirac point slow; thus, effectively, there is a large insulating gap present vs. $V_\mathrm{tg}$, and so the rest of the Landau levels would be shifted. The contribution of the midgap states in slowing down the gate tunability of the Fermi energy in graphene is also reflected in the slight non-linearity of the Landau fans close to the charge neutrality point ($|V_\mathrm{tg}-V_\mathrm{cnp}|<\SI{5}{V}$), as described in the main manuscript.\\

2.~The possible presence of a pseudo-magnetic field ($B_\mathrm{ps}$) in the heterostructure can also shift the Landau levels in energy. The finite $B_\mathrm{ps}$ could be originating from a possible uniaxial strain~\cite{zhu2015programmable} in this heterostructure that is expected from the lattice mismatch between graphene and CPS (also realized when forming the superlattice for the DFT calculations in Section 15), as well as from the anisotropic deformation of the CPS crystal upon cooling down (recently observed in~\cite{houmes2024highly}). Bulk CPS has shown an opposite sign for the thermal expansion coefficient along the $a$- and $b$-axes, with considerable enhancement (reduction) of the tensile strain along the a-axis (b-axis). This can increase the uniaxial strain gradient in the graphene channel in the proximity of the CPS. However, the $B_\mathrm{ps}$ originating from strain, is expected to be opposite for the K and K' valleys and this valley-polarization is not detected in the Landau fan diagram measurements in these heterostructures. Thus, it is less likely that the $B_\mathrm{ps}$ is the reason for the energy shift of the LLs.   \\

\textbf{\large{10.~Helical states in device A}}\\

\begin{figure*}[h]
    \centering
    \includegraphics[width=0.7\textwidth]{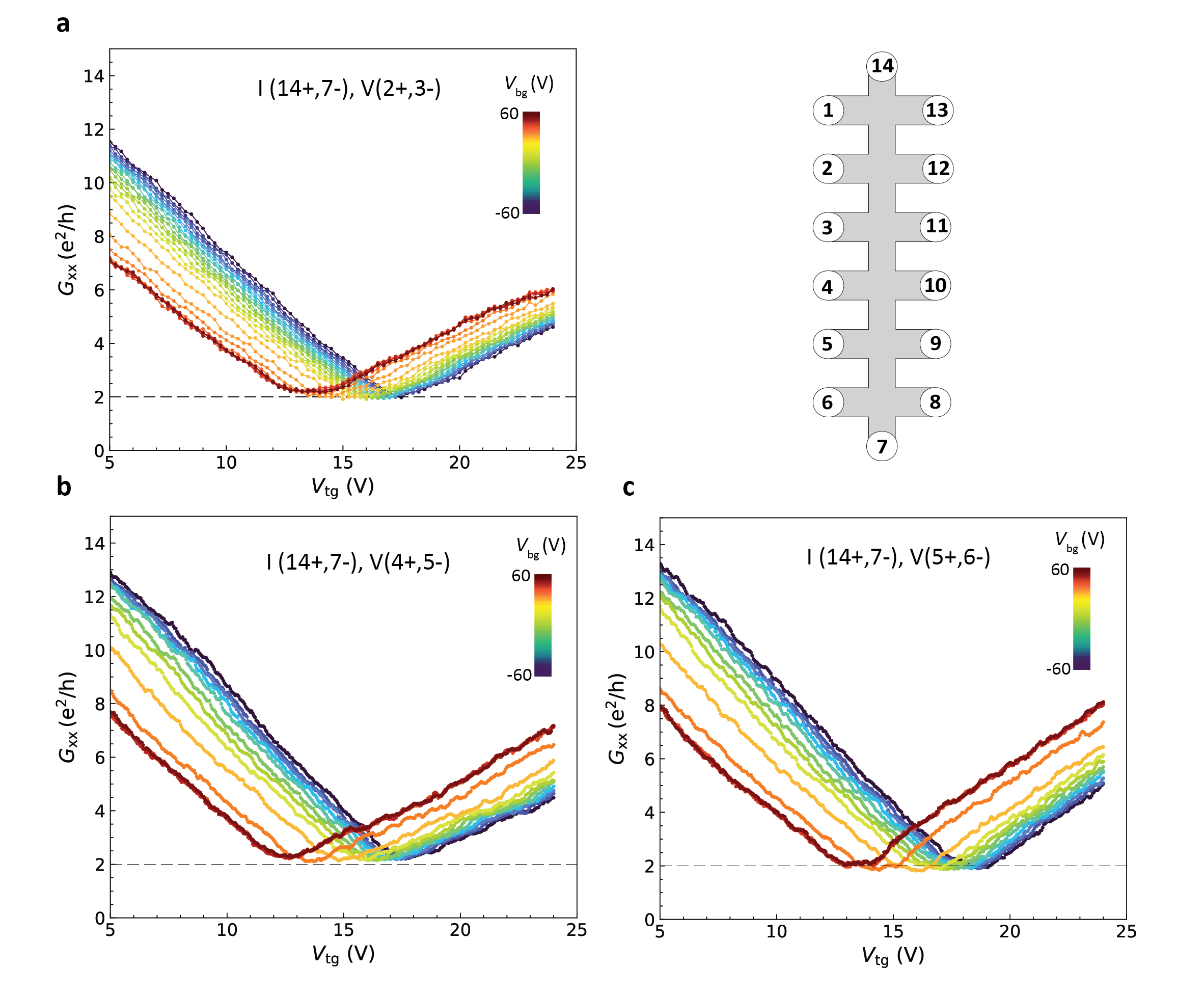}
    \caption{\textcolor{black}{(a-c) $V_\mathrm{tg}$-dependence of the $G_\mathrm{xx}= I/V_\mathrm{xx}$ measured at various $V_\mathrm{bg}$ at $T=\SI{1.8}{K}$ with the voltage probe numbers specified in the insets.}}
    \label{fig:QSHE_vs_VBG}
\end{figure*}

\begin{figure*}[h]
    \centering
    \includegraphics[width=0.6\textwidth]{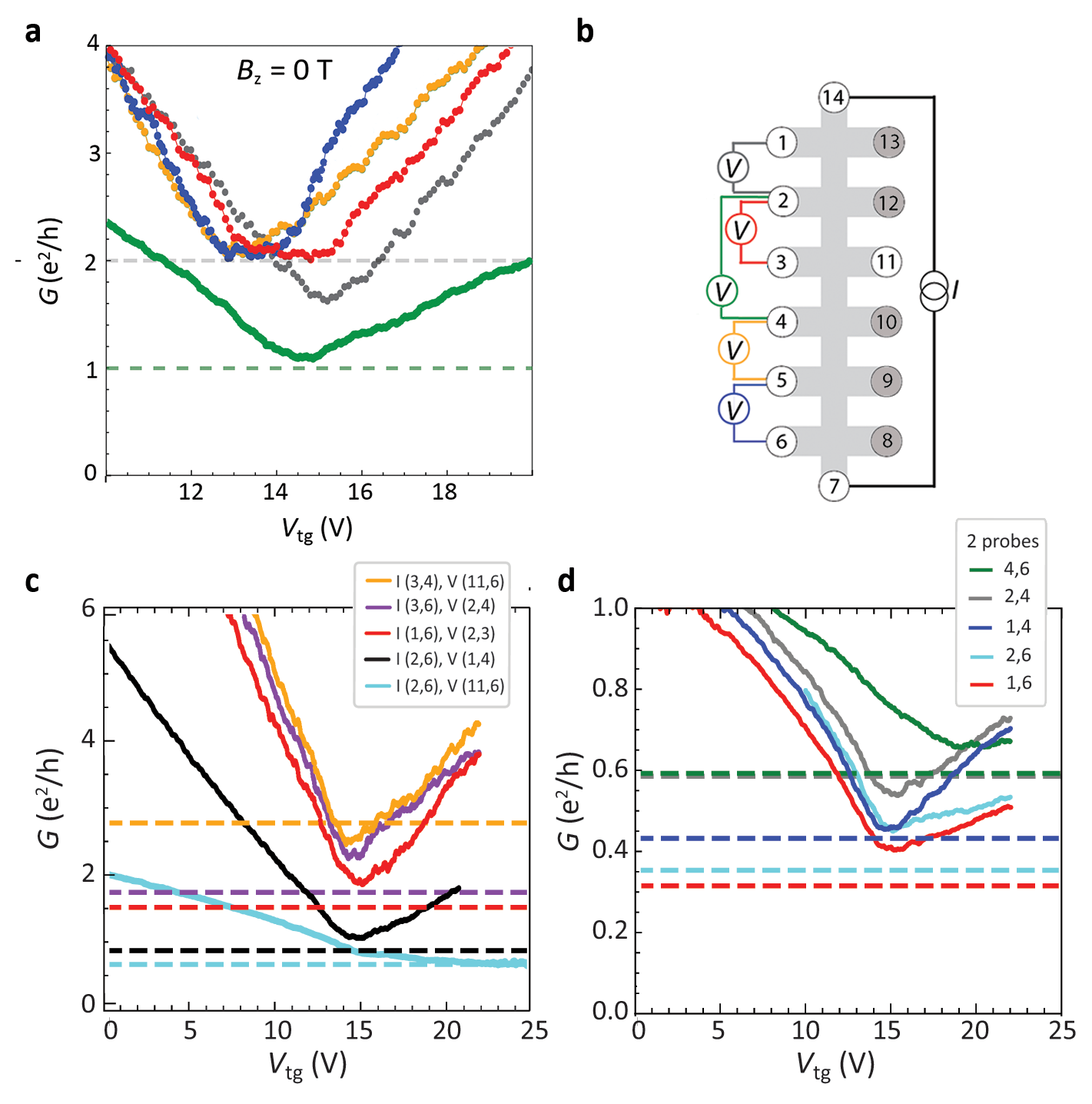}
    \caption{\textcolor{black}{(a) Four-terminal conductance measurement, shown for all various available voltage probes in device A. The corresponding measurement geometry for each plot is indicated in the device schematics in panel b. The probes shown with the gray circles (contacts 8, 9, 10, 12, and 13) were highly resistive. Thus, they have been avoided throughout the measurements.} (c,d) Various two-, three- and four-terminal conductance measured versus $V_\mathrm{tg}$ at various geometries at $B_\mathrm{z} = \SI{0}{T}$ at $T = \SI{1.8}{K}$. The dashed lines indicate the theoretically expected values for the conductance considering helical states, color-coded with respect to the measurement geometry indicated in the legend.}
    \label{fig:helicalstates_in_device A}
\end{figure*}

\textcolor{black}{In \Cref{fig:QSHE_vs_VBG}, we show the persistence of the plateaus of the conductance at 2e$^2$/h at various four-terminal voltage probes in sample A, measured versus top-gate voltage $V_\mathrm{tg}$ at various back-gate voltages $V_\mathrm{bg}$. }

In \Cref{fig:helicalstates_in_device A}, we evaluate the detection of helical states in various two-, three- and four-terminal geometries, at $B_\mathrm{z} = \SI{0}{T}$ in device A. We observe that the conductance at the Dirac point gets fairly close to the theoretical predictions for the presence of the helical states, following equations 1 and 2 in the main manuscript. Note that the contacts with the graphene channel in device A through the probes 8, 9, 10, 12, and 13 were highly resistive. Thus, they have been avoided throughout the measurements.

\begin{figure*}[h]
    \centering
    \includegraphics[width=\textwidth]{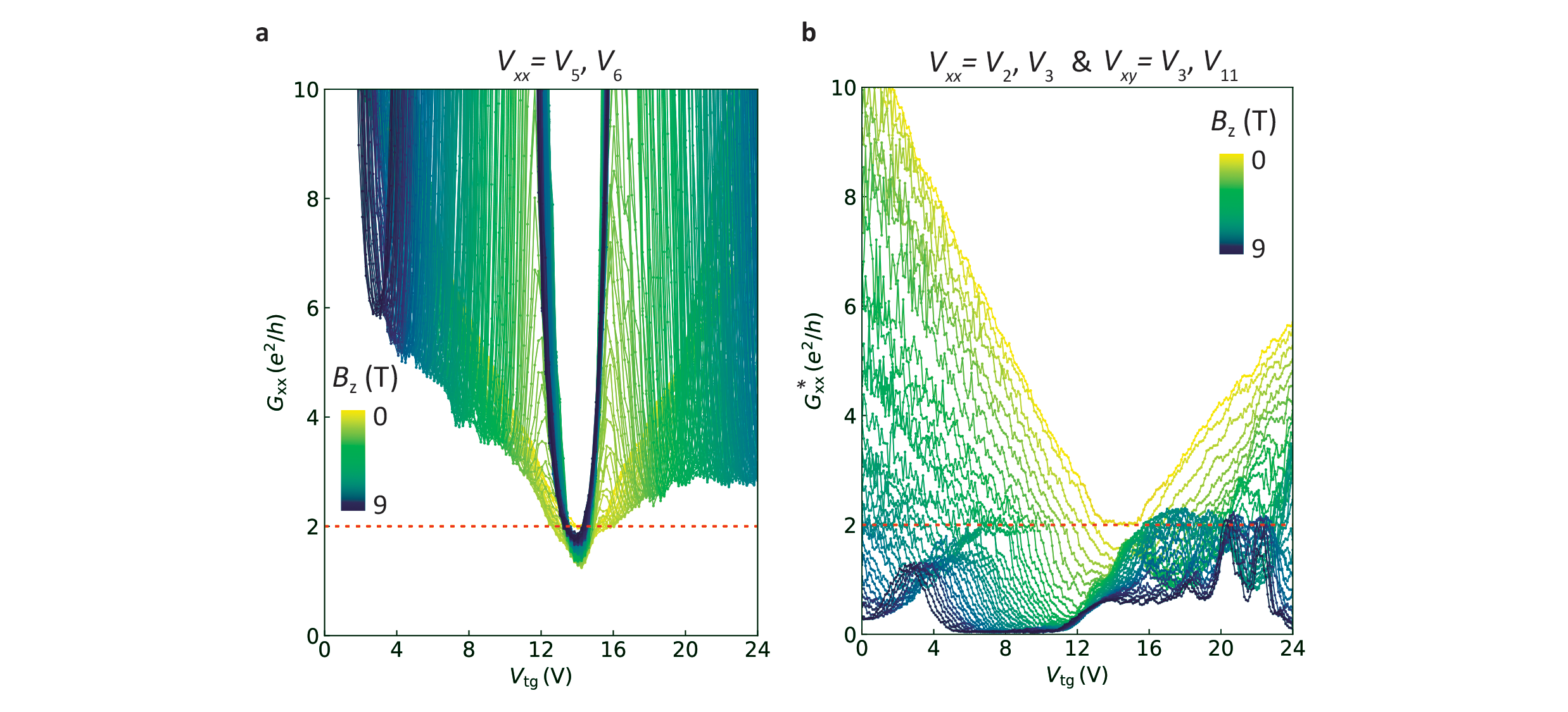}
    \caption{\textcolor{black}{Condutanctance measurements in device A, measured in four-terminal geometries at various magnetic fields. (a) Conductance is defined as $G_\mathrm{xx}=V_\mathrm{xx}/I$ with $V_\mathrm{xx}$ measured with $V_5$ and $V_6$. (b) Conductance is defined using conductivity matrix; $G_\mathrm{xx}^* = W\rho_\mathrm{xx}/L(\rho_\mathrm{xx}^2+R_\mathrm{xy}^2)$ with $V_\mathrm{xx}$ measured with $V_2$ and $V_3$, and $V_\mathrm{xx}$ measured with $V_3$ and $V_{11}$. See \Cref{fig:differentpairs} for the device schematics and the comparison of the QH transport in the various pairs.}}
    \label{fig:B_dep_device A}
\end{figure*}

\begin{figure*}[h]
    \centering
    \includegraphics[width=0.9\textwidth]{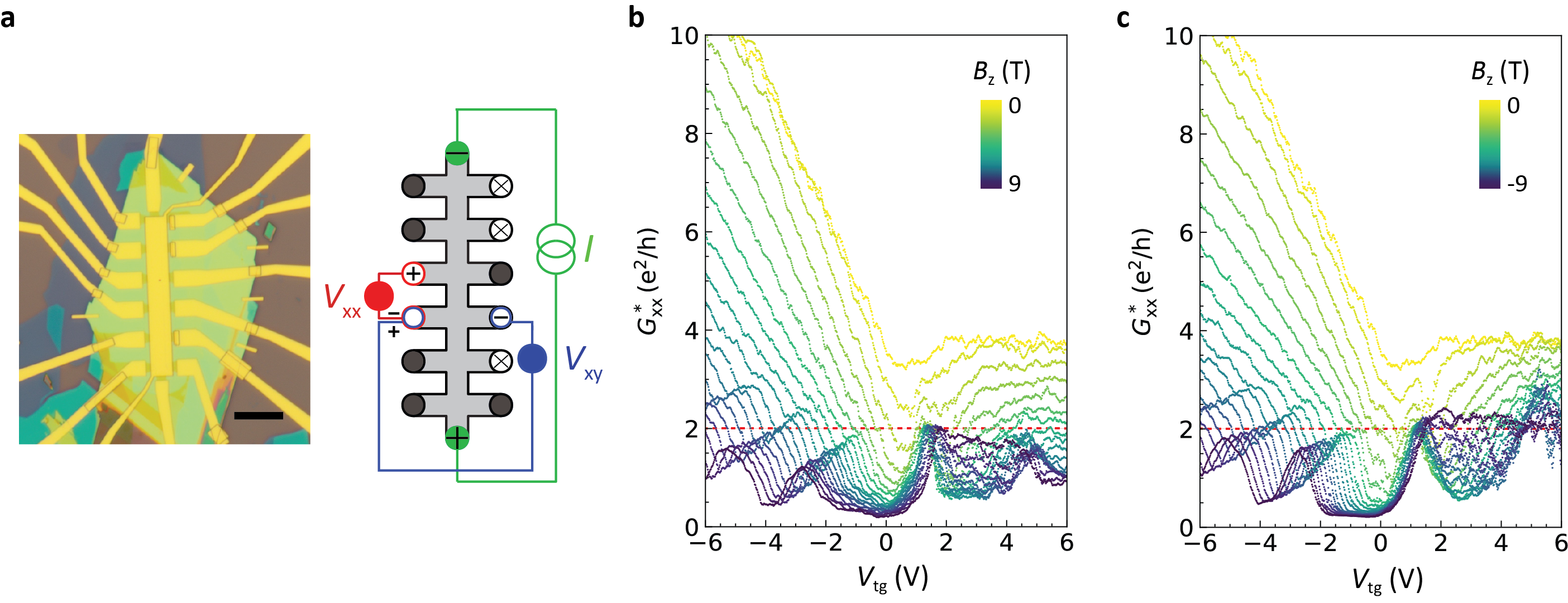}
    \caption{(a) Optical micrograph and schematics of device B, and the measurement geometry. The scale bar is 10~$\mu$m. The circles with `x' represent the electrodes through which there was no electrical connection. (b, c) Gate-dependence of the channel conductance $G_\mathrm{xx}^*$, measured in device B at positive and negative $B_\mathrm{z}$, shown per \SI{0.5}{T}. The $G_\mathrm{xx}^*$ is calculated considering the conductivity tensor that includes both $V_\mathrm{xx}$ and $V_\mathrm{xy}$ measured in a 4-terminal geometry.}
    \label{fig:B-dep of Gxx in deviceB}
\end{figure*}

\textcolor{black}{We now focus on the evolution of the helical edge states under finite $B_\mathrm{z}$. \Cref{fig:B_dep_device A} shows the gate-dependence of four-terminal conductance values measured for $B_\mathrm{z}$ up to \SI{9}{T} in device A. The $G_\mathrm{xx}= V_\mathrm{xx}/I$ in panel a is measured with probes $V_5$ and $V_6$, the voltage probes that have shown to pick up a minimal contribution from $V_\mathrm{xy}$, with the resistivity values reaching zero in the Landau gaps (see \Cref{fig:differentpairs}). In panel b, we show the conductance measurement performed using another longitudinal voltage pair, $V_2$ and $V_3$, which has shown to have contributions from the transverse voltage ($V_3$ and $V_{11}$), as shown in \Cref{fig:differentpairs} and \Cref{fig:Rxypair}. Having both longitudinal and transverse voltages available with this pair of contacts, we define the conductance by using the conductivity matrix: $G_\mathrm{xx}^* = W\rho_\mathrm{xx}/L(\rho_\mathrm{xx}^2+R_\mathrm{xy}^2)$. Note that the shift of the Dirac point in panel b, is seen in \Cref{fig:Rxypair}a for this pair of contacts, which is associated with the inhomogeneity in the channel. Despite the shift of the charge neutrality point by increasing the $B_\mathrm{z}$, we observe that the $G^*_\mathrm{xx}$ stays about $2e^2/h$, indicative of the presence of the helical states within the zeroth Landau level gap, which is consistent with the observation of the metallic temperature-dependence of the resistivity at the zeroth Landau level in device A, as shown in the main manuscript, Fig.~4b. Even though similar $B-$dependence and temperature-dependence is reproduced in device B (see \Cref{fig:B-dep of Gxx in deviceB}), we note that the persistence of the helical states at finite $B_\mathrm{z}$ depends on the evolution of the zLLs with applied $B_\mathrm{z}$ and can be disturbed by bands crossover at certain $B_\mathrm{z}$ ~\cite{bottcher2019survival}, thus it can vary depending on the various (proximity-related) parameters in the heterostructures.}

\FloatBarrier
\textbf{\large{11.~Helical states in device B}}\\

We also evaluate the presence of helical states in device B, by focusing on the gate dependence of the conductance close to the charge neutrality point. We start with four-terminal geometry as shown in the schematics of \Cref{fig:B-dep of Gxx in deviceB}a. The gate-dependence of the four-terminal longitudinal conductance ($G_\mathrm{xx}^* = W\rho_\mathrm{xx}/L(\rho_\mathrm{xx}^2+R_\mathrm{xy}^2)$) for positive and negative ranges of $B_\mathrm{z}$ is shown in panel b and c. The reason for using the conductance matrix in the determination of $G_\mathrm{xx}$ is to account for the mixing of the $R_\mathrm{xy}$ and $R_\mathrm{xx}$ measurements due to the device asymmetries. We realize that, unlike the observations in device A, the conductance close to the charge neutrality point, measured in this geometry at $B_\mathrm{z}= \SI{0}{T}$, does not reach the theoretical value of 2e$^2$/h expected for the helical states. This could be due to a larger contribution or dominance of the bulk transport as compared with edge transport at \SI{0}{T} in this device, \textcolor{black}{due to the presence of inhomogeneities and disorders in the channel that scatters the helical states (see \Cref{fig:Lfd_DeviceAB})}. However, by applying the $B_\mathrm{z}$ as small as $\pm\SI{1}{T}$ the channel is brought more in the QH regime \textcolor{black}{where the formation of the zeroth Landau level insulating gap can assist the detection of the helical states within the gap and allow for} the transport to be dominated by the edge states close to the charge neutrality point. With that, we observe that the $G_\mathrm{xx}^*$ reaches 2$e^2/h$ at the Dirac point and stays at that value for the whole range of $|B_\mathrm{z}|>\SI{1}{T}$, suggesting the presence of the helical states.

\begin{figure*}[h]
    \centering
    \includegraphics[width=0.9\textwidth]{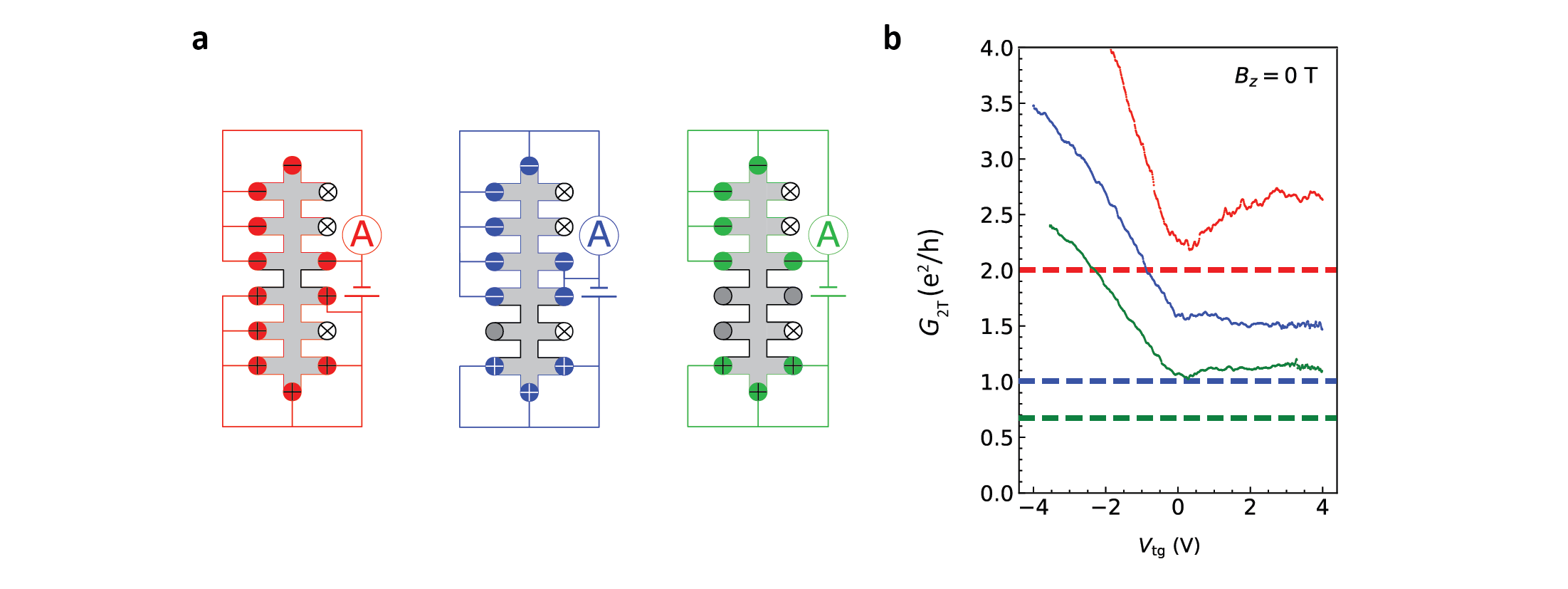}
    \caption{(a) Schematics of device B with different two-terminal measurement geometries used for evaluation of conductance in panel b. The circles with `x' represent the electrodes through which there was no/poor electrical connection. (b) Gate-dependence of the two-terminal conductance ($G_\mathrm{2T}$) at $B_\mathrm{z} = \SI{0}{T}$, color-coded with respect to the geometries shown in panel~a. The dashed lines are the expected values for the $G_\mathrm{2T}$, accounting for the presence of helical states.}
    \label{fig:sup:two_terminal}
\end{figure*}

We further conducted two-terminal measurements in device B, applying \SI{1}{mV} AC voltage and measuring current, shown in the schematics of \Cref{fig:sup:two_terminal}a. The 2T conductance ($G_\mathrm{2T}$) measured at $B_\mathrm{z}=\SI{0}{T}$ for different distances shown in the device schematics, indicates that the $G_\mathrm{2T}$ at the charge neutrality point gets close to the theoretically expected values for the helical states (shown by the dashed lines). Yet, the $G_\mathrm{2T}$ is not fully matching with the expected values, as also observed in device A (\Cref{fig:helicalstates_in_device A}) which once again indicates the finite contribution of bulk transport at $B_\mathrm{z}=\SI{0}{T}$. It is worth noting that the distance between the source and the drain in these configurations is relatively long and the device has shown to be inhomogeneously doped along the channel. With that and considering the low mobility of the charge carriers, coherent spin-polarized edge transport over such long distances is unlikely to occur, which can explain the discrepancy between the theory and the experiment.

\newpage

\textbf{\large{12.~\textcolor{black}{Quantum Hall plateaus}}}\\

\begin{figure*}[h]
    \centering
    \includegraphics[width=0.75\textwidth]{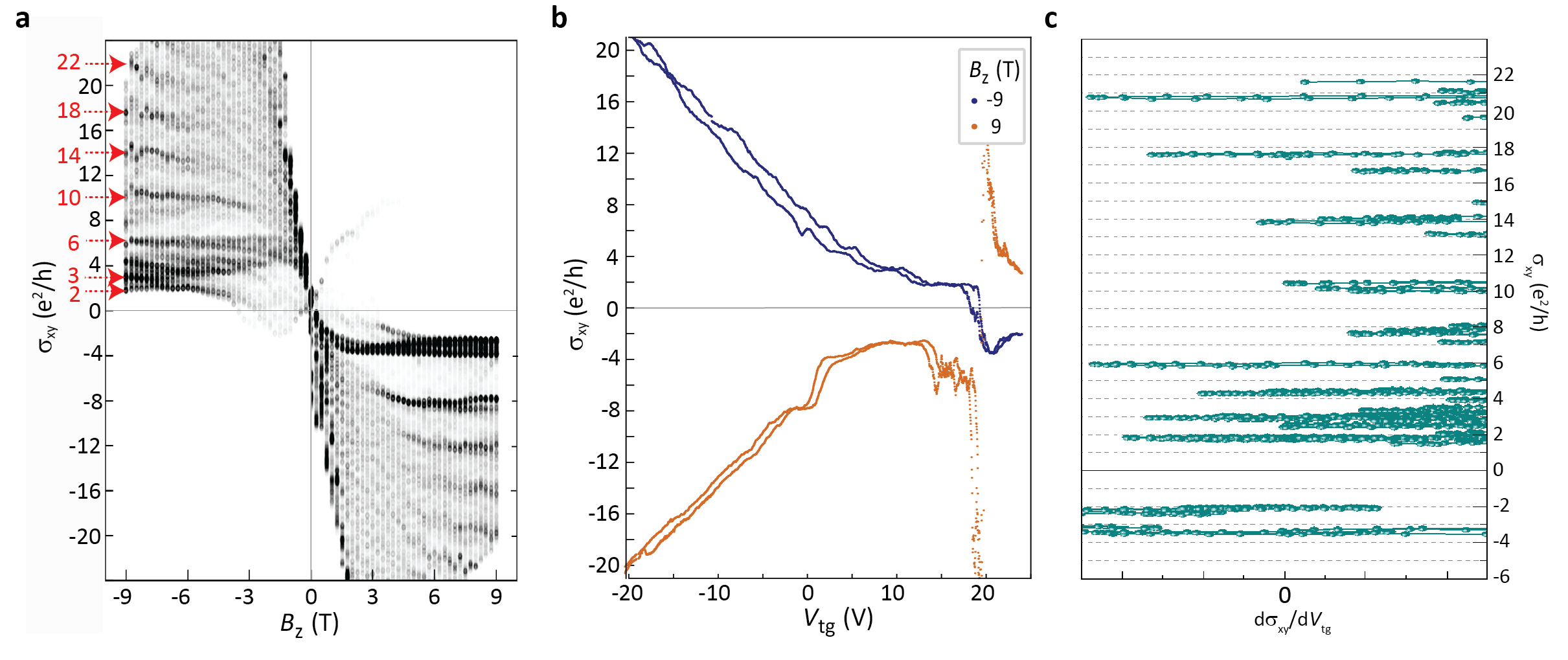}
    \caption{(a) Magnetic field dependence of $\sigma_\mathrm{xy}$, measured $-20 <V_\mathrm{tg}< 15\,$V. The data points are shown with transparency to visualize the data accumulation happening at the QH plateaus. The red arrows point to the formed QH plateaus at large negative magnetic fields. (b) Gate-dependence of $\sigma_\mathrm{xy}$, measured at $B_\mathrm{z} = \pm\SI{9}{T}$ shown for trace and retrace measurements. (c) Derivative of the $\sigma_\mathrm{xy}$ with respect to $V_\mathrm{tg}$ plotted against $\sigma_\mathrm{xy}$ for $B_\mathrm{z}= \SI{-9}{T}$.}
    \label{fig:QHplateaus_deviceA}
\end{figure*}

\textcolor{black}{We investigate transverse conductivity ($\sigma_\mathrm{xy}$) at $N_\mathrm{LL}\neq 0$, which is a measure of the number of chiral edge states in the channel in the QH regime. Considering conductivity matrix, we define $\sigma_\mathrm{xy}=R_\mathrm{xy}/(R_\mathrm{xy}^2+\rho_\mathrm{xx}^2)$. \Cref{fig:QHplateaus_deviceA}a shows the magnetic field dependence of the $\sigma_\mathrm{xy}$ for $-20 < V_\mathrm{tg}< \SI{15}{V}$, where the data is shown with transparency of the markers to visualize the accumulation of data points (resulting in the darker shades) around specific $\sigma_\mathrm{xy}$ values at the QH plateaus (indicated by the red arrows).}

\begin{figure*}[h]
    \centering
    \includegraphics[width=0.85\textwidth]{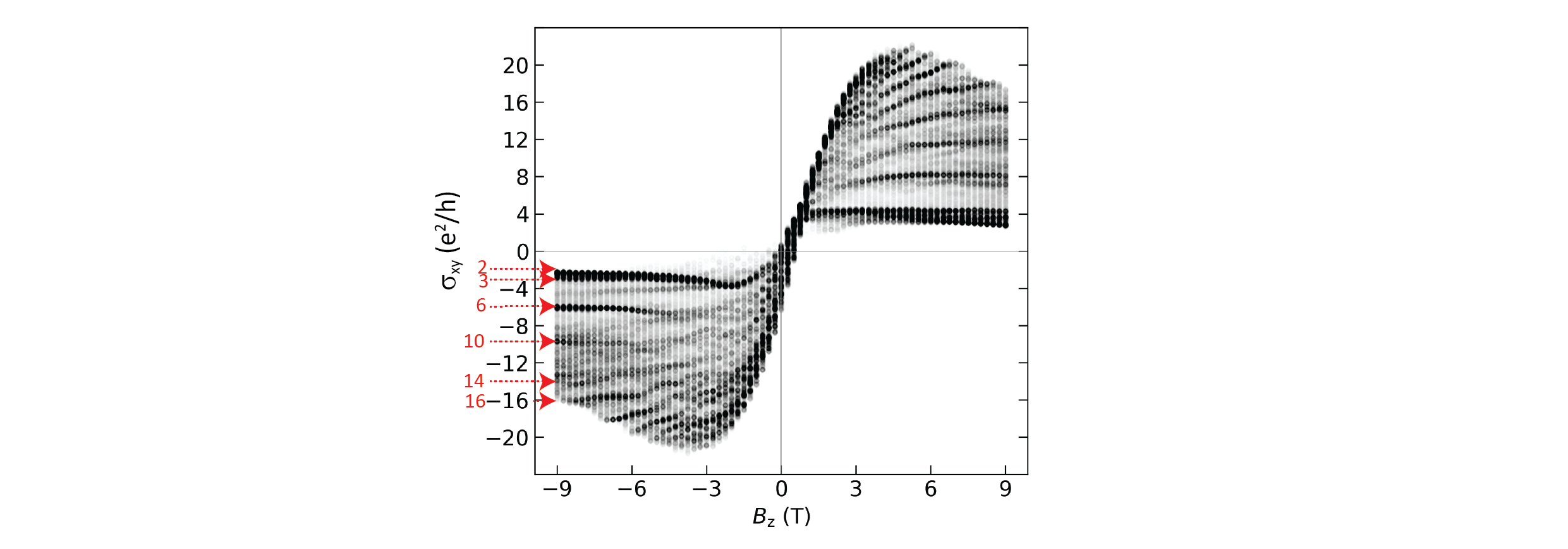}
    \caption{Dependence of $\sigma_\mathrm{xy}$ on $B_\mathrm{z}$ shown for $V_\mathrm{tg}=$ -10 to 1 V (hole doping). The data points are shown with transparency to visualize the data accumulation happening at the QH plateaus. Note that the sign of the transverse voltage probes in the measurements in device B is opposite to that of device A, which causes the sign of $\sigma_\mathrm{xy}$ to be opposite to that of device A for the same carrier type.}
   \label{fig:B-dependence in device B}
\end{figure*}

The QH plateaus appear to be asymmetric vs.~$B_\mathrm{z}$ which could be related to the fact that the values of the $\sigma_\mathrm{xy}$ are affected by non-zero values of the $\rho_\mathrm{xx}$ in the Landau gaps which are non-symmetric vs.~$B_\mathrm{z}$ (extracted from the Landau fan of \Cref{fig:Rxypair}a). The sequence of the quantum Hall plateaus (highlighted by the red arrows) seems to indicate that the degeneracy of the Landau levels is mostly preserved.~This is consistent we the fact that in the Landau fan diagrams the spin-splitting of the LLs with $N_\mathrm{LL}\neq0$ is not resolved within the broadening of the SdHOs. We attribute that to the high level of disorder in this system that hinders the resolution of the spin-split bands at higher energies. In \Cref{fig:QHplateaus_deviceA}b, we show the gate-dependence of the $\sigma_\mathrm{xy}$ for $\pm$\SI{9}{T}. In \Cref{fig:QHplateaus_deviceA}c we show the derivative of the QH conductance with respect to the $V_\mathrm{tg}$, plotted versus $\sigma_\mathrm{xy}$ for $B_\mathrm{z}= \SI{-9}{T}$. When there is a QH plateau in $\sigma_\mathrm{xy}$, the derivative of the Hall conductance vs. gate should approach zero. Thus, investigating $d\sigma_\mathrm{xy}/dV_\mathrm{tg}$ vs $\sigma_\mathrm{xy}$ allows for the determination of the QH conductance values, specifically in cases where the plateaus of the conductance are not pronounced. In \Cref{fig:B-dependence in device B} we show the transverse conductivity in device B, to be compared with that of device A.

\newpage
\textbf{\large{13.~Temperature dependence}}\\

\begin{figure*}[h]
    \centering
    \includegraphics[width=0.95\textwidth]{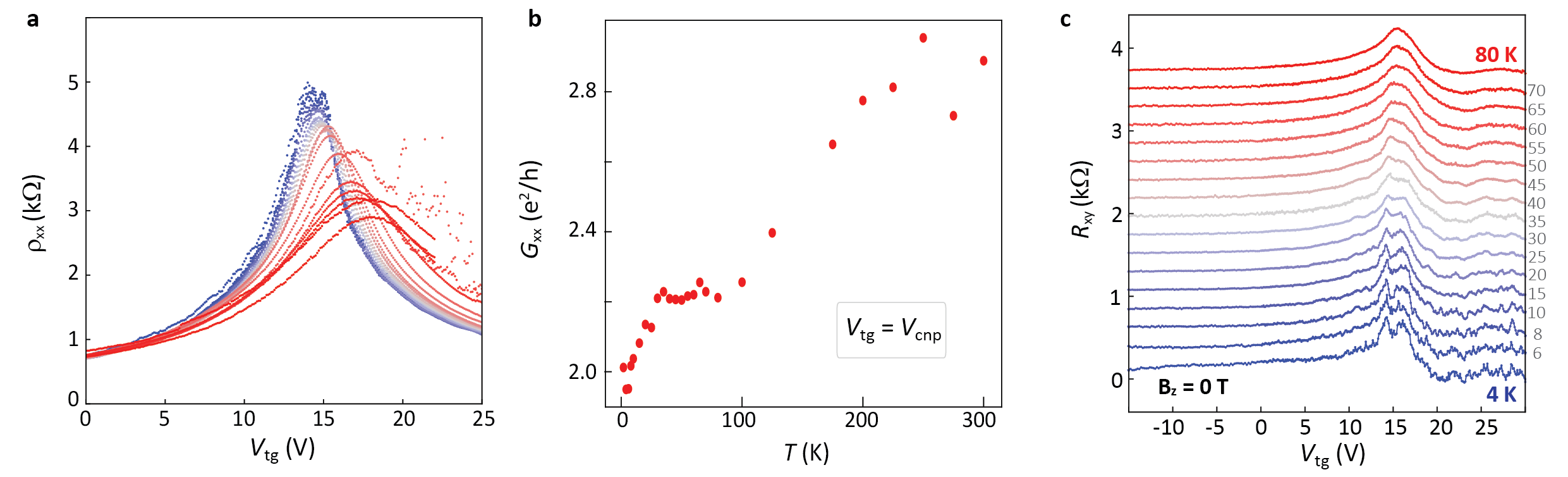}
    \caption{(a) Temperature dependence of the resistivity $\rho_\mathrm{xx}$, and (b) conductance values $G_\mathrm{xx}$ at the charge neutrality point ($V_\mathrm{tg} = V_\mathrm{cnp}$), measured in device A at $V_\mathrm{bg} = \SI{50}{V}$ and $B_\mathrm{z}= 0\,$T up to 300~K. Note that we account for the shift of the $V_\mathrm{cnp}$ vs. $T$. (c) Top-gate dependence of $R_\mathrm{xy}$ at $B_\mathrm{z}=0$, measured at various temperatures. The curves in panel c are shown with an offset along the y-axis for clarity. }  
    \label{fig:T-dep_fullRange}
\end{figure*}

In this section, first, we provide further information on the transport in devices A and B at $B_\mathrm{z} = \SI{0}{T}$ and then we elaborate on the measurements at $B_\mathrm{z} = \pm\SI{9}{T}$. In \Cref{fig:T-dep_fullRange}a and b, we show the $T-$dependence of the channel resistivity and conductance at the Dirac point in device A at $B_\mathrm{z} = 0$.  

\textcolor{black}{Simultaneous measurement of $R_\mathrm{xy}$ by probes $V_3$ and $V_{11}$ in Figure \Cref{fig:T-dep_fullRange}c shows that there is a finite transverse voltage at $B_\mathrm{z}= \SI{0}{T}$, which may contain some contribution from the AH effect as a result of the finite small $z$-component of the $M_\mathrm{CPS}$. The $R_\mathrm{xy}$ shows a double-peak feature at zero energy that merges into a single peak at higher temperatures, due to thermal broadening. The separation of the two peaks in the $R_\mathrm{xy}$ in terms of energy is similar in magnitude to the width of the plateau in $G_\mathrm{xx}$ at zLL ($\Delta V_\mathrm{tg}\sim$~2V for the bulk gap in Figure 3b of the main manuscript). The peaks of $R_\mathrm{xy}$ close to $V_\mathrm{cnp}$ can be related to the enhanced spin-polarization at lower energies. However, note that the $R_\mathrm{xy}$ signal at $B_\mathrm{z}=0$ measured by the $V_3$ and $V_{11}$ probes can also contain some contribution from $R_\mathrm{xx}$ due to possible geometrical asymmetries in the sample, as it is also reflected in the clear signatures of inhomogeneities in the Landau fan diagram measured in this region of the sample, shown in \Cref{{fig:Rxypair}}a. }

\begin{figure*}[h]
    \centering
    \includegraphics[width=0.8\textwidth]{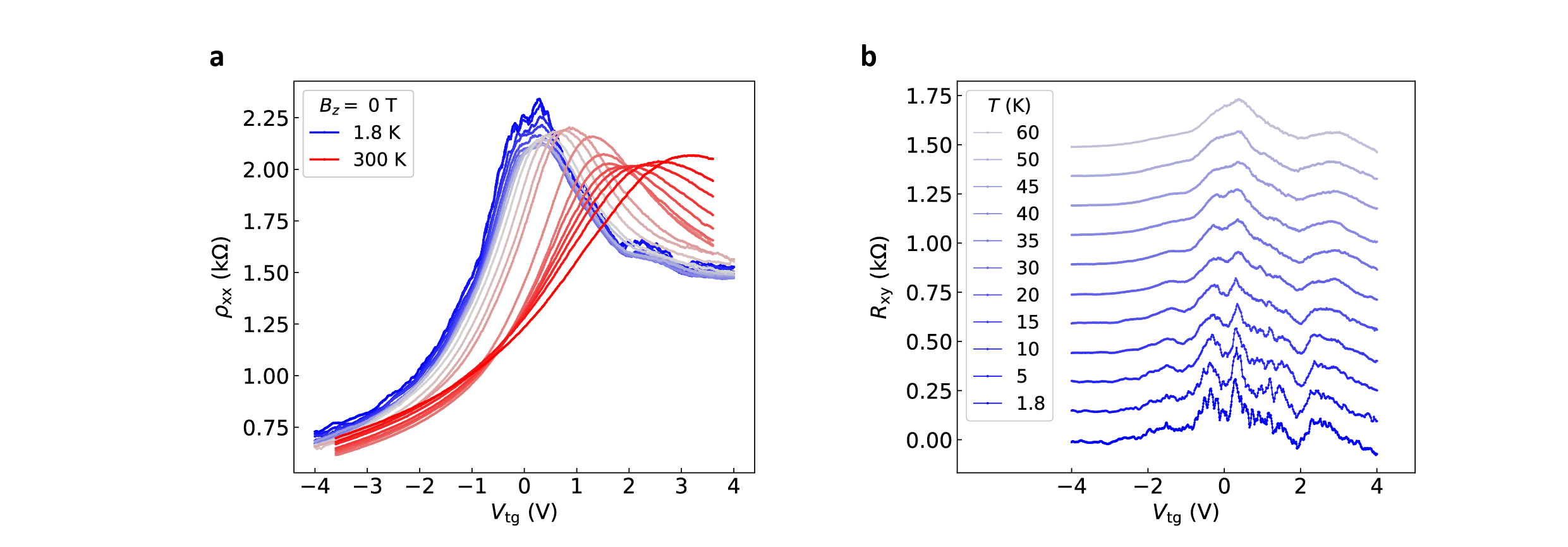}
    \caption{(a) Temperature-dependence of the longitudinal resistivity $\rho_\mathrm{xx}$ and (b) Hall resistance $R_\mathrm{xy} = V_\mathrm{xy}/I$ versus $V_\mathrm{tg}$ measured in device B at $B_\mathrm{z} = \SI{0}{T}$ and $V_\mathrm{bg}= \SI{50}{V}$. The curves in panel b are shown with an offset along the y-axis for clarity. }
    \label{fig:double_peak_new_sample}
\end{figure*}

We further investigate the temperature dependence of $\rho_\mathrm{xx}$ in device B, which is shown in \Cref{fig:double_peak_new_sample}a measured at $B = \SI{0}{T}$ for $T = $1.8 to \SI{300}{K}. The gate-dependence of $R_\mathrm{xy}$, shown in \Cref{fig:double_peak_new_sample}b also exhibits a double-peak feature at the Dirac point ($V_\mathrm{tg}\approx\SI{0}{V}$), similar to the same measurement conducted in device~A (shown in \Cref{fig:T-dep_fullRange}c). Note that the second bump in $R_\mathrm{xy}$ at $V_\mathrm{tg}\approx\SI{2.5}{V}$ is associated with the inhomogeneity in the channel. The shift of the Dirac peak versus temperature in device B is also attributed to the inhomogeneities in the channel. As shown in \Cref{fig:Lfd_DeviceAB}, there seems to be a second Dirac peak present at larger positive gate voltages at low $T$ associated with the regions with different doping. We believe that the sizable shift of the Dirac peak in device B vs. $T$ is due to the thermal broadening and merging of the two Dirac peaks.   

\newpage

\begin{figure*}[h]
    \centering
    \includegraphics[width=0.9\textwidth]{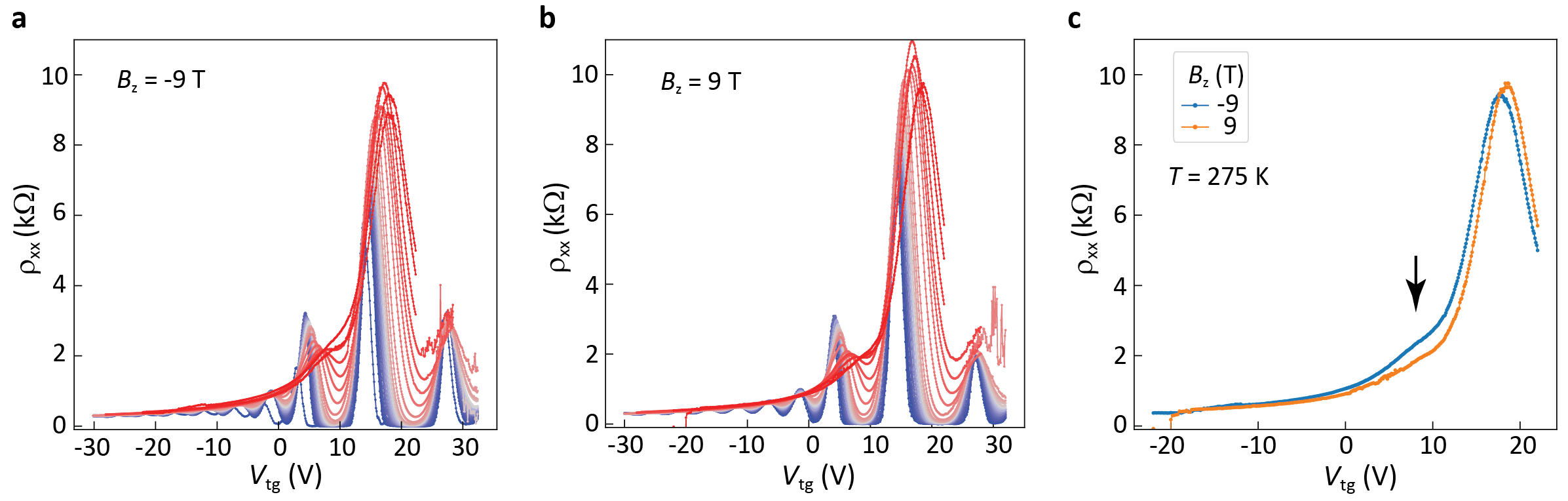}
    \caption{Temperature dependence of the channel resistivity $\rho_\mathrm{xx}$, measured up to 300 K versus $V_\mathrm{tg}$ at $B_\mathrm{z}= -9\,$T (a), and at $B_\mathrm{z}= 9\,$T (b) in device A with $V_\mathrm{5}$ and $V_\mathrm{6}$ voltage probes (see~\Cref{fig:differentpairs}). (c) $\rho_\mathrm{xx}$ measured at $T= \SI{275}{K}$, shown for $B_\mathrm{z}= \pm\SI{9}{T}$. The black arrow points at the shoulder associated with the SdHO related to the first LL, persisting up to elevated temperatures. All measurements are performed at $V_\mathrm{bg}= \SI{50}{V}$. } 
    \label{fig:T-dep_fullRange_rho_xx_Bpm9}
\end{figure*}

\begin{figure*}[h]
    \centering
    \includegraphics[width=0.85\textwidth]{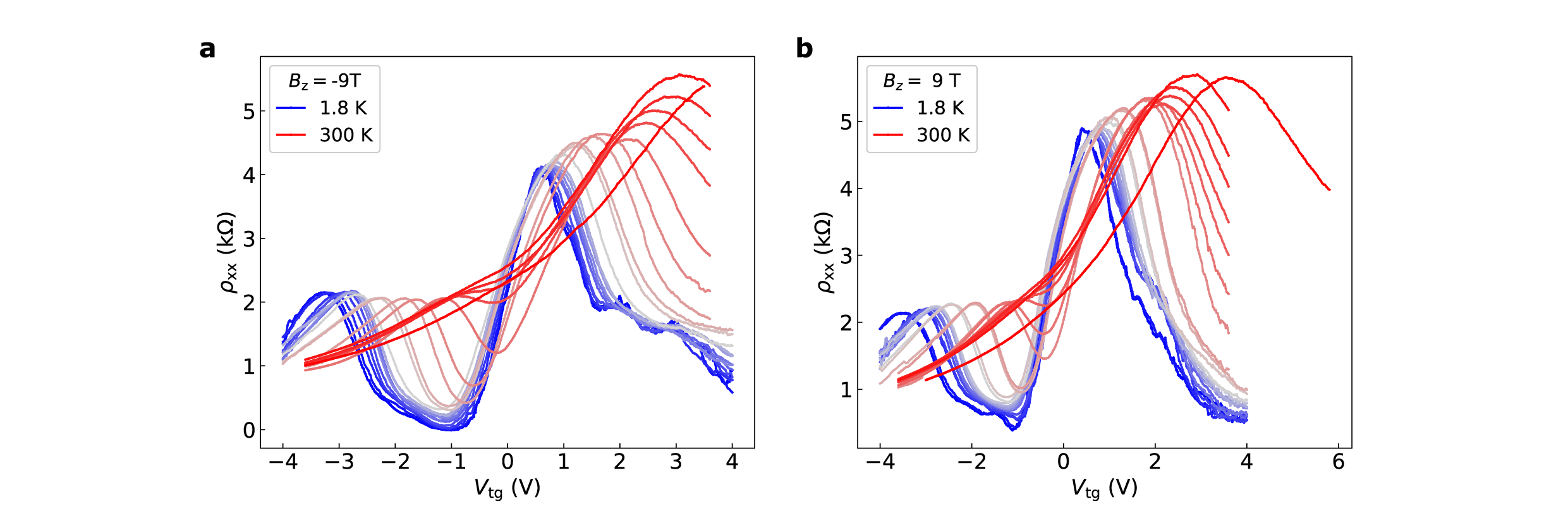}
    \caption{Temperature dependence of the longitudinal resistivity versus $V_\mathrm{tg}$ measured in device B, (a) at $B_\mathrm{z} = \SI{-9}{T}$ and (b) at $B_\mathrm{z} = \SI{9}{T}$. All measurements are performed at $V_\mathrm{bg}= \SI{50}{V}$.}
    \label{fig:rho_xx_vs_T_deviceB}
\end{figure*}

\begin{figure*}[h]
    \centering
    \includegraphics[width=0.9\textwidth]{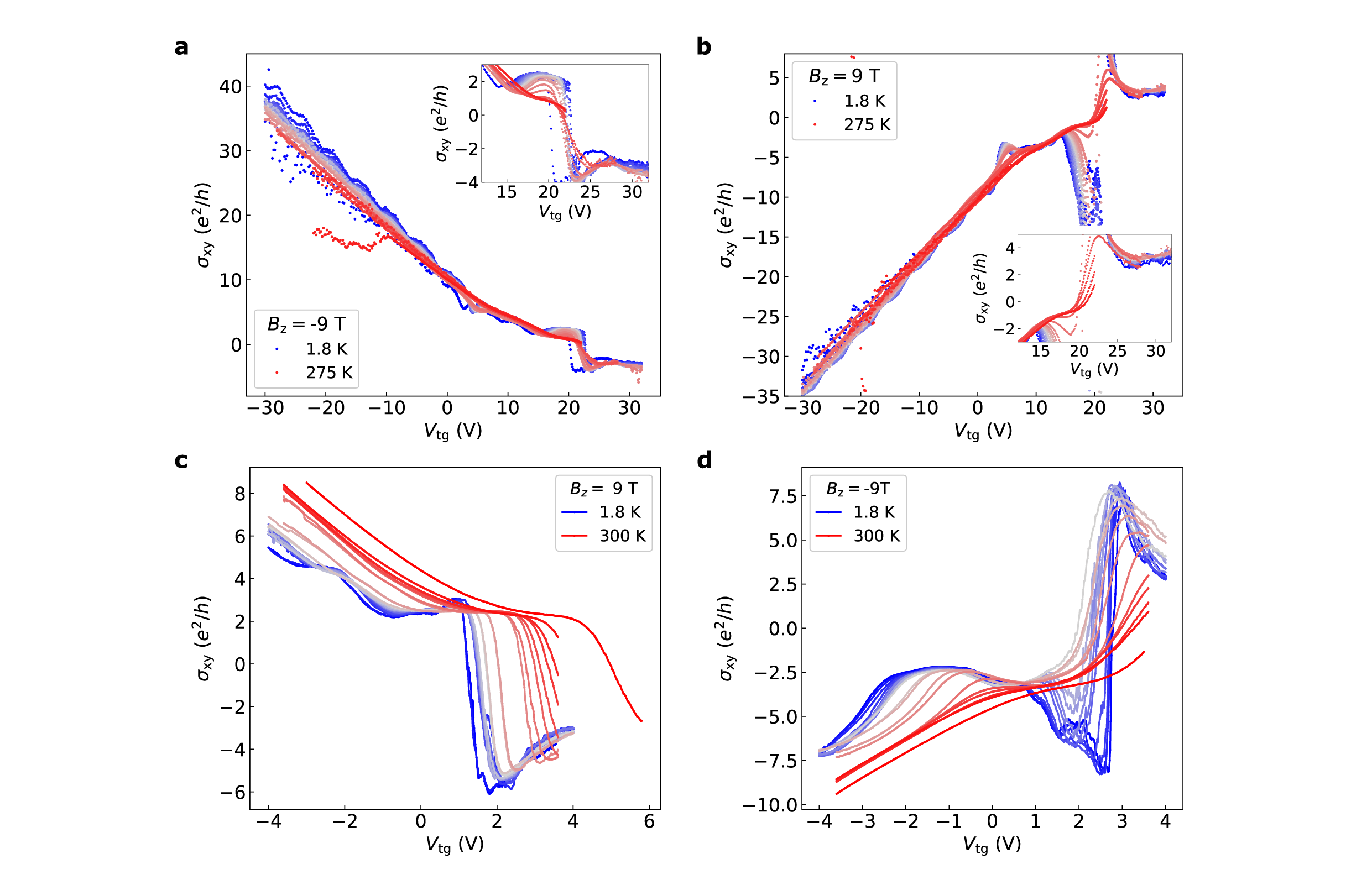}
    \caption{Temperature-dependence of the transverse conductivity versus $V_\mathrm{tg}$ measured in device A (a,b) and device B (c,d), at $B_\mathrm{z} = \pm \SI{9}{T}$. }
    \label{fig:sigma_xy_vs_Vtg_and_T}
\end{figure*}

\Cref{fig:T-dep_fullRange_rho_xx_Bpm9} and \Cref{fig:rho_xx_vs_T_deviceB} show the temperature dependence of $\rho_\mathrm{xx}$ versus $V_\mathrm{tg}$ for device A and B. We observe that the charge neutrality point slightly shifts to higher gate voltages with temperature, as the doping of the graphene channel slightly changes. The Landau levels become less pronounced as the temperature increases due to the thermal broadening. In both devices, at the Dirac point ($N_\mathrm{LL}= 0$), $\rho_\mathrm{xx}$ increases by increasing temperature. This metallic behavior is an indication of the presence of a gap-less zeroth Landau level which supports the presence of the helical states within the ZLL gap. Moreover, the oscillation in $\rho_\mathrm{xx}$ related to the first Landau level, in both devices, persists up to/close to room temperature. While the SdHOs merge into the background resistance (from the bulk of the graphene channel), the one related to $N_\mathrm{LL} = 1$ remains as a kink or shoulder near the Dirac point (e.g.~shown in \Cref{fig:T-dep_fullRange_rho_xx_Bpm9}c) which indicates the robustness of the first LL against temperature. This is further evaluated through the $T-$dependence of $\sigma_\mathrm{xy}$ in \Cref{fig:sigma_xy_vs_Vtg_and_T}.

\Cref{fig:sigma_xy_vs_Vtg_and_T} shows the temperature dependence of $\sigma_\mathrm{xy}$ versus $V_\mathrm{tg}$ for device A (a, b) and for device B (c, d) at $B_\mathrm{z} = \pm \SI{9}{T}$. For both devices, the plateaus corresponding to higher Landau level indexes become less pronounced as the temperature increases, yet the plateau-like features close to the Dirac point are robust in temperature. Note that the sign convention of the voltage probes in device A and B has been opposite to each other during the measurements, thus the results of device A at -9 T should be compared with that of device B at +9 T.

\newpage
\textbf{\large{14.~Landauer-Büttiker formalism}}\\

To model the role of the edge channels on the measured signals, we have used the Landauer-Büttiker equation \cite{buttiker1986}:
\begin{equation}
    I_i=\frac{\mathrm{e^2}}{\mathrm{h}}\sum_j\left(T_{ji}V_i-T_{ij}V_j\right),
    \label{eq:landauer}
\end{equation}
where $I_i$ is the current in terminal $i$, $T_{ij}$ the transmission probability from electrode $i$ to electrode $j$, and $V_j$ the voltage at electrode $j$.
To account for all the electrodes in device A, for instance, we have modeled for a device geometry with 14 electrodes that are connected in a closed loop where contact 14 is between contacts 13 and 1 (as shown in the schematics of \Cref{fig:helicalstates_in_device A}). In this case, the $i=14$ component of Equation~\ref{eq:landauer} is linearly dependent on the other components ($i=$ 1 to 13). %The $i=14$ component relates $V_{14}$ with $V_{13}$ and $V_{1}$ but, since one can find a linear combination of the $i>14$ cases that already correlates $V_{13}$ and $V_{1}$, this information can be already found in the previous components. 
To determine $V_1$ to $V_{14}$, we have written Equation~\ref{eq:landauer} for the terminals 1 to 13 and replaced the $i=14$ component with the condition $V_\mathrm{I_-}=0$, where $V_\mathrm{I_-}$ is the potential at the grounded contact. 
The resulting system is shown here:
\begin{equation*}
    \begin{bmatrix}
        I_1\\
        I_2\\
        \vdots\\
        I_{14}
    \end{bmatrix}
=
    \begin{bmatrix}
        m+n & -n & 0  & \cdots & 0 & -m\\
        -m & m+n & -n & \cdots & 0 & 0\\
        \vdots & \vdots &\vdots & &\vdots& \vdots \\
        0   & 0 & 0 & \cdots & 0 & 0
    \end{bmatrix}
    \begin{bmatrix}
        V_1\\
        V_2\\
        \vdots\\
        V_{14}
    \end{bmatrix},
\end{equation*}
which we write as $I_\mathrm{mat}= M V_\mathrm{mat}$, where
%\begin{equation*}
%I_\mathrm{mat}=
%    \begin{bmatrix}
%        I_1\\
%        
%        \vdots\\
%        I_{14}
%    \end{bmatrix},
%    V_\mathrm{mat}=
%    \begin{bmatrix}
%        V_1\\
%        
%        \vdots\\
%        V_{14}
%    \end{bmatrix},
%\end{equation*}
$I_\mathrm{mat}$ describes the input currents, $V_\mathrm{mat}$ the output voltages, and $M$ is the matrix obtained from the above considerations. $m$ and $n$ are the number of edge modes propagating from electrode $i$ to electrode $i-1$ and to $i+1$, respectively. The last row of $M_{14,j}=0$ $\forall j\neq 6$, $M_{14,6}=1$ and $I_i=0$ $\forall i\neq 1$ or $6$, $I_1=-I_6=I_0$. Finally, we calculate $V_\mathrm{mat}$ using $M^{-1}I_\mathrm{mat}=V_\mathrm{mat}$, where $M^{-1}$ is the inverse of $M$, and obtain the potentials in the 14 terminals as a function of $m$ and $n$. Using the Landauer-Büttiker formalism, we acquire the equations 1 and 2 in the main text for two- and four-terminal conductance at the Dirac point where $m=1$ and $n=1$, to consider for a co-presence of an electron-like and hole-like band.    \\

%\textbf{\large{TO-DO}
%\begin{itemize}
%    \item Figure showing the Dirac curve for extracting the mobility. 
%    \item Show the back-gate dependence
 %   \item Show how n vs. Vg is extracted
 %   \item Rewrite all numerical values with units using the SI unit package, such as $B_\mathrm{z} = \SI{9}{T}$, \SI{1e16}{m^{-2}} etc.
%\end{itemize}

\textbf{\textcolor{black}{\large{15.~DFT calculations of the graphene band structure in the proximity of CrPS$_4$}}}\\

%------------------------------------------------------------
%------------------------------------------------------------
\subsection{Structural Setup}
%------------------------------------------------------------
%------------------------------------------------------------------------
\begin{figure}[!htb]
	\includegraphics[width=0.5\columnwidth]{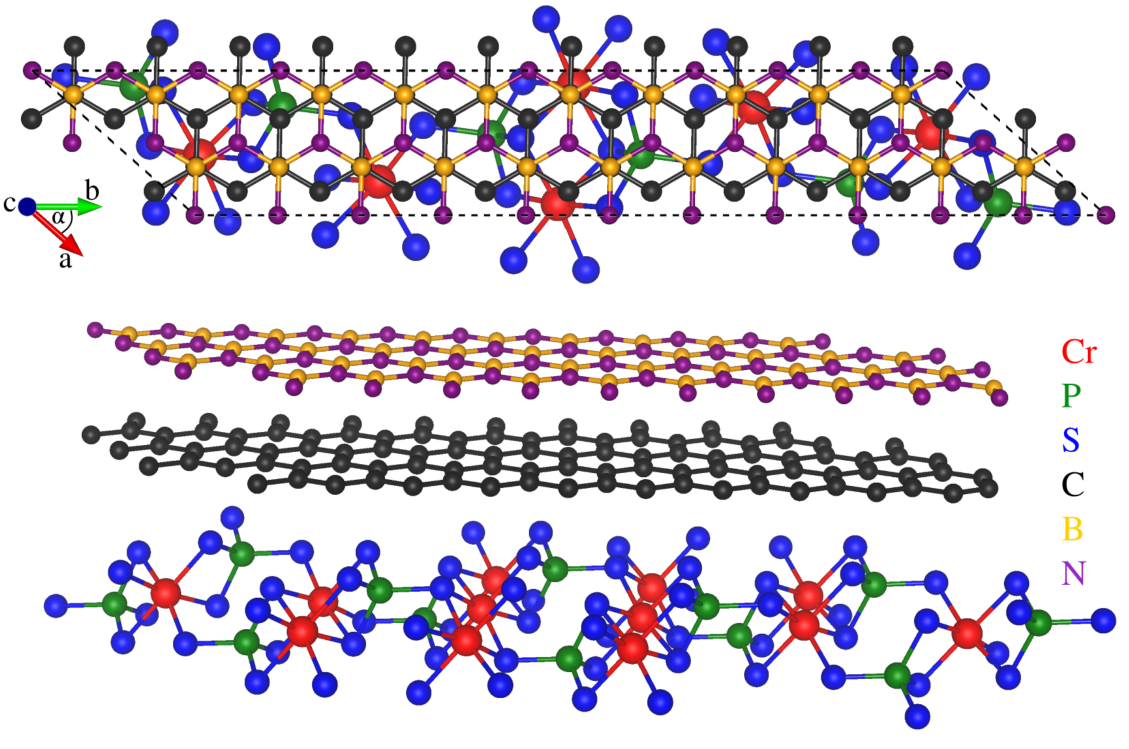}
	\caption{Top and side view of the CrPS$_4$/graphene/hBN heterostructure. The supercell has 124 atoms, with the lattice parameters $|a|=6.562$~\AA, $|b|=27.453$~\AA, $|c|=34.461$~\AA, and $\alpha = 41.653^{\circ}$. The average interlayer distance between C atoms of graphene and outermost S atoms of CrPS$_4$ is $d = 3.418$~\AA, while the interlayer distance between graphene and hBN is $d = 3.346$~\AA. 
 \label{Fig:Structure2}}
\end{figure}
%------------------------------------------------------------------------
 The CrPS$_4$/graphene/hBN heterostructure was set up with the {\tt atomic simulation environment (ASE)} \cite{ASE} and the {\tt CellMatch} code \cite{Lazic2015:CPC}, implementing the coincidence lattice method \cite{Koda2016:JPCC,Carr2020:NRM}. The lattice parameters of CrPS$_4$ within the heterostructure are about $a = 7.336$~\AA~and $b = 10.882$~\AA, following the literature values~\cite{Lee2017:ACS}, while the graphene and hBN lattices are minimally deformed with the new lattice parameters of about $a = 2.5341$~\AA, $b = 2.4957$~\AA~ and $\gamma = 59.362^{\circ}$. In order to simulate quasi-2D systems, we add a vacuum of about $24$~\AA~to avoid interactions between periodic images in our slab geometry. The resulting heterostructure is shown in Fig.~\ref{Fig:Structure2}. 

 %------------------------------------------------------------
\subsection{Computational Details}
%------------------------------------------------------------

The electronic structure calculations and structural relaxations of the CrPS$_4$/graphene/hBN heterostructure are performed by DFT~\cite{Hohenberg1964:PRB} 
with {\tt Quantum ESPRESSO}~\cite{Giannozzi2009:JPCM}. Self-consistent calculations are carried out with a $k$-point sampling of $36\times 12\times 1$. 
We perform open shell calculations that provide the
spin-polarized ground state of the CrPS$_4$ monolayer. 
A Hubbard parameter of $U = 2.0$~eV is used for Cr $d$-orbitals \cite{Joe2017:JP,Zhuang2016:PRB2}.
We use an energy cutoff for charge density of $560$~Ry and the kinetic energy cutoff for wavefunctions is $70$~Ry for the (scalar) relativistic pseudopotentials
with the projector augmented wave method~\cite{Kresse1999:PRB} with the 
Perdew-Burke-Ernzerhof exchange correlation functional~\cite{Perdew1996:PRL}.
For the relaxation of the heterostructures, we add DFT-D2 vdW corrections~\cite{Grimme2006:JCC,Grimme2010:JCP,Barone2009:JCC} and use 
quasi-Newton algorithm based on trust radius procedure. 
Dipole corrections \cite{Bengtsson1999:PRB} are also included to get correct band offsets and internal electric fields.
 To get proper interlayer distances and to capture possible moir\'{e} reconstructions, we allow all atoms to move freely within the heterostructure geometry during relaxation. 
The graphene/hBN interlayer distance is kept fixed. Relaxation is performed until every component of each force is reduced below $1\times10^{-3}$~[Ry/$a_0$], where $a_0$ is the Bohr radius.

We note that due to the lattice mismatch between graphene and CPS, there is finite deformation of the lattices indicative of the presence of a finite uniaxial strain in graphene. The strength of the strain is dependent on the twist angle between the two lattices. In the DFT model, the crystallographic alignment is considered such that the strain on the layers is minimized.   
 %------------------------------------------------------------
\subsection{Results I: spin-polarized case without spin-orbit coupling}
%------------------------------------------------------------

\begin{figure}[!htb]
	\includegraphics[width=0.7\columnwidth]{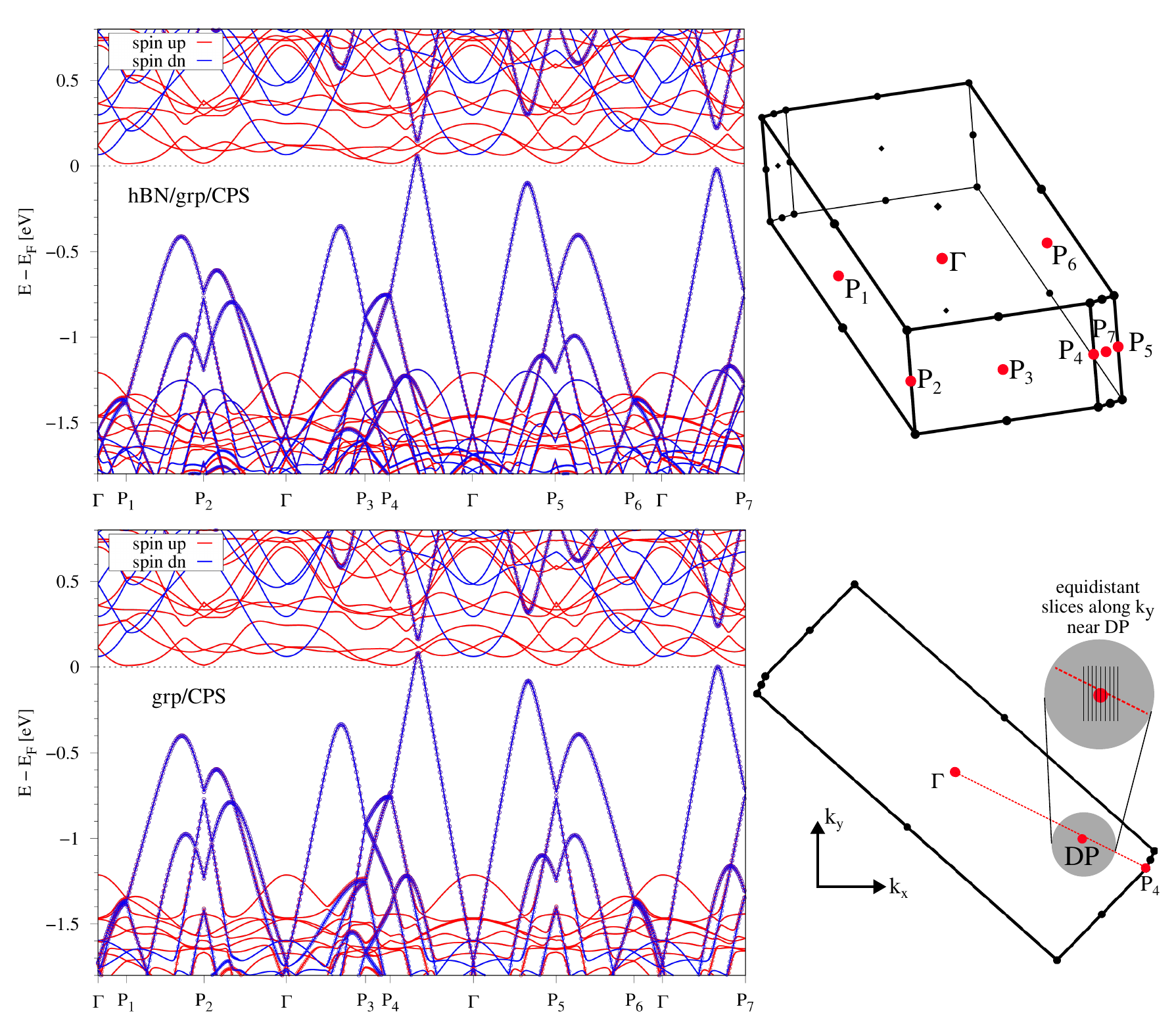}
	\caption{DFT-calculated band structure of CrPS$_4$/graphene/hBN (top) and CrPS$_4$/graphene (bottom) heterostructures. Red (blue) lines correspond to spin up (down) and the open spheres are projections onto graphene states. Right: The first Brillouin Zone, where we indicate the selected $k$ points. The Dirac point (DP) is located near the $\Gamma$-P$_4$ line as sketched. 
 \label{Fig:Bands2}}
\end{figure}

In Fig.~\ref{Fig:Bands2} we show the calculated band structure of the CrPS$_4$/graphene/hBN and the CrPS$_4$/graphene heterostructures. The latter case is constructed by deleting the hBN layer of the initial structure. 
The semiconducting nature of CrPS$_4$ and the Dirac states of graphene are preserved within the heterostructure. However, there is quite some charge transfer present, as the Dirac point is located within the conduction band states of CrPS$_4$ and about 120~meV above the heterostructure Fermi level. Within the heterostructure, the Dirac point gets folded near the $\Gamma$-P$_4$ line as sketched in Fig.~\ref{Fig:Bands2}. The overall global band structure is not much affected by the hBN layer and therefore we focus on the geometry without the hBN.

\begin{figure}[!htb]
	\includegraphics[width=0.6\columnwidth]{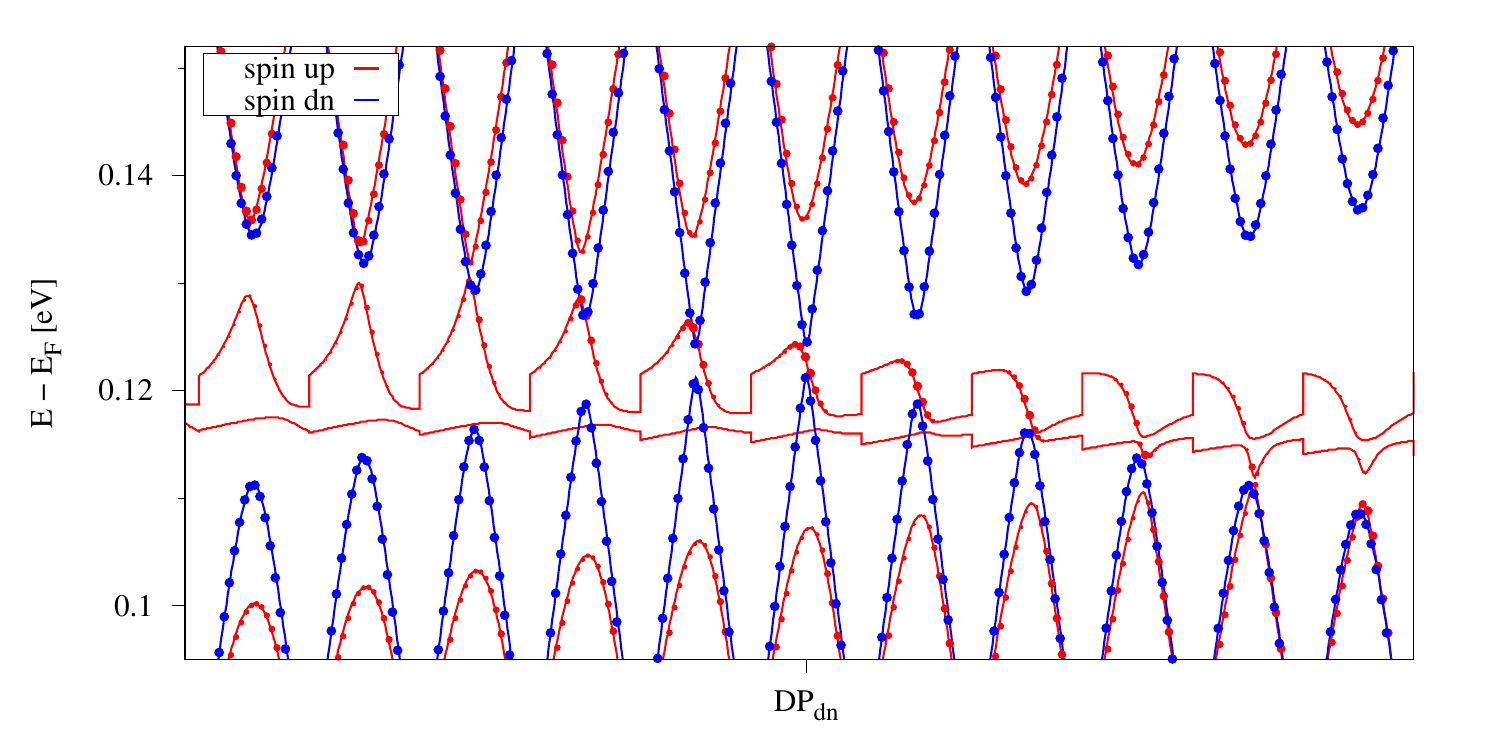}
	\caption{Band structure of CrPS$_4$/graphene calculated for equidistant slices along $k_y$ direction, as sketched in Fig.~\ref{Fig:Bands2}, in the vicinity of the Dirac point (DP). Red (blue) lines correspond to spin up (down) and the solid spheres are projections onto graphene states. The Dirac point of the spin-down component is identified as DP$_{\textrm{dn}}$. The spin-up Dirac states hybridize with the substrate bands and the corresponding Dirac point cannot be uniquely identified.
 \label{Fig:Dirac_bands2}}
\end{figure}

In the following, we take equidistant slices along $k_y$ direction in the vicinity of the Dirac point, in order to resolve the low energy Dirac bands and possible proximity effects. The resulting band structure is shown in Fig.~\ref{Fig:Dirac_bands2}. 
We find that the spin-down Dirac states are not affected by any substrate bands, while the spin-up Dirac states strongly hybridize with substrate bands. This creates a strong spin-splitting in the graphene band structure near the Dirac point. However, due to the strong hybridization, a model Hamiltonian description of the proximitized Dirac bands is not possible.

\begin{figure}[!htb]
	\includegraphics[width=0.6\columnwidth]{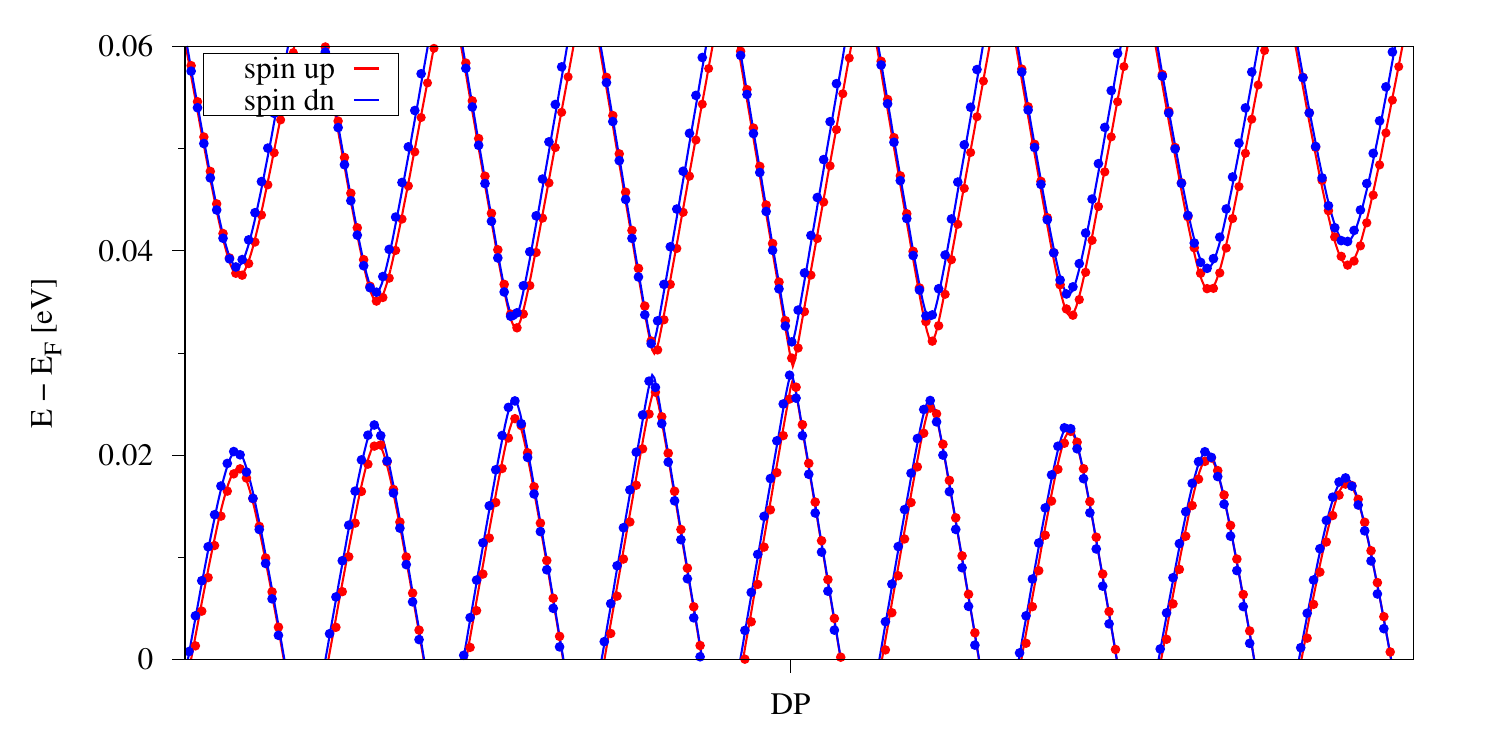}
	\caption{Band structure of CrPS$_4$/graphene calculated for equidistant slices along $k_y$ direction, as sketched in Fig.~\ref{Fig:Bands2}, in the vicinity of the Dirac point (DP). Here, we have applied a transverse electric field of $-1$~V/nm across the heterostructure to shift the Dirac states away from the hybridizing substrate bands. Red (blue) lines correspond to spin up (down) and the solid spheres are projections onto graphene states. 
 \label{Fig:Dirac_bands_Efield}}
\end{figure}

\begin{figure}[!htb]
	\includegraphics[width=0.3\columnwidth]{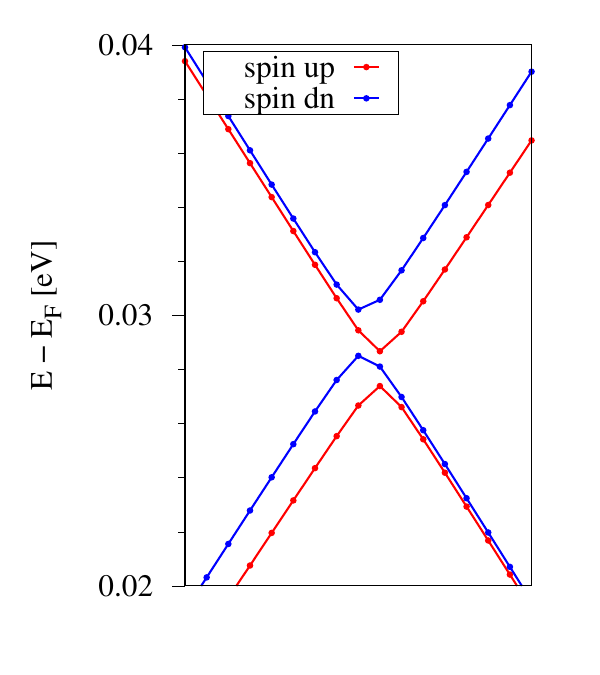}
	\caption{Same as Fig.~\ref{Fig:Dirac_bands_Efield}, presented with greater detail. From the band energies, we can extract a staggered potential of about 0.74~meV, and proximity exchange coupling parameters for the conduction (valence) band of 0.75~meV (0.55~meV).
 \label{Fig:Dirac_bands_Efield_detail}}
\end{figure}

We apply a transverse electric field of $-1$~V/nm across the heterostructure to shift the Dirac states away from the hybridizing substrate bands. The resulting low-energy Dirac states are shown in Fig.~\ref{Fig:Dirac_bands_Efield}. With the field applied, the Dirac point is about 50~meV above the heterostructure Fermi level and well separated from any hybridizing substrate bands. Due to symmetry breaking, spin-up and spin-down Dirac bands are shifted in momentum space with respect to each other.
This can be seen in greater detail in Fig.~\ref{Fig:Dirac_bands_Efield_detail}, which has been seen also in other graphene/2D-magnet systems, where symmetry breaking appears~\cite{Zollner2022:PRB}. From the dispersion, we can extract proximity-induced spin splittings on the order of 0.5-2~meV.

%------------------------------------------------------------------------
%------------------------------------------------------------------------

%------------------------------------------------------------------------
%------------------------------------------------------------------------

%------------------------------------------------------------------------
%------------------------------------------------------------------------

%------------------------------------------------------------------------
%------------------------------------------------------------------------
\newpage

 %------------------------------------------------------------
\subsection{Results I: spin-polarized case with spin-orbit coupling}
%------------------------------------------------------------

\begin{figure}[!htb]
	\includegraphics[width=0.7\columnwidth]{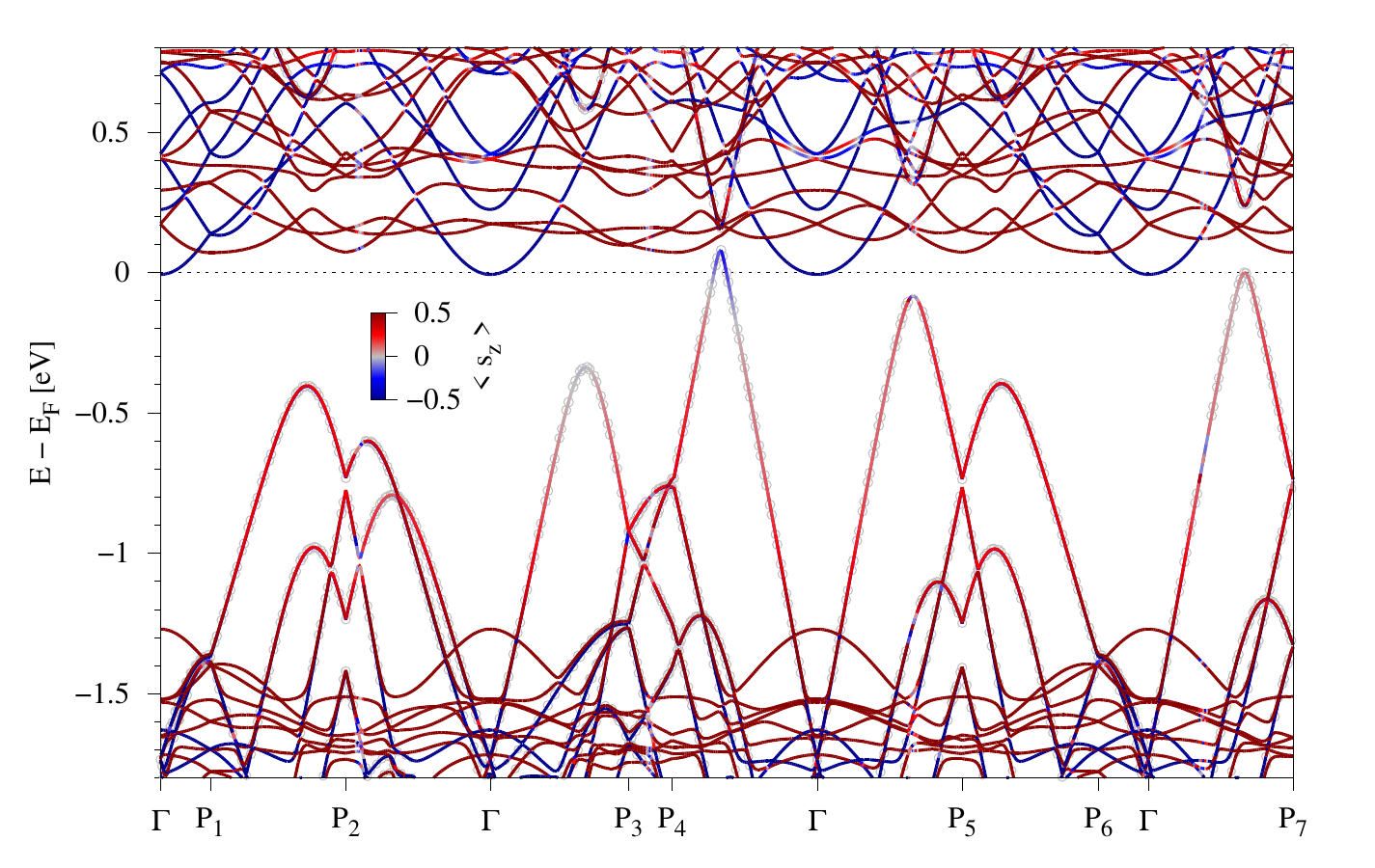}
	\caption{DFT-calculated band structure of CrPS$_4$/graphene heterostructure with spin-orbit coupling. The open spheres are projections onto graphene states, while the color code represents the $s_z$ spin expectation value. 
 \label{Fig:Bands_SOC}}
\end{figure}

\begin{figure}[!htb]
	\includegraphics[width=0.7\columnwidth]{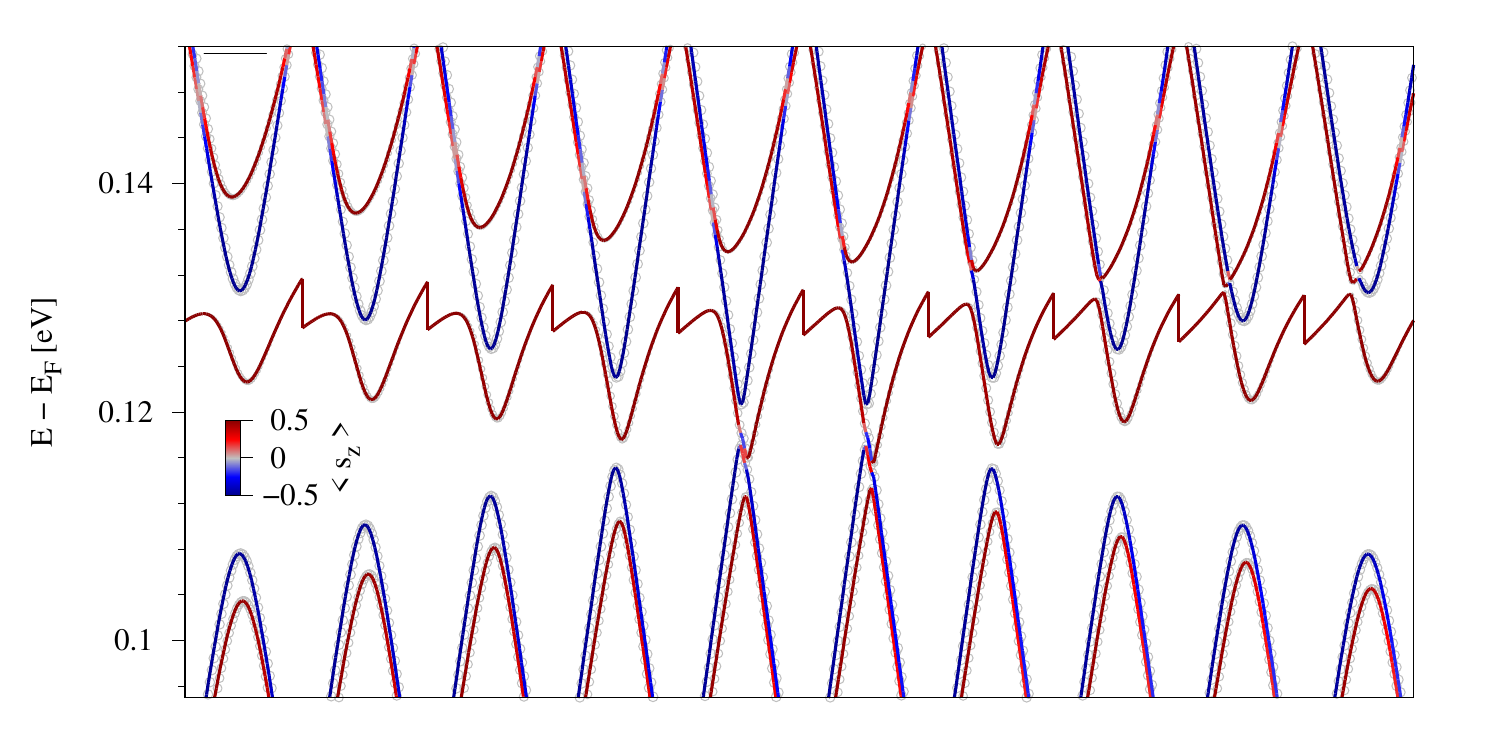}
	\caption{Band structure with spin-orbit coupling of CrPS$_4$/graphene calculated for equidistant slices along $k_y$ direction, as sketched in Fig.~\ref{Fig:Bands2}, in the vicinity of the Dirac point. 
 \label{Fig:Dirac_bands3}}
\end{figure}

\begin{figure}[!htb]
	\includegraphics[width=0.7\columnwidth]{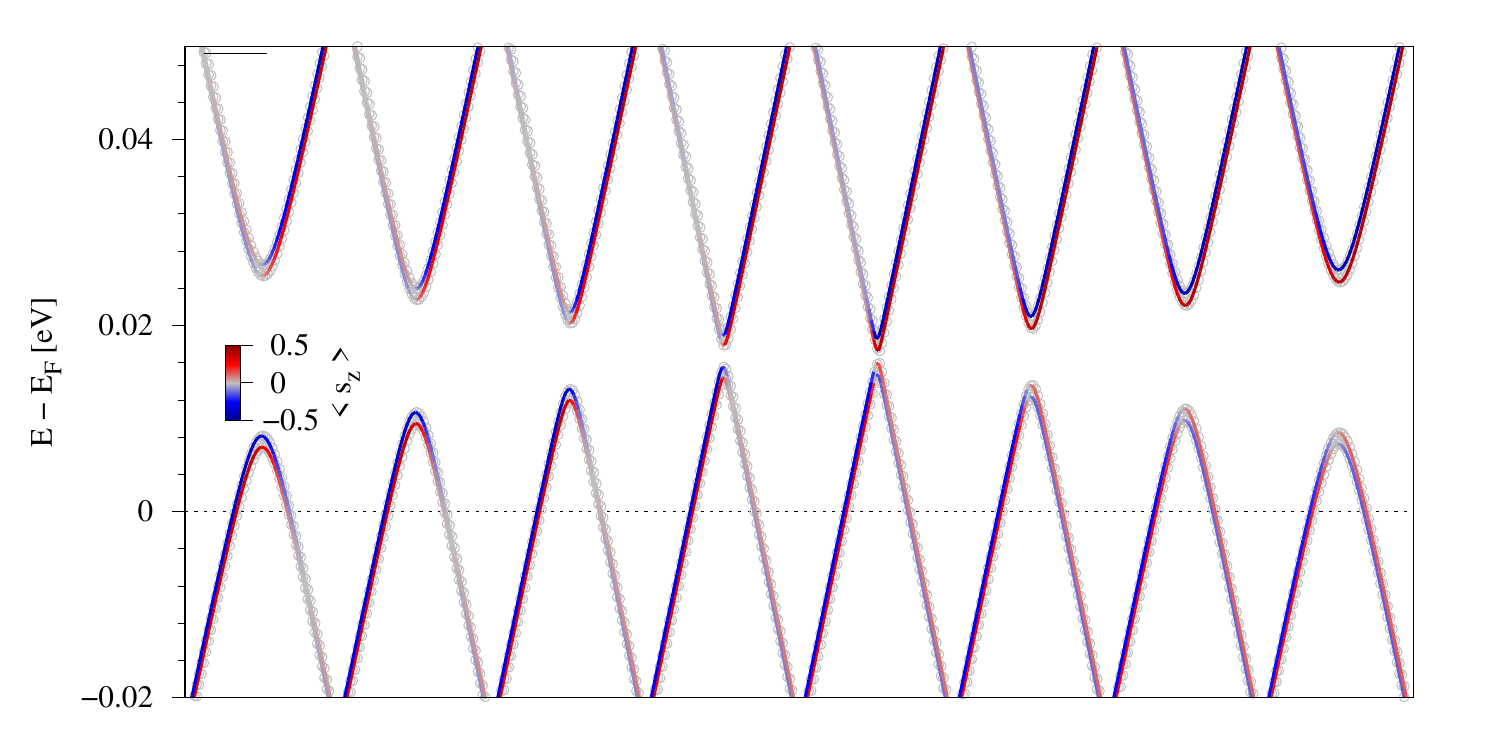}
	\caption{Band structure with spin-orbit coupling of CrPS$_4$/graphene calculated for equidistant slices along $k_y$ direction, as sketched in Fig.~\ref{Fig:Bands2}, in the vicinity of the Dirac point. Here, we have applied a transverse electric field of $-1$~V/nm across the heterostructure to shift the Dirac states away from the hybridizing substrate bands.
 \label{Fig:Dirac_bands4}}
\end{figure}

Next, we include the effects of spin-orbit coupling and recalculate the dispersion. The results can be seen in Fig.~\ref{Fig:Bands_SOC} and in Fig.~\ref{Fig:Dirac_bands3}. Similar to the case without spin-orbit coupling, the Dirac bands strongly hybridize with substrate states, which permits a reliable extraction of proximity spin-orbit coupling-related band splittings. Therefore, we again apply an electric field of -1~V/nm to separate the Dirac and substrate bands in energy. The result can be seen in Fig.~\ref{Fig:Dirac_bands4}.
We realize that the hybridization is strongly dependent on the transverse electric field, as the field pushes the Dirac point further away from the conduction band of the CrPS$_4$ (towards the bandgap), which diminishes the spin-splitting.

\begin{figure}[!htb]
	\includegraphics[width=0.6\columnwidth]{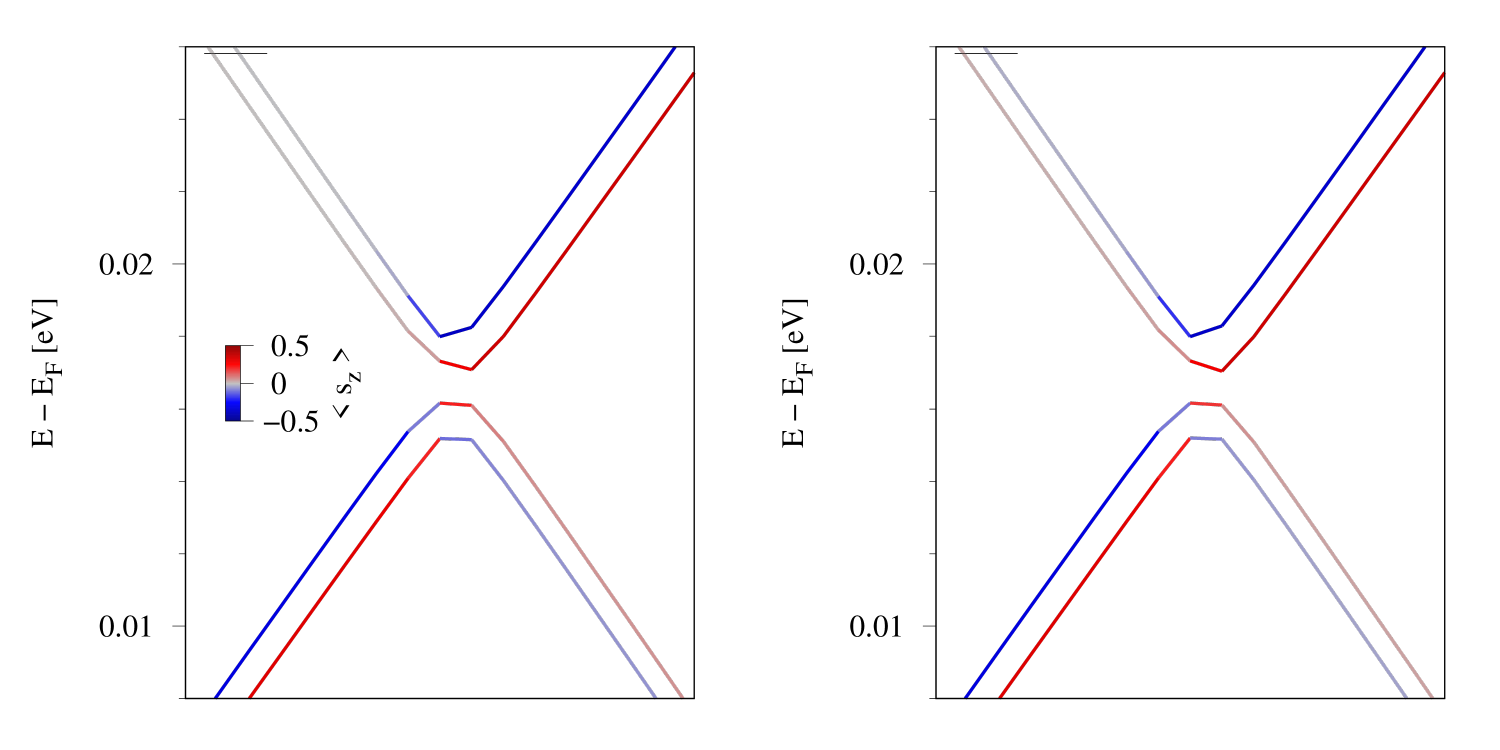}
	\caption{Same as Fig.~\ref{Fig:Dirac_bands4}, but with greater detail. Left (Right) band structure corresponds to the K (K') valley. Comparing the bands with the ones without spin-orbit coupling, see Fig.~\ref{Fig:Dirac_bands_Efield_detail}, we can estimate the Rashba coupling to be about 0.4~meV.
 \label{Fig:Dirac_bands5}}
\end{figure}

Since time-reversal symmetry is broken, when spin-orbit coupling and magnetism are combined, both graphene valleys, K and K', become important in the analysis of the Dirac bands. In Fig.~\ref{Fig:Dirac_bands5}, we show a detailed view of the bands near DP (see Fig.~\ref{Fig:Bands2}), which say corresponds to the K valley, and at the states near -DP, which corresponds to the K' valley. We find that the low-energy Dirac bands look nearly identical at the two valleys, indicating that graphene experiences proximity exchange combined with a Rashba coupling. Note that it is not excluded to find finite Kane-Mele spin-orbit coupling in these heterostructures at other twist angles than the one considered for this calculation. While the proximity spin exchange physics is well described by Zeeman-like Hamiltonian terms, the phenomenological model description based on the simple functional forms offered by k-independent Rashba, Kane-Mele, and valley-Zeeman spin-orbit fields
appears too simplistic to realistically capture the spin-orbit physics of this structurally nontrivial hybrid-lattice heterostructure. More advanced modeling, including higher-order (in momentum) terms, could provide further insight into the physics described in the experiment.

%------------------------------------------------------------------------
%------------------------------------------------------------------------

%------------------------------------------------------------------------
%------------------------------------------------------------------------

%------------------------------------------------------------------------
%------------------------------------------------------------------------

%------------------------------------------------------------------------

\end{document}